\documentclass[letterpaper,11pt]{article}

\usepackage{jheppub}
\usepackage{graphicx} 
\usepackage{amsmath}
\usepackage{amssymb}
\usepackage{xspace}

\newcommand{\nn}{\nonumber}

\newcommand{\beq}{\begin{equation}}
\newcommand{\eeq}{\end{equation}}
\newcommand{\bea}{\begin{eqnarray}}
\newcommand{\eea}{\end{eqnarray}}

\newcommand{\e}{\epsilon}

\newcommand{\oXn}{\mathbb{X}_{\hat n}}
\newcommand{\Xn}{X_{\hat n}}
\newcommand{\oPpb}{ \bar{\mathbb P}_{\!\! n \! \perp} }
\newcommand{\oPp}{ {\mathbb P}_{\!\! n \! \perp} }
\newcommand{\sceti}{SCET$_{\rm I}$}
\newcommand{\scetii}{SCET$_{\rm II}$}
\newcommand{\half}{\frac{1}{2}}
\newcommand{\slashed}{\hspace{-0.4cm}\not\,\,\,}
\newcommand{\LN}{\text{ln}}
\def\nslash{n\hspace{-2mm}\slash}
\def\nbarslash{\bar n\hspace{-2mm}\slash}
\def\nslashinline{n\!\!\!\slash}
\def\nbarslashinline{\bar n\!\!\!\slash}
\def\nbar{\bar n}
\def\dd {{\rm{d}}}

\newcommand{\plusf}[2]{\left [ \frac{#1}{#2} \right ]_{\!+}}
\newcommand{\infplus}[2]{\left[\frac{#1}{#2}\right]_{\!+}^{\!^\infty}}
\newcommand{\sqrts}{\sqrt{s}}

\allowdisplaybreaks[2]

\title{ A Formalism for the Systematic Treatment  of Rapidity Logarithms in Quantum Field Theory}

\author[a]{Jui-yu Chiu,}
\author[a]{Ambar Jain,}
\author[a]{Duff Neill,}
\author[a,b]{Ira Z. Rothstein}

\affiliation[a]{Department of Physics, Carnegie Mellon University, Pittsburgh, PA~15213, U.S.A.}
\affiliation[b]{California Institute of Technology, Pasadena CA, 91125, U.S.A.}

\emailAdd{jychiu@andrew.cmu.edu}
\emailAdd{ambar@andrew.cmu.edu}
\emailAdd{dneill@andrew.cmu.edu}
\emailAdd{izr@andrew.cmu.edu}

\abstract{

Many observables in QCD rely upon the resummation of perturbation theory to retain predictive power. Resummation
follows after one factorizes the cross section into the relevant modes. The class of  observables which are sensitive to
soft recoil effects  are particularly challenging to factorize and resum since they involve rapidity logarithms.
Such observables include: transverse momentum distributions  at $p_T$  much less then the high
energy scattering scale, jet broadening,  exclusive hadroproduction and decay, as well as the  Sudakov form factor.
In this paper we will present
a formalism which allows one to factorize and resum the perturbative series for such observables in a systematic fashion through the notion of 
a ``rapidity renormalization group".  That is, a Collin-Soper like equation is realized as a renormalization group equation, but has a more universal applicability to observables beyond the traditional transverse momentum dependent parton distribution functions (TMDPDFs) and the Sudakov form factor. This formalism has the feature that it allows one to track the (non-standard)  scheme
dependence which is inherent  in any scenario where one performs a resummation of rapidity divergences. 
We present a pedagogical introduction to the formalism by applying it to the well-known massive Sudakov form factor.
The formalism is then used to study observables of current interest. A factorization theorem for the transverse momentum distribution of Higgs production is presented along with the result for the resummed cross section at NLL.  Our  formalism allows one to define gauge invariant  TMDPDFs  which
are independent of both the hard scattering amplitude and the soft function, i.e. they are universal.
We present details of the factorization and resummation of the jet broadening cross section 
including a renormalization in $p_\perp$ space.
We furthermore show how to regulate and renormalize  exclusive processes which are plagued
by endpoint singularities in such a way as to allow for a consistent resummation.

}

\begin{document}
{\flushright INT-PUB-11-057 \\
CALT 68-2864
\\[-6ex]}

\maketitle

\section{Introduction and Motivation}
Many observables in high energy collisions suffer from poorly behaved perturbative expansions due to the existence of large logarithms. Logarithms of fixed scales, such as masses, are easily handled by standard renormalization group procedures within the confines of effective field theories. However, when one is interested in less inclusive observables, it is often the case that large logarithms of kinematic factors can arise. In particular, when one is interested in studying corners of phase space large hierarchies can induce large logarithms. In such cases  resummations allow us to maintain control over theoretical errors \cite{Sterman:1995fz}.

Within the vast class of observables which require resummation there is a large sub-class which are technically more challenging to handle than others. These observables correspond generically to those for which the recoil of a collinear jet due to soft radiation is an order one effect. We will use the acronym SRSO for Soft Recoil Sensitive Observables. A classic example of an SRSO is jet broadening \cite{Dokshitzer:1998kz,Catani:1992jc}. 
Another set of SRSO's  are transverse momentum distributions when $p_T/Q\ll 1$, where\footnote{We will always assume that $p_T\gg \Lambda_{QCD}$, so we are away from the ``forward region".} logarithms of the ratio $p_T/Q$ can invalidate fixed order results.

In this paper we will be employing effective field theory (EFT) techniques to factorize and resum large logarithms in these SRSO's.  Traditional  EFT's allow for resummation in a systematic fashion by morphing these large logarithms into logarithms associated with UV divergences and then applying standard renormalization group techniques. However, such traditional methods where one sums logs of invariant mass scales, are insufficient for SRSO's. Indeed, as we will see below for such observables, not all the large logarithms are associated with UV divergences in the effective theory.

Standard SCET methods breakdown when the modal decomposition in the EFT involves multiple fields with the same invariant mass scalings (such theories fall under the rubric of what is known as \scetii\, \cite{Bauer:2002aj}). These cases exactly correspond to SRSO's since the soft radiation, in light-cone coordinates, has momentum scaling $(\lambda,\lambda,\lambda)$, where $\lambda$ is the small power counting parameter, while the collinears scale as ($1,\lambda^2,\lambda$). For such observables one can run into a new type of divergence which is not associated with singular behavior in the  UV or IR, but with limits of large rapidities, as was elucidated in \cite{Manohar:2006nz}. These ``rapidity divergences'' have been studied extensively outside the realm of EFT, albeit perhaps with differing nomenclature, especially within the context of transverse momentum distribution functions and Sudakov form factor \cite{Collins:1992tv,Collins:1981uk,Collins:2008ht}.

 The purpose of this paper is to present  a renormalization group program for the treatment of rapidity logarithms.  In particular,  we show how one can regulate and renormalize rapidity divergences, and then using a renormalization group technique resum the associated logarithms, all in a systematic fashion within SCET. To facilitate this procedure we introduce a regulator  that  necessarily breaks boost invariance in order to distinguish between modes which share a mass shell hyperbola. The regulator preserves eikonal exponentiation and manifest gauge invariance in each sector.
 However, our formalism is also applicable using other regulators which break boost invariance, as will be discussed below.
  Once one sums over the soft and collinear sectors, the rapidity divergences cancel at each order in perturbation theory. The regulator introduces a new scale, $\nu$, which leads to a rapidity renormalization group (RRG) flow. The solution to the ensuing differential equation has the effect of summing the  large rapidity logarithms which existed in the full theory calculation.

\subsection*{Outline of  this paper}
We begin in (\ref{def}) by defining the notion of rapidity logarithms (divergences) and present the necessary and sufficient criteria
for their existence.  In (\ref{sec:intro}) we  present a physical argument which allows us to isolate the type
of generic observable where one expects them to arise. In (\ref{rapdiv}) we demonstrate how to regulate  the rapidity divergences in 
context of the massive Sudakov form factor with massless external lines. This section also illustrates how 
one can resum the rapidity logarithms using the rapidity renormalization group (RRG).  We then show how the regulator
is applied to generalized soft and collinear jet functions. In (\ref{Higgs}) we apply the formalism to the transverse momentum
distribution in Higgs production at small $p_T$, resumming logarithms of $p_\perp/m_h$ to next to leading log order.
This section includes the definition of a gauge invariant transverse momentum dependent parton distribution function (TMDPDF).
At the end of this section we compare our results with previous works on the subject.  
Section (\ref{JB}) gives a factorization theorem for jet broadening including a next to leading log result for
the resummed cross section.  Since we have attempted to make this section available to readers not interested in the Higgs production section there is some formal overlap with the previous section.
Finally in (\ref{ex}) we show how our formalism can be utilized to renormalize end-point divergences 
in exclusive decays. 
We conclude with a summary.

\section{Rapidity Divergences}
\label{def}
We define a rapidity divergence as arising from momentum region where the invariant mass  $k^2$ is held fixed but the ratio $k_{+}/k_{-}$ (or $k_{-}/k_{+}$ ) diverges, where $k_\pm$ are light-cone momenta. Rapidity divergences  are not IR in origin, as they do not show up in the full theory, nor should they be thought of as UV divergences since they can arise from either the upper or lower limit of an integral, as will be shown later. The existence of the divergence stems from the fact that to preserve manifest power counting the EFT must be multipole expanded \cite{Grinstein:1997gv} which leads to eikonal propagators. Furthermore, while rapidity divergences arise in factorized IR sectors of the theory, i.e. collinear or soft, the sum of EFT sectors will have no rapidity divergences. Which is to say that  rapidity divergences  arise as an artifact of factorization. However, in order to resum logarithms in attempting to save perturbation theory, factorization, and thus rapidity divergences are inevitable for SRSO's. In this sense, they are entirely analogous to the traditional UV divergences which arise in factorization and are necessary for resummations.

 To demonstrate how rapidity divergences arise, let us consider an integral of the following form
 \beq
 I= \int_{\mu_L}^{Q} \frac{dk_+}{k_+},
 \eeq
which may arise when transverse momentum is measured in the real radiation. $Q$ is the scale of hard scattering and $\mu_L$ is the relevant low energy scale. Suppose that the $n$-collinear and  soft modes have $k_+/Q$  or order $1$  and $\lambda\sim \mu_L/Q$, respectively. Then this integral ranges over  both mode regions. To factorize the integral  into rapidity regions we introduce a set of cut-offs
 \beq
 \label{one}
 I= \int_{\mu_L}^{\Lambda} \frac{dk_+}{k_+}+ \int_{\Lambda}^{Q} \frac{dk_+}{k_+},
 \eeq
 corresponding to soft and collinear contributions respectively. In the effective theory,  the cut-off should not be finite, to preserve power counting. Or put differently, each sector should depend only upon one relevant scale, which follows after performing the multipole expansion. Taking the limit, $Q >> \Lambda >> \mu_L$, the EFT result reads
\beq
\label{two}
 I= \int_{\mu_L}^{\infty} \frac{dk_+}{k_+}+ \int_{0}^{Q} \frac{dk_+}{k_+},
 \eeq
We see that we generate a set of divergences which only cancel in the sum of the sectors.
It should also be emphasized here that these divergences are not regulated by dimensional regularization, a necessary
but not sufficient criteria for rapidity divergences.

 
Obviously not all observables in QCD will generate rapidity divergences in SCET. A necessary condition for their appearance is that the observable under consideration receive contributions from modes with parametrically distinct rigidities,  but whose invariant mass are of the same order. The prototypical observables of this type are transverse momentum distributions. When $p_T\ll Q$,  $p_T$ plays the role of $\mu_L$ in the example above and rapidity logarithms arise which must be resummed. As such, we will consider in this paper two observables, namely  Higgs production  at small transverse momentum, and Jet broadening. Rapidity divergences also occur in exclusive processes. In \cite{Manohar:2006nz} the authors show that  the end point singularity problem  arises as a consequence of rapidity divergences. We apply our formalism to this issue and show how one can systematically renormalize the divergences in these exclusive decays.

\section{What Theories Give Rise to Rapidity Divergences: \sceti\, vs. \scetii\, }
\label{sec:intro}
SCET \cite{Bauer:2000ew,Bauer:2000yr,Bauer:2001yt} (Soft Collinear Effective Theory) is  a formalism designed to separate scales in  high energy scattering processes  for which the  hard scattering scale ($Q^2$) is much greater then the scale of hadronic physics. Here we will not review SCET but only illuminate the points that are germane to the main thrust of this paper. In particular we are interested in the critical issue of (not) double counting regions of phase space. That is, how does one cleanly separate (factorize) the modes which give rise to IR singularities.

 As in any well defined EFT, the scale separation is  made manifest at the level of the action which systematizes the power counting. Power corrections can be  included by adding operators which have definite scalings in powers of $\mu_L/Q$, where $\mu_L$ is some low scale of interest. SCET, like its cousin NRQCD, is a ``modal" theory whereby fields are decomposed into a set of sub-fields each of which has momenta with  definite scalings.  For instance, in \sceti\, the gluon field is written as
 \beq
 \label{decomp}
 A_{\mu}= A^{c,n}_
 \mu + A^{c,\bar n}_ \mu 
 +A^{US}_\mu+....
 \eeq
 where $A^{c,(n,\bar n)}_\mu$  are collinear fields whose momenta  scale as $Q(1,\lambda^2,\lambda)$ and $Q(\lambda^2,1,\lambda)$ respectively. While $A^{US}_\mu$ is an ultra-soft (US)  field whose momentum scales as $Q(\lambda^2,\lambda^2,\lambda^2)$, where the power counting parameter is $\lambda\equiv \mu_L/Q$ and $\mu_L$ is the relevant low energy scale\footnote{The $...$ represent non-linear terms in the fields that are needed to insure that gauge transformations do not mix orders in the power counting\cite{Bauer:2003mga}.}. 
   To cleanly separate the scales it is helpful to use a dynamical label formalism  \cite{Luke:1999kz} also used in SCET \cite{Bauer:2000yr}. 
We re-write the full QCD field as 
\beq
\label{sum}
\psi(x)= \sum_{n\cdot p, p_\perp} e^{-i n \cdot p \,  \bar n \cdot x + i p_\perp \cdot x_\perp} \xi_{ n\cdot p, p_\perp} (x) \, .
\eeq
The purpose of this rescaling is to insure that all derivatives acting on $\xi$ scale as $\lambda^2$. Note that (\ref{sum}) is written as a sum not an integral. One tessellates the space of possible large momenta into bins whose dimensions scale with the size of the residual momentum  of order $\lambda^2$ .

The Lagrangian interactions can change the large momentum components of fields. In particular, collinear gluons can split, changing their large light cone momentum. This implies that there are loops in which one must sum over labels. It is then natural to ask  what happens in the label sum \footnote{The fact that these are sums and not integrals is a consequence of separating momenta into labels and residual momenta. This can be thought of as a grid in which labels give the coordinates of a box (or ``bin'') whose size is of order of the residual momentum. One can combine the sum over bins with the residual momenta integrals when performing loop calculations. See \cite{Rothstein:2003mp} for a discussion.} when one of the labels becomes parametrically small\footnote{In \scetii\, labels between modes can also overlap by becoming large.} and modes begin to overlap. The existence of these overlap regions, when not treated properly, obscures the physics underlying the effective theory calculations. This is perhaps simplest to see in NRQCD, where the existence of the overlaps leads to pinch singularities, as well as the inability to clearly distinguish between IR and UV divergences. These points were made clear in a seminal paper by Manohar and Stewart \cite{Manohar:2006nz}. In this paper the authors show that one may exclude these overlap regions   by taking a diagram involving a particular mode,  Taylor expanding it around the region of the complementary mode one is trying  to exclude, and subtracting this contribution from the original diagram.   Doing so eliminates the aforementioned problems. It has been shown that in \sceti\, this ``zero bin'' subtraction \cite{Lee:2006nr,Idilbi:2007yi,Idilbi:2007ff} is {\it often} equivalent to dividing by matrix element of Wilson lines in non-SCET perturbative QCD factorization formulae \cite{Collins:1989bt}, which can often be identified with the inverse of the soft-function in the factorization theorem. In this paper we will also have occasion to comment on the role of the what we term the ``soft-bin'' subtractions, and draw a distinction between the soft-bin and zero-bin, where by zero-bin we always mean an ultra-soft scaling (see appendix (\ref{soft_bin_subtractions})). 

 In \sceti\, all divergences can be regulated using  dimensional regularization and/or off-shellness, and all modes have distinct virtualities (collinear modes in differing directions being the exception \footnote{The soft zero bin will eliminate any such overlap. Given the non-existence of a soft mode in \sceti\, such an overlap is necessarily absent.}).  Zero bin subtractions are relatively simple to utilize to insure that there is no double counting. Thus we should expect that rapidity divergences should not be an issue in \sceti\,, and indeed this is  in fact the case. Note this is not to say that integrals of the form of (\ref{one}) will not arise in \sceti\,. In fact they are ubiquitous, however, they should not be interpreted as rapidity divergences, as they will cancel within each sector. That is to say, there can, and will be such divergences in the collinear (or US) sector, but when one sums over graphs, including zero bins, these divergences will cancel as they must {\it within each sector}. A classic example arises in the one loop correction to the parton distribution function (PDF). In this case in calculating a real correction one encounters an integral of the form
\beq
\int \frac{dz}{1-z},
\eeq
where $z$ is the momentum fraction carried by the struck parton. The integral diverges at the upper end point of integration, where the incoming parton and and the struck parton carry the same momentum. To regulate this integral one must introduce a new regulator since  dimensional regularization  is insufficient. Nonetheless the divergence arises when the emitted gluon goes ultrasoft, and thus by the Kinoshita, Lee and Nauenberg (KLN) theorem we expect it to cancel with the corresponding virtual diagram. Thus the divergence cancels within the collinear sector itself.
 Note  that in this case the zero bin did not play a role, since they actually cancel in this calculation, again by the KLN theorem. An illustrative example of the cancellation of divergences, which one might have thought were rapidity divergences in \sceti\, can be found in \cite{Jain:2011xz}. In summary, \sceti\, rapidity divergences do not pose a problem because there is no issue in distinguishing modes of identical virtuality.

For certain observables, such as transverse momentum  distributions with $p_{\perp}\ll Q$,  \sceti\, is not the proper effective theory. The reason is that we must account for real radiation with momenta that scale as $(p_\perp,p_\perp,p_\perp)$. Given that $p_\perp$ is the IR scale of the theory the collinear modes scale as either $(Q, \frac{p_\perp^2}{Q},p_\perp)$ or $(\frac{p_\perp^2}{Q},Q,p_\perp)$,   and both the soft and collinear mode have the same invariant mass. As such, these modes can be interchanged by a boosts and  the only real distinction between them is in their relative rapidities. In such cases,  the EFT is called \scetii\,. The need for \scetii\,  was first pointed out in the context of exclusive $B$ decays \cite{Bauer:2002aj} where it was necessary to introduce  a second effective theory   below \sceti .  The equality of invariant masses of the modes in \scetii\, leads to complications in the factorization of physical observables since one must break the boost invariance of operators to cleanly distinguish between sectors. As was first pointed out in \cite{Beneke:2003pa}, the process of factorization in \scetii\, can lead to additional divergences in sectors that can not be regulated by dimensional regularization, or off-shellness.  These divergences will not cancel within each sector as they do in \sceti. They will cancel only when we sum over sectors, but the lack of cancellation within a sector changes the RG structure of theory. In fact it is the lack of cancellation that allows for the resummation of large rapidity logarithms
.
 \subsection*{Boost Invariance in \sceti~ and \scetii  }
 
In SCET we make  a convenient  choice of frames  where relevant modes are (ultra)soft or collinear, breaking the full Lorentz symmetry of QCD. However, SCET is invariant under boosts along the light-cone direction. This residual boost symmetry of SCET is called RPI$_{\rm III}$ \cite{Manohar:2002fd}.  For problems relevant to \sceti~ this symmetry is preserved in each matrix element belonging to the collinear or ultrasoft sectors. The natural distinction between the collinear and ultrasoft sectors of \sceti~ comes from their parametrically differing invariant masses\footnote{Distinction between two collinear sectors arise essentially from different light-cone directions, so they need not have different invariant mass for distinction.}. Since dimensional regularization breaks dilatation symmetry, it is sufficient to distinguish between collinear and ultrasoft modes\footnote{An ultrasoft mode can be transformed into a collinear mode by a dilatation and a boost.}. Since dimensional regularization preserves boost invariance and is the only regularization required to separate modes in \sceti, each sector individually remains RPI$_{\rm III}$ invariant.
  
 In \scetii ~ there is no distinction  between the invariant masses of the soft and collinear modes and they can be  interchanged via boosts. Dimensional regularization does not distinguish between these modes, so  we  should introduce a regulator which accomplishes this goal by breaking   the boost invariance along the light-cone direction.  Due to the soft recoil, ``jet functions'' describing collinear radiation are not exactly aligned with the preferred light-cone direction used in factorization, in that they carry transverse momentum {\it w.r.t.} the light-cone direction\footnote{For example, jet functions in the jet broadening factorization theorem \cite{Chiu:2011qc} carry non-zero transverse momentum {\it w.r.t.} the thrust axis.}. A jet function will usually depend upon the transverse momentum and the large light cone component of the momentum $Q_{\pm}$ carried by all the collinear particles constituting the jet. While the transverse momentum is boost invariant, $Q_{\pm}$ is not and hence one shouldn't expect the jet function describing collinear radiation in a physical process to be boost invariant. In contrast, the jets in \sceti\, are aligned with the preferred light-cone direction since ultrasofts cannot recoil jets in transverse momentum.
 
In a back-to-back jet scenario for SRSO's, like the jet broadening event shape, a small boost will reduce the number of particles in one jet and increase the number of particles in the other jet while keeping the number density unchanged in the soft region, when averaged over all events. Thus it is expected that 
in a renormalized factorized cross-section boost invariance will be broken in each sector and will only be restored when all sectors are added. In problems pertaining to \sceti, boosts alone cannot interchange  ultrasoft and collinear excitations and  hence RPI$_{\rm III}$ must be preserved in each sector. We see  that in order to  factorize one must distinguish  between soft and collinear radiation, and  RPI$_{\rm III}$ must be broken via regularization of \scetii\, matrix elements. This point was emphasized in \cite{Becher:2003qh}.


\section{ Rapidity Divergences in \scetii\,}
\label{rapdiv}
To understand the nature of rapidity divergences we  consider how they arise in the effective theory. Thus we will begin by considering perhaps the simplest case where such divergences arise, the Sudakov form factor. In particular  we will renormalize and resum the logarithms in   the on-shell, massive gauge boson, space-like Sudakov form factor \cite{Collins:1989bt} \footnote{The logarithms in the Sudakov form factor are distinguished from ``Sudakov logarithms" which can arise in running currents in that they contain rapidity divergences.} \footnote{ The Sudakov form factor can be regulated by going off-shell in which case the systematics may change \cite{Korchemsky}.}.  As opposed to the massless form factor, which in isolation,  is unphysical, the massive case   is IR safe and in principle observable. 
This form factor  is also relevant to for summing large electro-weak corrections at energy far above the gauge boson masses.

We defined our power counting parameter as $\lambda\equiv M/Q$.
The factorization formula is composed of decomposition in terms of modes which can contribute to the
non-analytic structure of the matrix element.  The relevant modes for this observable are the soft with light-cone  momenta scaling as  $(\lambda,\lambda,\lambda)$, and  collinear and anti-collinear with momenta scaling as  $(1,\lambda^2,\lambda)$ and $(\lambda^2,1,\lambda)$ respectively.  There are no ultra-soft
contributions since their momentum $(\lambda^2,\lambda^2,\lambda^2)$ will decouple from all the other lines.
Given these modes, this defines an \scetii\, process. The jets will recoil against
the soft virtual emissions and hence, despite the exclusive nature of this process, it is still an SRSO. However,
in the Breit frame the net transverse momentum exchanged between the jets must vanish.


The factorization of the massless Sudakov form factor in SCET  was performed in  \cite{Bauer:2010cc}\footnote{This  factorization was for the massless case where gauge invariance uniquely fixes the form of the factorization theorem.
However, the gauge boson mass does not alter the result. Note that this result was  formal in that  the IR scales were not clearly delineated. As such, it was not sharply defined to be living in \sceti~ or \scetii and the factorization formula contained both soft and ultra-soft Wilson lines.}. At leading power we have
\begin{eqnarray}
\label{scetII_current_form_factor}J_{\mu}\equiv \bar u(p_{n}) \gamma_{\mu}^{\perp} u(p_{\bar n})\, F(Q^2,M^2) &\approx &   \langle p_{n}| \bar \xi_n W_n S^\dagger_n \gamma^\perp_\mu C(n \cdot {\cal P} , \nbar \cdot {\cal P})S_{\nbar} W^\dagger_{\nbar} \xi_{\nbar}|p_{\bar n} \rangle\\ \nn
&\equiv& H(Q^2,\mu) J_n(M;\mu,\nu/Q)\gamma_{\mu}^\perp J_{\bar n}(M;\mu,\nu/Q) S(M;\mu,\nu/M) \, . \nn
\end{eqnarray}
In the last line, we have factorized the form factor $F$ in terms of \scetii~ matrix elements  $J_n,J_{\nbar},S$
which are defined as
\bea
\label{defs}
S(M;\mu,\nu/M)&=& \langle 0 \mid S^\dagger S \mid 0 \rangle \nn \\
J_n(M;\mu,\nu/Q)&=&\langle p_{n}| \bar \xi_n W_n |0 \rangle \nn \\
J_{\nbar}(M;\mu,\nu/Q)&=&  \langle 0 | \bar  W^\dagger_{\nbar} \xi_{\nbar}\mid p_{\bar n} \rangle
\eea
$S$ and $W$ are Wilson lines  composed of soft and collinear lines respectively.

Finally we must mention the cumbersome issue of the Glauber modes. These modes, which scale as $(\lambda^2,\lambda^2,\lambda)$ will in general contribute IR singular pieces at the level of the amplitude.
While they have been shown to cancel in certain processes  outside the realm of effective field theory \cite{Collins:1988ig,Bodwin:1984hc},
a systematic treatment of such modes \footnote{For a discussion of these modes within the context of SCET see \cite{Bauer:2010cc,Idilbi:2008vm,D'Eramo:2010xk}.} within a self-consistent EFT treatment is still lacking. Here we will assume, as do all SCET
treatments, that Glaubers will not contribute. 

Let us now see how factorization of the soft from collinear modes leads to rapidity divergences.
Consider the full theory one loop vertex correction. The relevant scalar integral is given by
\beq
I_f= \int [d^nk] \frac{1}{(k^2-M^2)}\frac{1}{(k^2-n\cdot k \nbar \cdot p_1+i \epsilon)}\frac{1}{(k^2-\nbar\cdot k n \cdot  p_2-i \epsilon)}
\eeq

This integral is finite in UV as well as the IR. In the effective theory  there are three contributions. A soft integral coming from taking the limit $k^\mu \rightarrow (M,M,M)$
\beq
I_S= \int [d^nk] \frac{1}{(k^2-M^2)}\frac{1}{(-n\cdot k+i \epsilon)}\frac{1}{(-\nbar\cdot k+i \epsilon)}
\eeq
and two collinear integrals  $(I_n,I_{\nbar} )$ of the form
\beq
I_n=\int [d^nk] \frac{1}{(k^2-M^2)}\frac{1}{(k^2- n \cdot k\, \nbar \cdot p_1+i \epsilon)}\frac{1}{(-\nbar\cdot k+i \epsilon)}.
\eeq

\begin{figure}
  \label{fig1}
    \centering
    \includegraphics[width=5 cm]{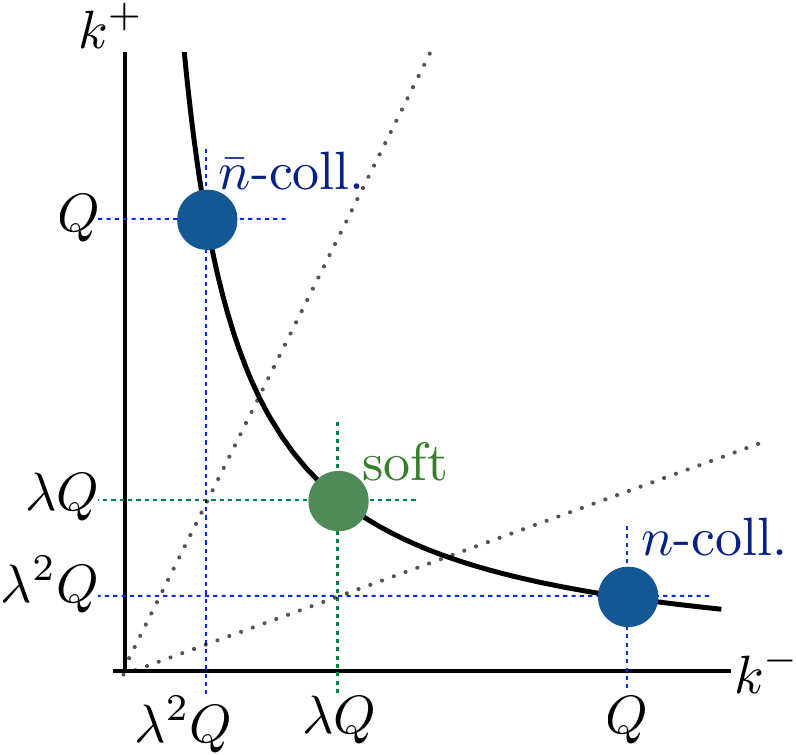}
\caption{The mass-shell hyperbolae showing the distinction between the different
sectors \cite{Manohar:2006nz}. The separation between soft and collinear modes is arbitrary and leads to
rapidity divergences.
 The soft sector has two distinct rapidity (UV) divergences
that must cancel with rapidity (IR) divergences arising from the collinear sector.
}
\label{binz}
\end{figure}

Given that the full theory graph is IR finite, so must be the sum of the effective theory graphs. Let us consider the soft graph integrating over $k_\perp$.
\bea
I_S&\sim & \int  [d^2 k] (n \cdot k \, \nbar \cdot k -M^2)^{-2 \epsilon}\frac{1}{(-n\cdot k+i \epsilon)}\frac{1}{(-\nbar\cdot k+i \epsilon)} \nn \\
\eea
We see that the relevant region of phase space lives on the hyperbola $n\cdot k\, \bar n \cdot k \sim M^2$, shown in figure 1.  Off the hyperbola the integral becomes scaleless. Given this restriction, we note that the integral diverges when the rapidity $(n\cdot k /\nbar \cdot k)$ approaches infinity or zero. These divergences are not regulated by dimensional regularization and correspond to the rapidity divergences that arise when  the soft integral overlaps with the two collinear rapidity regions. This is illustrated in figure (\ref{fig1}).
On the other hand,  if we consider the collinear $n$ diagram we see that it only has divergences associated with the limit where $(n \cdot k /\nbar \cdot k)$ approaches infinity, and similarly with $(n \rightarrow \nbar)$ for the $I_{\nbar}$ collinear integral, since there is only one border between a collinear sector and the neighboring soft sector.

There are multiple ways of regulating these rapidity divergences. One can go off the light cone by setting $n^2 \neq 0$ \cite{Collins:1981uk,Collins:1989bt}, use an analytic regulator \cite{Smirnov:1990rz}, or a ``delta''  regulator as was done in \cite{Chiu:2009yx}. Choosing a regulator determines how much of an overlap there is between modes. For instance, in the case of a delta regulator, where one shifts  the eikonal propagators 
\beq
\frac{1}{n \cdot k }\rightarrow \frac{1}{n \cdot k +\Delta}\, ,
\eeq
one must perform a soft-bin subtraction to generate the correct result in the effective theory. In fact, the authors of \cite{Chiu:2009yx} showed that the sum of the integrands, once properly soft-bin subtracted leaves a finite integral with no rapidity divergences. With an analytic regulator the soft function vanishes explicitly. In this case there is no double counting as half of the soft contribution comes from each of the collinear sectors, and thus there is no zero bin.

While physically it seems clear that a sensible rapidity regulator should cancel in the sum over sectors, we should have a proof of this assertion. A direct proof follows noting that if all of the regulated EFT diagrams arise from an asymptotic expansion of the full theory diagrams, then given that the full theory has no rapidity divergence the finiteness of the EFT sum then follows. By this reasoning  the delta regulator must also cancel in the sum over sectors, as will any rapidity regulator if we assume that it preserves the equality between the full theory  integrals and their asymptotic expansion in regions. It is important however, to recall that the method of regions is distinct from EFT in that, in the latter, it is not necessarily true that there is a one to one correspondence between the IR of a full theory diagram and a corresponding diagram in the EFT. All that is necessary is that the sum of the EFT graphs reproduces the IR of the full theory. Such cases arise when one uses the equations of motion in the effective theory to remove redundant operators.
\subsection{Regulating the Rapidity Divergences}
\label{RRD}
 Given that EFT's are created to sum logarithms we would like to be able to regulate the theory in a way that makes an  RG treatment manifest. 
 There are  multiple ways in which to regulate the rapidity divergences, and the formalism developed here can
 be applied using any sensible choice, such as the delta regulator \cite{Chiu:2009yx}.
 Here we will concentrate on the regularization introduced  
 in \cite{Chiu:2011qc}, where we utilized a rapidity regulator which is closely related to dimensional regularization. 
 It is implemented by  modifying the momentum space Wilson lines in the following fashion.
\beq
W_n=\sum_{\rm perms}  \exp\left[-\frac{g w^2}{\nbar \cdot { \cal P}} \frac{\mid \nbar \cdot {\cal P}_g\mid^{-\eta}}{\nu^{-\eta}}\nbar \cdot A_n \right]
\eeq
\beq
S_n=\sum_{\rm perms}  \exp\left[-\frac{gw}{ n \cdot  { \cal P}} \frac{\mid 2 {\cal P}_{g3}\mid^{-\eta/2}}{\nu^{-\eta/2}}n\cdot A_s \right]
\eeq
We have introduced a new dimensionful parameter $\nu$ which will play the role of an effective rapidity cut-off. Here ${\cal  P}^\mu$ is the momentum operator and we have essentially regulated the longitudinal momenta, and  since $\vert 2{\cal P}_3\vert  \to |\nbar \cdot {\cal P}|$ in the collinear limit. 
Note the differing powers of $\eta$ in the soft and collinear Wilson lines. 
The appropriate power is fixed by ensuring that the rapidity divergences cancel to all orders which we shall show below. Alternatively, and equivalently, the power is fixed by regulating the full theory diagram and taking limits of the integrand. The relative factor of two comes from that fact that for a given gluon line in the full theory  there are two soft eikonal vertices (connecting the two eikonal lines) relative to the one collinear eikonal vertex. We have also introduced the bookkeeping parameter $w$ for convenience, which eventually will be set to one. It will play a role when we derive RG equations. 
The $g$ subscript on the momentum (label) operator will only play a role when we consider going to
 higher orders as is explained in section (\ref{higherorders}) and appendix (\ref{A}).

With this regulator the effective theory will have divergences in both the $\eta$ and $\epsilon$ go to zero limits.  The order of the limits is crucial to sensibly renormalize the theory. Given our physical arguments regarding the nature of the rapidity divergences, the proper order of limits must be: $\eta \to 0$, then $\epsilon \to 0$ with $\eta/\epsilon^n \to 0$ for all $n > 0$.
The physical reason for this ordering is clear since we must remain on the invariant mass hyperbola when we take the rapidity cut-off to its limit. 
To see how this works in practice let us evaluate the integrals $I_S$ and $I_n$ using this regulator.

The $I_S$ integral is most simply evaluated by  first doing the $k_0$ integral by contours. The result, after repristinating the expression with the coupling, group theory factor and the relevant numerator for the Sudakov form factor, in Feynman gauge, is given by \footnote{$w$ has been set to one, and is utilized below when we derive the renormalization group equation. We have also absorbed the $\overline {MS}$ factor into $\mu$ to simplify the expressions.}
\beq
\label{soft}
I_S=-g^2C_F(e^{\gamma_E \epsilon} 2^{-\eta -2}\pi^{-5/2}) \left(\frac{\mu}{M}\right)^{2 \epsilon}\left(\frac{\nu}{M}\right)^\eta\frac{\Gamma(1/2-\eta/2) \Gamma(\epsilon+\eta/2)}{\eta}
\eeq
Expanding first in $\eta$ and then in $\epsilon$ we find
\beq
\label{softs}
I_S=g^2 C_F \left[-\frac{e^{\gamma_E \epsilon}\Gamma(\epsilon)  \left(\frac{\mu}{M}\right)^{2 \epsilon}}{4\pi^2\eta}
+\frac{1}{4\pi^2}\left( \frac{\ln (\frac{\mu}{\nu})}{\epsilon} 
+\ln ^2(\frac{\mu}{M}) -2\ln (\frac{\mu}{M})\ln (\frac{\nu}{M})+ \frac{1}{2\epsilon^2}\right)-\frac{1}{96} \right]
\eeq
Similarly, the collinear integral $I_n$ is given by
\beq
\label{cols}
I_n=g^2 C_F \left[\frac{e^{\gamma_E \epsilon}\Gamma(\epsilon)  \left(\frac{\mu}{M}\right)^{2 \epsilon}}{8\pi^2\eta}+\frac{1}{4\pi^2}\left( \ln (\frac{\mu}{M}) \ln (\frac{\nu}{\bar n \cdot p_1}) +\ln (\frac{\mu}{M}) +\frac{1}{2\epsilon} \left(1+\ln (\frac{\nu }{\bar n \cdot p_1}) \right)+\frac{1}{2}\right)-\frac{1}{48}\right] \, ,
\eeq
and $I_{\bar n}$ by replacing $\bar n \cdot p_1$ with $n\cdot p_2$.
Summing the sectors we find 
\beq
I_S+I_{\nbar}+I_n=g^2 C_F \left[\frac{1}{4\pi^2}\left(\frac{1}{2\epsilon^2}+ \frac{\ln (\frac{\mu}{Q})}{\epsilon}+\frac{1}{\epsilon}+\ln ^2(\frac{\mu}{M})+2\ln (\frac{\mu}{M})+2 \ln  \frac{M}{\mu} \ln  \frac{Q}{M}+1 \right)-\frac{5}{96}\right] \, ,
\eeq
where we have used $\bar n \cdot p_1 = n\cdot p_2 = Q$. We see that the $\eta$ (rapidity) divergences vanish, there is no dependence on the scale $\nu$ and the answer is  boost \cite{Manohar:2002fd}  invariant.

In addition, note that the  soft-bins are all scaleless and vanish. However, this does not mean that they should be ignored, as explained in appendix (\ref{soft_bin_subtractions}). Indeed, as emphasized in \cite{Manohar:2006nz}, these subtractions can play a crucial role in being able to discern IR and UV singularities. In the case of the $\eta$ regulator this scaleless, vanishing, soft-bin contribution has the effect shifting the rapidity cut-off to its proper place. That is, if we are regulating the a collinear integral the effect of the soft-bin will (formally) to shift the cut-off to its proper place separating the collinear from the soft.

\subsection{The Rapidity Renormalization Group}\label{RRG}
An advantage of the regulator we have introduced is that it allows one to write down a
renormalization group equation in a rather straightforward manner. We begin by examining the Sudakov form factor of the space-like current in terms of the \scetii~ fields,
\begin{eqnarray}\label{scetII_current_form_factor} 
J_{\mu}&=&  H(Q^2,\mu) J_n(M;\mu,\nu/Q)\gamma_{\mu}^\perp J_{\bar n}(M;\mu,\nu/Q) S(M;\mu,\nu/M)
\end{eqnarray}
The one loop values of   matrix elements  $J_n,J_{\nbar},S$ defined in (\ref{defs}),
are given by  
 (\ref{softs},\ref{cols}).
The renormalizaton group follows from the set of equations
\bea
\frac{d}{d\ln [\mu]}(J_{n},S)^{\rm bare}=\frac{d}{d\ln [\nu]}(J_{n},S)^{\rm bare}=0.
\eea
Moreover the independence of $\mu$ and $\nu$ leads to
\beq
\label{commute}
[\frac{d}{d\ln [\mu]},\frac{d}{d\ln [\nu]}]=0 \, ,
\eeq
which is of course true for any observable not just the Sudakov form factor.

Defining the anomalous dimension under $\mu$ and $\nu$ variations 
as $(\gamma_\mu,\gamma_\nu)$ respectively,
such that
\begin{align}
\gamma_\mu^{n,S} = - Z_{n,S}^{-1} (\frac{\partial}{\partial \ln [\mu] }+\beta \frac{\partial}{\partial g}) Z_{n,S} \, ,
\end{align}
\beq
\gamma_\nu^{n,S} = -Z^{-1}_{n,S}  \frac{\partial}{\partial \ln\nu} Z_{n,S} \, ,
\eeq
 equation (\ref{commute}) imposes the
constraint
\beq
\label{con}
(\frac{\partial}{\partial \ln [\mu] }+\beta \frac{\partial}{\partial g}) \gamma_\nu= \frac{d}{d\ln [\nu] }\gamma_\mu \, = \mathbb{Z} \Gamma_{\rm cusp}\, ,
\eeq
which holds for any observable of interest.
 $\mathbb{Z}$ is an  integer whose value   depends upon whether we are considering an amplitude or the
 square of an amplitude.  For the Sudakov
 form factor $\mathbb{Z}$ is either $1$ or $2$ (see below). The last equality comes from the consistency of $\mu$-anomalous dimension with the hard anomalous dimension which is linear in the logarithm with coefficient $\Gamma_{\rm cusp}$. The universal relation between the collinear $c$ and
soft $S$ anomalous dimension 
\beq -2 \mathbb{Z}_c= \mathbb{Z}_S 
\eeq
follows automatically from the $\nu$ independence of the hard function, as will be discussed below.

Let us now apply the RRG to the Sudakov case we studied above.
Since our regulator allows us to define the jet and soft functions independently
we may renormalize them in standard fashion by absorbing $\frac{1}{\epsilon}$ and $\frac{1}{\eta}$ divergences in the renormalization constants, and then run renormalized quantities individually. We define
the renormalization factor $Z_{n},~Z_S$ via
\beq
J_n^R= Z_{\psi}^{1/2}Z^{-1}_{n}J_n^B~~~S^R= Z^{-1}_{{S}}S^B \,
\eeq
where $I^B$ corresponds to bare quantities and $I^R$ to renormalized.
Then using our result from above, at one loop we have
\begin{align}
Z_S&=1-\frac{g(\mu)^2 w^2 C_F}{4\pi^2} \left[\frac{e^{\epsilon \gamma_E}\Gamma(\epsilon)  \left(\frac{\mu}{M}\right)^{2 \epsilon}}{\eta}-\frac{1}{2\epsilon^2}-\frac{\ln \frac{\mu}{\nu}}{\epsilon}\right]  \, , \nn \\
Z_{n}&=1+\frac{g(\mu)^2 w^2 C_F}{4\pi^2} \left[\frac{e^{\epsilon\gamma_E }\Gamma(\epsilon)  \left(\frac{\mu}{M}\right)^{2 \epsilon}}{2\eta}+\frac{1}{2\epsilon}\bigg(1+\ln \frac{\nu}{\nbar \cdot p_1} \bigg)\right]  \, ,
\end{align}
where $Z_\psi$ is wave function renormalization which is the same as in full QCD.
\beq
Z_{\psi}=1-\frac{g(\mu)^2C_F}{16\pi^2 \epsilon}.
\eeq
The $\mu$ anomalous dimensions 
 are given by
\bea
\gamma_\mu^n&=& \frac{g^2(\mu)C_F}{4\pi^2}\left(\frac{3}{4}+\ln \frac{\nu}{\nbar \cdot p_1} \right) \, , \nn \\
\gamma_\mu^{ \bar n}&=& \frac{g^2(\mu)C_F}{4\pi^2}\left(\frac{3}{4}+\ln \frac{\nu}{n \cdot p_2} \right) \, ,  \\
\gamma_\mu^{S}&=& \frac{g^2(\mu)C_F}{4\pi^2}\ln \frac{\mu^2}{\nu^2}\nn \, .
\eea
As a consistency check see that
\beq
\gamma_\mu^n+\gamma_\mu^{ \bar n}+\gamma_\mu^S=-\gamma_H=
\frac{g^2(\mu)C_F}{4\pi^2}\left(\ln \frac{\mu^2}{Q^2}+\frac{3}{2}\right) \, ,
\eeq
where $\gamma_H$ is the anomalous dimension of the hard matching coefficient.

The calculation of the $\nu$ anomalous dimensions
necessitates care.  The bare book keeping parameter is $\nu$ independent and thus, in analogy with
the coupling $g$, the ``renormalized"\footnote{It is important to remember that $w$ is not a coupling, but strictly a calculational tool.} $w$ obeys 
\beq\label{nu-beta}
\nu \frac{\partial}{\partial \nu}w= -\frac{\eta}{2} w 
\eeq
we find at one loop
\bea \label{eq:nu-AD-SFF}
\gamma_\nu^{n}&=&\frac{g^2(\mu)C_F}{8\pi^2}\ln \frac{\mu^2}{M^2}  \, , \nn \\
\gamma_\nu^{S}&=&- \frac{g^2(\mu)C_F}{4 \pi^2} \ln \frac{\mu^2}{M^2} \, .
\eea
These correctly obey the consistency equation
\begin{align}\label{eq:nu-consistency}
\gamma_\nu^{n}+\gamma_\nu^{ \bar n}+\gamma_\nu^{S}=0 \, .
\end{align}

\begin{figure}
   \centering
     \includegraphics[width=3 cm]{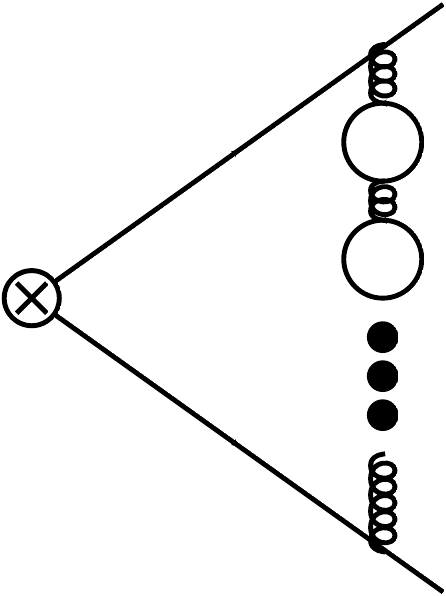}
   \caption{ \label{bubbles} Coupling renormalization (Abelian)  contributes to $\gamma_\nu$ and is missed in the fixed order one loop result.}
\end{figure}

\begin{figure}
   \centering
     \includegraphics[width=8 cm]{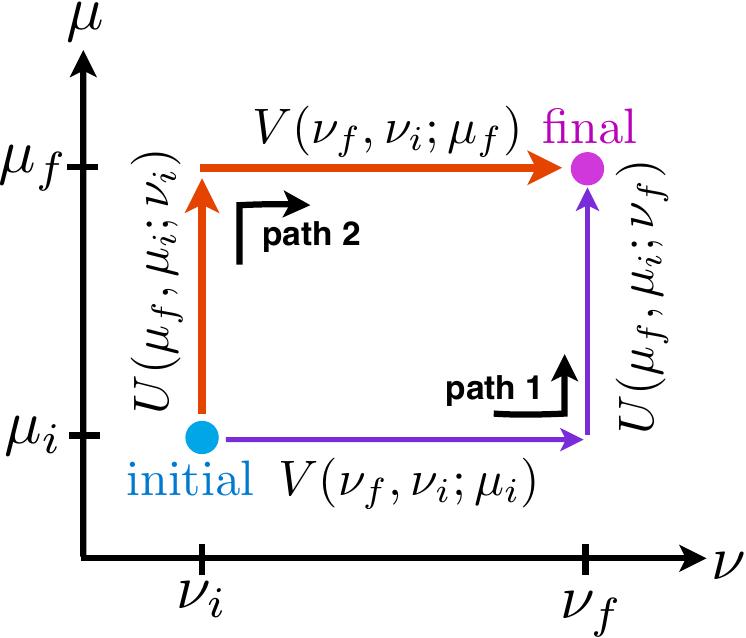}
   \caption{ \label{fig:commute} Two alternate paths are shown for evolution in $\mu$-$\nu$ plane. Due to independence of $\mu$ and $\nu$ scales evolution along either path will yield the same result.}
\end{figure}

Both the large logarithms, due to large invariant mass ratio and large rapidity ratio, can be resummed by the RG equations
\begin{align}
\mu\frac{d}{d\mu }(J_{n},S) &= \gamma_\mu^{n,S}(J_{n},S) \, , \nn \\
\nu\frac{d}{d\nu }(J_{n},S) &= \gamma_\nu^{n,S} (J_{n},S) \, .
\end{align}
The relation (\ref{con}) guarantees that the $\mu$ and $\nu$ evolutions commute, hence, the evolution in $\mu$-$\nu$ plane is path independent. However, care must be taken when solving the $\nu$-RG equation. $\gamma_\nu$ contains terms of form $\alpha_s^n(\mu) \ln^m(\mu/M)$ with $m\le n$. For instance, one can see from Fig. \ref{bubbles} that the one loop result
will be multiplied by a series of logarithms of the form
$\sum_n[\beta_0 \alpha_s \ln(\mu/M)]^n$. These logarithms can be large if $\mu \gg M$, for example, and would require resummation. This is easily obtained by solving the consistency relation (\ref{con}) up to the required order in perturbation theory,
\begin{align}
\label{eq:full-log}
 \gamma_\nu
 &=\int^{\ln\mu} d\ln
 (\mu^\prime) \frac{d}{d\ln(\nu)} \gamma_\mu(\mu^\prime) + {\rm const.}\nn \\ 
 &\propto \int^{\ln\mu} d\ln
 (\mu^\prime) \Gamma_{\rm cusp}(\mu^\prime) + {\rm const.} \, ,
\end{align}
where integration constant is fixed by the fixed order calculation of anomalous dimension and corresponds to its non-cusp piece. From eqn. (\ref{eq:nu-AD-SFF}) we see that non-cusp piece is zero at one loop. Eqn. (\ref{eq:full-log}) completely fixes the logarithmic ($\mu$) structure of $\gamma_\nu$ to all orders in perturbation theory when expanded in $\alpha_s(\mu)$. If we had calculated $\gamma_\nu$ to higher orders we would see these logarithms explicitly. Thus, it constitutes a check on the higher order calculations.
In its integrated form $\gamma_\nu$ resums the set of diagrams which renormalize the coupling, which in the Abelian case, arise from the bubble chain  shown in figure \ref{bubbles}, thus taking into account the running of $\alpha_s$. Fixed order form of $\gamma_\nu$ suffices when evolution is done along path 1 shown in figure \ref{fig:commute} with $\mu_i \sim \nu_i \sim ~ M \ll \mu_f \sim \nu_f$.  However, the  integrated form (\ref{eq:full-log}) is required when evolution is done along path 2. Since $\mu_f \gg M$ there are large logarithms in $\gamma_\nu$ that require resummation in addition to the rapidity logs.
In figure \ref{fig:commute}, $U$ and $V$ are the evolution factors in $\mu$ and $\nu$ respectively and $\mu_i,\nu_i$ are
the scales for the initial conditions.
The notation $U(\mu_f,\mu_i;\nu_a)$ implies running $\mu$ from $\mu_i$ to $\mu_f$ at fixed $\nu=\nu_a$; similarly for $V(\nu_f,\nu_i;\mu_a)$.
Along path 1, we have chosen to run first in $\nu$ and then in $\mu$. Path 2 shows the alternate choice and should yield the same result, thus
\beq
\label{eq:commute}
V(\nu_f,\nu_i;\mu_f) U(\mu_f,\mu_i;\nu_i) = U(\mu_f,\mu_i;\nu_f) V(\nu_f,\nu_i;\mu_i) \, .
\eeq
To ensure this in practice, we must use the resummed form of $\gamma_\nu$ when calculating $V(\nu_f,\nu_i;\mu_f)$.

Notice that these anomalous dimensions depend upon the ``low'' energy parameter, $M$,
which normally would, and should, not show up in the expression for an anomalous
dimension. However, we must recall as far as the rapidity divergences are concerned
$M$ is not a low energy parameter, but just the invariant mass of the hyperbola
along which the rapidity renormalization group flows.

To sum the large logarithms we  first identify the natural scales for the Hard, Soft and Jet Function which are given by
$(\mu_H), (\mu_S,\nu_S)$ and $(\mu_J,\nu_J$) respectively. Numerically they can be read off from (\ref{softs},\ref{cols})
\beq
\mu_H \sim Q, \mu_S \sim \nu_S \sim \mu_J \sim M, \nu_J \sim Q.
\eeq 
To eliminate the large logarithms we may run in both $\mu$ and $\nu$ to some fixed scale, while evaluating the fixed order functions at their natural scales. That is, we may write
\bea
\label{eq:one-AJ}
S(\mu,\nu)&=& V_S(\nu,\nu_S;\mu)( U_S(\mu,\mu_S;\nu_S)S(\mu_S,\nu_S))  \nn \\
J_n(\mu,\nu) &=& V_J(\nu,\nu_J;\mu)( U_J(\mu,\mu_J;\nu_J)J_n(\mu_J,\nu_J) )\nn \\
H(\mu) &=&H(\mu_H)U(\mu,\mu_H) \, ,
\eea
where $U_{n,S}$ and $V_{n,S}$ are respectively $\mu$ and $\nu$ evolution factors for jet and soft functions. In (\ref{eq:one-AJ}) we have chosen to run first in $\mu$ and then in $\nu$. We could equally well have switched the order leading to the same result. Note that in the ordering of eqn. (\ref{eq:one-AJ}) we are required to use the integrated form of $\gamma_\nu$ of eqn. (\ref{eq:full-log}) in order to resum all the large logs due to the running coupling. We get,
\bea
\label{eq:U_S}
U_S(\mu,\mu_S;\nu_S)&=& exp\left[- \frac{8\pi C_F}{\beta_0^2} \left( \frac{1}{\alpha(\mu)}-\frac{1}{\alpha(\mu_S)}
-\frac{1}{\alpha(\nu_S)} \ln  \frac{\alpha(\mu)}{\alpha(\mu_S)} \right)  \right] \\
V_S(\nu,\nu_S;\mu)&=& exp\left[ \frac{2C_F}{\beta_0}\ln \left( \frac{\alpha(\mu)}{\alpha(M)}\right)
 \ln \left( \frac{\nu^2}{\nu_S^2}\right) \right] \\ 
U_J(\mu,\mu_J;\nu_J) &=&exp\left[ -\frac{2 C_F}{\beta_0} \left( \frac{3}{4}+\frac{1}{2}\ln \left( \frac{\nu_J^2}{Q^2}\right) \right)
\ln \frac{\alpha(\mu)}{\alpha(\mu_J)}  \right]  \\
\label{eq:V_J}
V_J(\nu,\nu_J;\mu)&=& exp\left[-\frac{C_F}{\beta_0} \ln \left( \frac{\alpha(\mu)}{\alpha(M)} \right) \ln \left( \frac{\nu^2}{\nu_J^2}\right) \right] \\ 
U_H(\mu,\mu_H)&=&exp\left[- \frac{8\pi C_F}{\beta_0^2} \left( \frac{1}{\alpha(\mu_H)}-\frac{1}{\alpha(\mu)}
-\frac{1}{\alpha(Q)} \ln \frac{\alpha(\mu)}{\alpha(\mu_H)} \right) \right]
\eea
with 
\bea
\label{eq:SJ-FO}
S(\mu_S,\nu_S)&=&1+\frac{\alpha(\mu_S) C_F}{\pi} \left[ \ln ^2(\frac{ \mu_S}{M}) -2\ln (\frac{ \mu_S}{M})\ln (\frac{\nu_S}{M})-\frac{\pi^2}{24} \right] \\
J_n(\mu_J,\nu_J)&=&1+\frac{\alpha(\mu_J) C_F}{\pi} \left[ \ln (\frac{ \mu_J}{M}) \ln (\frac{\nu_J}{n \cdot p_1}) +\frac{3}{4}\ln (\frac{ \mu_J}{M})-\frac{\pi^2}{12}+\frac{1}{2} \right]. 
\eea

\begin{figure}
   \centering
     \includegraphics[width=8 cm]{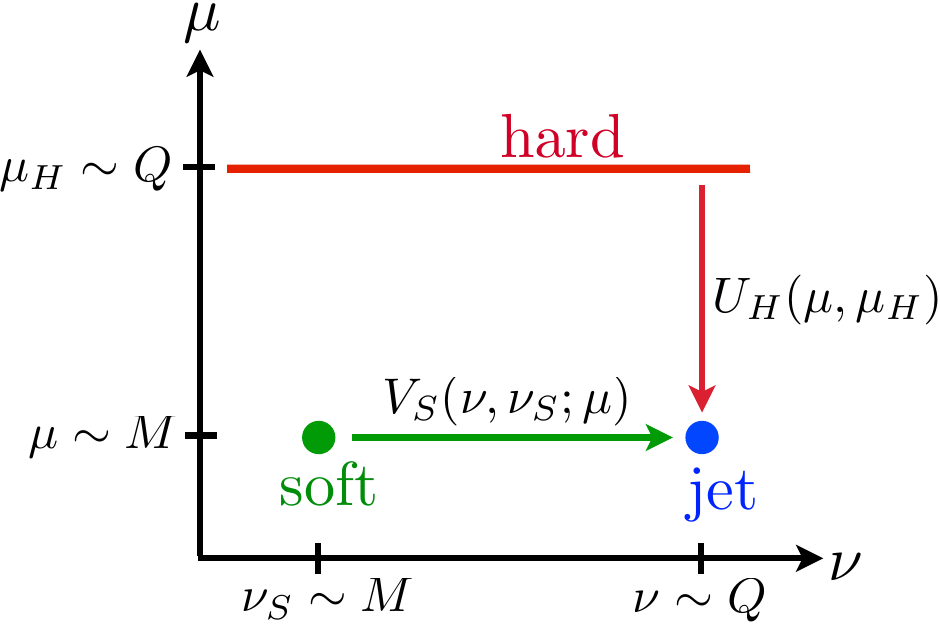}
   \caption{ \label{fig:strategy} Simplest running strategy to resum all the large logarithms in the Sudakov Form Factor.
}
\end{figure}

Using relations (\ref{eq:U_S}) to (\ref{eq:V_J}) we can explicitly verify the commutation relation (\ref{eq:commute}) at the order we are working. Equations (\ref{eq:one-AJ}) to (\ref{eq:SJ-FO}) give the resummation for the most general choice of scales $\mu$ and $\nu$. However, in order to resum all the logarithms, the  most convenient choice of scales is $\mu = \mu_J = \mu_S \sim M$ and $\nu = \nu_J  \sim Q$. Running with this choice of scales only requires running the hard function in $\mu$ and soft function in $\nu$ to the natural scales of the jet function. This strategy is shown in figure \ref{fig:strategy}. With this strategy, it is not required to use the  integrated form (\ref{eq:full-log}) and the  fixed order form of $\gamma_\nu$ suffices.

The physics of the RRG  flow can be understood from figure \ref{RRG}.
A change in the scale $\nu$ corresponds  to a flow between the
soft and collinear regions. The natural scale for the soft function is
$n\cdot k \sim  \bar n \cdot k \sim M$ whereas the collinear functions sit at the scale
$Q$. To sum the logarithms we may slide the cut-off(s)  of the soft function up the
hyperbola, such that the scale $\nu$ minimizes the logarithms in the collinear sectors.
 \begin{figure}
   \centering
     \includegraphics[width=5 cm]{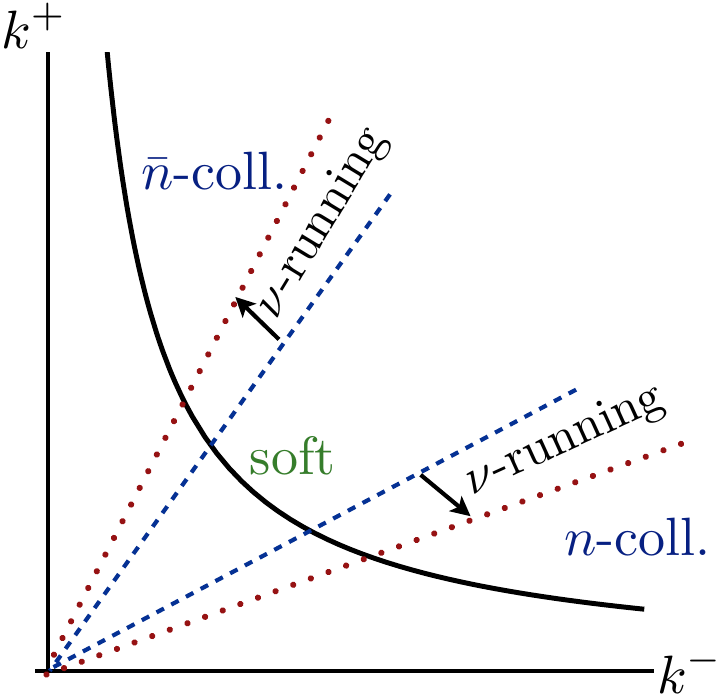}
   \caption{ \label{RRG} Running in $\nu$ corresponds to flow along the mass-shell hyperbola.
}
\end{figure}

\subsection{The Necessity for RRG}
\label{needRRG}
The RRG is critical in establishing the $\mu$ independence of the resummed form factor. To illustrate this we can combine the evolution factors, and present the completely resummed form factor as
\begin{multline}
F(Q^2,M^2)=E(\mu,\mu_H,\mu_J,\mu_S;\nu_J,\nu_S)H(Q^2,\mu_{H}^2)
J_{n}(\mu_{J},\nu_{J};M;Q)J_{\bar n}(\mu_{J},\nu_{J};M;Q)S(\mu_{S},\nu_{S};M)
\end{multline}
where we have made clear that the jet and soft function depend explicitly on the scales $M$ and $Q$.
The $\mu$ dependence in $E$  is always sub-leading, and would cancel in the exact result. We will
keep $\mu$ to be arbitrary to show how the $\mu$ dependence cancels to any given working order.
Minimizing the logarithms in all sectors can be accomplished by the choices:
\begin{align}
\mu_{H}= Q& &\mu_{J}= \mu_S=M \sim \mu\\
\nu_{J}= Q& &\nu_{S}= M.\end{align}
Then the total evolution factor then at one loop is:
\begin{multline}
E=exp\left[ -\overbrace{\frac{8\pi C_F}{\beta_0^2} \left( \frac{1}{\alpha(Q)}-\frac{1}{\alpha(\mu)}
-\frac{1}{\alpha(Q)} \ln \left( \frac{\alpha(\mu)}{\alpha(Q)} \right)\right) }^{\text{$\mu$ RG}} +2 \underbrace{\frac{ C_F}{ \beta_0} \ln \left( \frac{\alpha(\mu)}{\alpha(M)} \right) \ln \left( \frac{Q^2}{M^2}\right)}_{\text{RRG}}\right],
\end{multline}
To the logarithmic order we are working, the exponent of the form factor is $\mu$-independent. We achieved this critically important feature because of the rapidity renormalization group. The variation of the hard double logarithms must be canceled by the variation of the jet and soft sectors. Given the two scale nature of our \scetii\, problem,  it is not possible  to cancel the variation of the hard double log without the presence of  a large logarithm in the low-scale matrix elements.

 Schematically we can write for a generic soft-collinear factorization of a physical process featuring large double logarithms:
\begin{align}
\label{generic_double_log_resum}\sigma_{resum}&=exp\Big[\Gamma[\alpha]L^2-2\Gamma[\alpha]L\,\tilde{L}+...\Big]f(\tilde{L})\\
L&=\text{ln}\Big(\frac{Q}{\mu}\Big)\\
\tilde{L}&=\text{ln}\Big(\frac{M}{\mu}\Big)
\end{align}
Where $M$ is the infra-red scale, $Q$ is the hard scale, and $\mu$ is assumed to be of order $M$ (thus $L\gg\tilde{L}$). The function $f$ is the low scale matrix elements of the collinear and soft sector, and should have no large logarithms. We have neglected the running in $\alpha$, but its inclusion does not change the substance of our argument. Varying $\mu$ we find that the leading $\mu$ variation in the logarithmic power counting (i.e., terms that scale as $L\delta L$) is given by:
\begin{align}
\delta\sigma_{resum}&=\sigma_{resum}\delta\Big(\Gamma[\alpha]L^2-2\Gamma[\alpha]L\,\tilde{L}\Big)+...\\
&=0+...
\end{align}
The exponent has the required form to cancel the leading variation, since $\delta L=\delta \tilde{L}$. In general terms in the resummed exponent such as $\Gamma_{n-1}\alpha^n L^2$ get canceled by an RRG resummation of the form $2\Gamma_{n-1}\alpha^nL\tilde{L}$. There are further sub-leading variations that scale as $\alpha^n\tilde{L}$ or $\alpha^n$, but these variations are canceled by the matrix elements as the low scale since they involve no large logarithm.

This form of the exponent is found in both \sceti\, and \scetii\, as the hard double logarithmic terms appear generically when there are collinear and soft modes, irrespective of the scaling of the soft modes themselves.  \sceti\,  differs from \scetii\,  in  that the corresponding $L\tilde L$ term in \sceti\, comes solely as a consequence of traditional RG techniques.  One can see that the low/high-scale mixed double logarithms can be generated through traditional resummations since the virtuality of the various modes are separated by an equivalent amount, namely $\frac{Q_J^2}{Q^2_H}\sim\frac{Q_{US}^2}{Q^2_J}\sim\lambda^2$. Hence in the running from the jet to the ultra-soft sectors, a double log of the form $L\tilde{L}$ is generated \footnote{This statement is of course contingent on the arbitrary choice of which functions to run. Different choices would have these logs appear in other evolution factors. But their appearance is intimately tied to the invariant mass scale separation of the effective theory sectors.}.  Since \scetii\, is bereft of the invariant mass separation in the infra-red physics, one can conclude that something like the RRG must exist in all \scetii\, factorizations to generate the required $L\tilde{L}$ term in the exponent.

It is important to note that the inclusion of the rapidity logs in the exponent, necessary for insuring $\mu$ independence,
also leads to an ambiguity in the scale of the rapidity logs. In previous SCET treatments,  $\mu$ independence is achieved    by explicitly exponentiating the large logarithm found in the low scale matrix elements \cite{Chiu:2009mg}. \footnote{ In this work it was shown that there is  at most one log in the
low scale matching.} However, doing things in this way obscures the errors
at higher orders in perturbation theory, due to an ambiguity in the choice of low scale in the exponentiated single logarithm. 
$\mu$ independence only tells us that we must
have a $\ln(\mu/A)$ in the exponent, but  the scale $A$ is arbitrary. The dependence on $A$  should of course
cancel between the matrix element of the exponent  (at any given order), which is achieved automatically in the RRG.
One can exponentiate the rapidity logs by hand, without recourse to the RRG, but then it becomes difficult to
track the scheme dependence (i.e. how varying the choice of A affects the sub-leading pieces). 
 In contrast, RRG provides an independent scale $\nu$ to vary so that size of all the sub-leading logarithms is properly captured. We propose to  quantify error estimates by independently varying in $\mu$ and $\nu$ in a suitable range and then adding the  errors in quadrature.
 
 Finally,  in earlier, non-EFT,  treatments of the form factor, the $\mu$ independence was achieved via the Collins-Soper equation \cite{Collins:1981uk}. In the CSS approach to the form factor the resummed exponent is written as \cite{Collins:1989bt}
\begin{align}
\frac{d}{d\ln Q^2}\ln F(Q^2,M^2)=G(\alpha(\mu),\ln(Q/\mu))+K(\alpha(\mu),\ln(M/\mu)).
\end{align}
By running in $Q$, this effectively mixes rapidity and $\mu$ running since the hard function depends
upon the scale $Q$. By introducing the scale $\nu$ we avoid this issue, which allows us to cleanly
separate the rapidity logs from the invariant mass logs.
However, note that this methodology will lead to two independent integration constants just as in our formalism.
Having two such integration constants allows one to systematically control scale variation errors.
\subsection{Gauge Invariance and the Structure of Rapidity Divergences at Higher Orders }
\label{higherorders}
In any approach to renormalization, one does not want the procedure to violate gauge invariance. That the total contribution, i.e. the sum of the soft and collinear pieces, is gauge invariant follows from the same argument used to show the rapidity divergences cancel in the effective theory. One  first introduces the regulator in the full theory, where the
regulator is not needed to make integrals well defined.  Since the rapidity regulator is inherited by each sector (in  appendix \ref{A}, we show how this is specifically accomplished for the $\eta$-regulator) from the full theory, it is guaranteed to cancel in the sum of sectors. Gauge invariance follows similarly. That is, the full theory is gauge invariant, and under the assumption that the effective theory  is properly reproducing the infrared physics, then the sum of the effective theory diagrams must also be gauge invariant.

Given the intimate connection between Lorentz invariance the gauge symmetry is might seem surprising
that the regulated sectors are {\it themselves} gauge invariant. Nonetheless, as we prove in appendix (\ref{A})
the regulated sectors are indeed gauge invariant to all orders, in covariant gauges. This result follows once one introduces the notion of  non-abelian exponentiation \cite{Gatheral:1983cz,Frenkel:1984pz}, (see \cite{Laenen:2008gt} for a nice modern approach)  which strongly constrains the structure of the rapidity divergences. The diagrammatic expansion of any generalized soft function can be rewritten as the exponential of a distinct subset of diagrams contributing to the series. That is, a generalized soft function can be written as the exponential of two eikonal line irreducible graphs with a particular color weight. The sum of these graphs is known as a CWEB. Since only a single rapidity divergence can appear in the logarithm of a generalized soft function \cite{Korchemsky:1987wg,Manohar:2003vb}, it follows  that only a single rapidity divergence can appear in a CWEB, regardless of the number of loops involved.  Hence CWEBs are minimally divergent with respect to rapidity divergences \footnote{For rapidity divergences, this statement is expressing the fact that the anomalous dimensions is at most linear in logarithm associated with the rapidity divergence. There are of course other UV renormalization point dependent logarithms in the rapidity anomalous dimension, since a CWEB at higher orders have multiple sub-loops. But these logs are predicted by the UV divergences of QCD: see \eqref{eq:full-log}.}. Indeed, this fact is critical in establishing the gauge independence of the rapidity anomalous dimension.  Given this marginal divergence, the gauge dependent piece of the covariant gauge polarization tensor, which is proportional to $k_\mu k_\nu$, will lead to an integral with no rapidity divergence. 

To see this, consider an $n$ loop  CWEB. Since the marginal nature of the divergence implies that  there are no sub-divergences, we may perform the $n-1$ loop integrals leaving one loop integral left over with a gluon attaching to an eikonal line. Given that we have performed all the loop integrals except one, the  gauge dependent piece of the polarization of this gluon must be proportional to the momentum carried by the Wilson line itself. Contraction with the eikonal vertex will then cancel the denominator which is the cause of the rapidity divergence.

In appendix (\ref{A}) we give details of this argument, and show how one needs to generalize the Wilson line regulator at higher orders. This appendix also contains a discussion of the delta regulator. In either case, organizing the calculation in terms of CWEBs greatly reduces the amount of work  necessary.

\section{Transverse Momentum Spectrum in Higgs Production}
\label{Higgs}
The differential cross-section for producing the Higgs boson with fixed momentum in hadronic collisions 
is an observable of obvious relevance.  It has been shown that the $p_\perp$ distribution can
be used as a smoking gun for new physics \cite{Arnesen:2008fb}. Thus having a reliable theoretical
prediction is a worthwhile enterprise. In the limit where $p_\perp\ll m_h$ this cross section 
provides another important instance where factorization proceeds in \scetii, involving rapidity divergences. The Higgs $p_\perp$ spectrum is kinematically and formally very similar to the Drell-Yan $p_\perp$ spectrum. Much effort has been devoted to both understanding factorization in SCET framework \cite{Becher:2010tm,Mantry:2009qz,Gao:2005iu,Idilbi:2005er} and resummation of logarithms of the form $\ln( p_\perp^2/m_h)$, usually in the context of the CSS resummation formalism \cite{Collins:1984kg,PhysRevD.38.3475,PhysRevD.44.1415,deFlorian:2000pr,deFlorian:2001zd,Catani:2010pd}.

To see why this observable fits into \scetii, let us consider the kinematics.
We impose the kinematical constraint  that the transverse momentum of the Higgs relative  to the colliding beam be small compared to the Higgs mass, $\lambda\sim\frac{p_\perp}{m_h}\ll 1$. Thus all final state radiation recoiling  against the Higgs must fulfill this same condition. Taking the
Higgs momenta to scale as \begin{equation}
p_h\sim m_h(1,1,\lambda),
\end{equation}
it is simple to see that the on-shell radiation that can recoils against the Higgs scales as:
\begin{align}
p_h=p_c+p_{\bar c}+p_s+\sum p_{CJ} \, , \nn\\
p_c\sim m_h(1,\lambda^2,\lambda) \, , \nn\\
p_{\bar c}\sim m_h(\lambda^2,1,\lambda)\, , \nn\\
p_s\sim m_h(\lambda,\lambda,\lambda) \, ,\\
p_{CJ}\sim m_h(1,1,1), \nn
\end{align}
where 
$(p_c,p_{\bar c},p_s)$ stand for collinear, anti-collinear and soft momentum respectively.
We have allowed for  the possibility of jets in the central region with momentum $p_{CJ}$.
 These jets each have large transverse momenta, but their net transverse momentum must scale as $\lambda$, and they  impart little transverse momentum to the Higgs itself. At fixed order in QCD, these jets do not appear until NLO in the $p_\perp$ spectrum (or NNLO in total Higgs production). In what follows we will prove a factorization theorem that robustly accounts for all such radiation, and calculate the resummation to NLL. 

\subsection{QCD cross-section}
Incorporating the most recent bounds from the LHC \cite{ATLAS, CMS}  we will assume the Higgs is sufficiently light that its dominant production mechanism is gluon fusion.
Given this assumption,   we may work within the Higgs effective theory where the top quark is integrated out, generating
the  dimension six operator
\beq{\cal H}(x) = h(x) {\rm Tr} [G^{\mu\nu}(x)G_{\mu\nu}(x)].
\eeq
The  matching coefficient for this operator  is known to two loops and is given by \cite{Inami:1982xt,Djouadi:1991tka}
\begin{align}
C_t&=\frac{\alpha_s}{12\pi}+\frac{\alpha_s^2}{64\pi^2}\Big(\frac{5}{3}C_A-C_F\Big).\end{align}
The differential  cross-section in Higgs boson transverse momentum ($p_{\perp}$) and rapidity ($y$) is given by
\begin{align}\label{eq:QCD-X-sec-Higgs}
\frac{d\sigma}{d p_{\perp}^2 dy}&= \frac{C_t^2}{8v^2S}\int d^4x \sum_{\text{spins}}\langle p_{n} p_{\bar{n}} \vert \, {\cal H}(x)\,  \delta\!\!\left(y-\half \ln\frac{\mathcal{P}_h^+}{\mathcal{P}_h^-}\right) \delta(p_{\perp}^2 - \vert \vec {\mathcal{P}}_{h\perp}\vert^2) {\cal H}(0) \vert p_{n} p_{\bar{n}} \rangle.\end{align}
   $\mathcal{P}_h$ is the momentum operator that picks out the Higgs momentum.  
 consider it a derivative. 
  $\vert p_{n} p_{\bar{n}}\rangle$  is the incoming proton state with momenta $p_n$ and $p_{\bar{n}}$. $v$ is electro-weak symmetry breaking scale, and $\sqrt{S}$ is center of mass energy.
When there exists central jets the cross section  will become sensitive to higher dimensional operators, but as we will
see below, this region of phase space is power suppressed.
 We can  simplify \eqref{eq:QCD-X-sec-Higgs} by writing 
 \begin{multline}
\langle 0|h(x)\delta\!\!\left(y-\half \ln\frac{\mathcal{P}_h^+}{\mathcal{P}_h^-}\right) \delta(p_{\perp}^2 - \vert \vec {\mathcal{P}}_{h\perp}\vert^2)h(0)|0\rangle\\=\int \frac{d^4p_h}{(2\pi)^4}(2\pi)\delta^{+}(p_h^2-m_h^2)e^{-ip_h.x}\delta\!\!\left(y-\half \ln\frac{p_h^+}{p_h^-}\right) \delta(p_{\perp}^2 - \vert \vec {p}_{h\perp}\vert^2),
\end{multline}
the cross-section then becomes
\begin{align}
\label{eq:QCD-X-sec-Higgs-II}\frac{d\sigma}{d p_{\perp}^2 dy}&= \frac{C_t^2}{8v^2S}\int \frac{d^4p_h}{(2\pi)^4}(2\pi)\delta^{+}(p_h^2-m_h^2)\delta\!\!\left(y-\half \ln\frac{p_h^+}{p_h^-}\right)\delta(p_{\perp}^2 - \vert \vec {p}_{h\perp}\vert^2)\nonumber\\
&\int d^4x e^{-ip_h.x}\sum_{\text{spins}} \langle p_{n} p_{\bar{n}}\vert  {\rm Tr} [G^{\mu\nu}(x)G_{\mu\nu}(x)] {\rm Tr} [G^{\alpha\beta}(0)G_{\alpha\beta}(0)] \vert p_{n} p_{\bar{n}} \rangle \, ,
\end{align}
where $\delta^+(p_h^2-m_h^2) = \theta(p_h^0)\delta(p_h^2-m_h^2)$.

\subsection{Factorization in \scetii}
\subsubsection{Central Jets are Power Suppressed}
To match to the effective theory, one should perform an OPE that matches the full theory operator in \eqref{eq:QCD-X-sec-Higgs-II} to a product of effective theory operators at the hard scale, as is done in the case of inclusive Drell-Yan \cite{Bauer:2002nz}.  As long as one can show that the contributions due to hard colored particles crossing the cut in a given full theory diagram is power-suppressed in the $\frac{p_\perp}{m_h}$ expansion, then
effectively there is no OPE since the currents are still separated by distance scales large compared to $m_h$.
 If there are no hard partons crossing the cut, then matching at the hard scale reduces to matching the  full theory operator $\text{Tr}[GG](x)$ onto effective theory currents  ${\cal B}_{n\perp}^{\mu a}{\cal B}_{\bar n\perp\mu}^{a}(x)$ , where  ${\cal B}_{n\perp}^{\mu a}$ will be defined below.  In what follows, we will establish that the central jets are power suppressed, and so one can simply match currents.
 \begin{figure}
    \includegraphics[width=9cm]{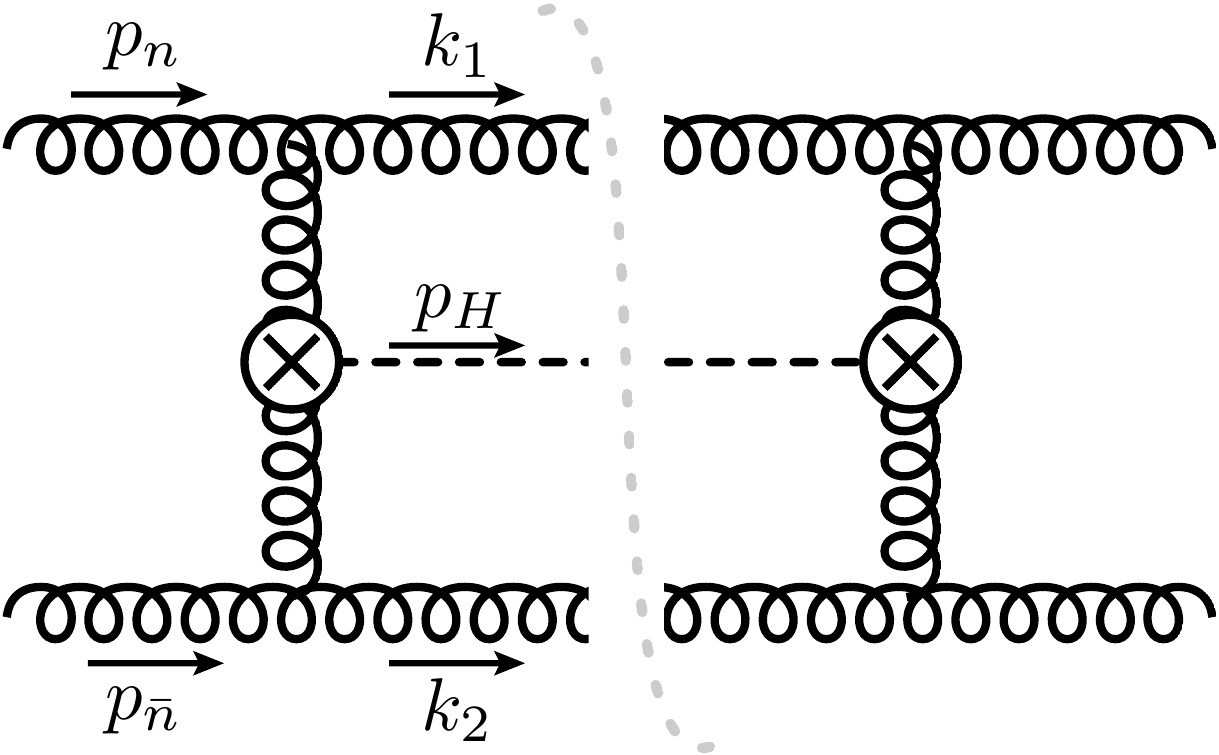}
   \centering
\caption{ \label{central}A contribution to the production cross section with central jets.}
\end{figure}

At leading order in the $p_\perp/m_h$ expansion, the only full theory diagrams that contribute must come with a $\delta(p_\perp^2)$ or the $\frac{1}{p_\perp^2}$ singularity, as both of these distributions are leading order in the power counting. 
Delta function  and power like singularities are associated with virtual and collinear contributions respectively,
neither of which can arise from central jets.  
When matching at the high scale we expand in powers of $p_\perp$.
Since the central jets only depend on the hard scales, one can set transverse momentum to zero in diagrams containing the central jets. 
Thus the part of the full QCD amplitude-squared that produce any modification of the Higgs \emph{transverse momentum} by central jets is power suppressed.

More formally (pedantically), consider a full theory diagram that contains central jets, such as in Figure \ref{central}. The diagram has the form:
\begin{multline}
I_{CJ}=\int \frac{d^dk_1}{(2\pi)^d}\delta^{+}(k_1^2)\frac{d^dk_2}{(2\pi)^d}\delta^{+}(k_2^2)\frac{d^dp_h}{(2\pi)^d}\delta^{+}(p_h^2-m_h^2)\delta^{(d)}(p_n+p_{\bar n}-k_1-k_2-p_h)P(p_n,p_{\bar n},k_1,k_2,p_h).
\end{multline}
The hard contribution to the matching from this diagrams can be obtained by simply considering the $k_1$ and $k_2$ momenta to be large, and asymptotically expanding the integrand accordingly (along with the power counting of the external momenta $p_n,p_{\bar n},p_h$). Then the above integral becomes:
\begin{align}
I_{CJ}|_{hard}=&\int d\Omega \delta(\bar{n}.p_n-\bar{n}.k_1-\bar{n}.k_2-\bar{n}.p_h)\delta(n.p_{\bar{n}}-n.k_1-n.k_2-n.p_h) \nonumber\\
&\,\,\,\,\,\,\,\,\,\,\,\,\delta^{(d-2)}(\vec{k}_{1\perp}+\vec{k}_{2\perp})P_{asym}(p_n,p_{\bar n},k_1,k_2,p_h)\\
d\Omega=&\frac{d^dk_1}{(2\pi)^d}\delta^{+}(k_1^2)\frac{d^dk_2}{(2\pi)^d}\delta^{+}(k_2^2)\frac{d^dp_h}{(2\pi)^d}\delta^{+}(p_h^2-m_h^2)
\end{align}
Since all propagators in the integrand $P_{asym}$ have a hard scaling and the momentum conservation delta function lacks the scale $p_\perp$, the integral is completely independent of $p_\perp$. Hence it has no contribution that scales as $\frac{1}{p_\perp^2}$. 
 Thus the hard contribution from this diagram is power suppressed. The argument easily generalizes to more complicated diagrams.
\subsubsection{Matching}
Having established that central jets are power-suppressed,  we first match the non-local operator ${\cal H}(x){\cal H}(0)$ onto the product of SCET currents:
\begin{multline}\label{eq:GGtoBB-current-matching}
\langle p_{n} p_{\bar{n}} \vert   {\rm Tr} [G^{\mu\nu}(x)G_{\mu\nu}(x)] {\rm Tr} [G^{\alpha\beta}(0)G_{\alpha\beta}(0)]\vert p_{n} p_{\bar{n}} \rangle \\
= \sum_n H(m_h)\langle p_{n} p_{\bar{n}} \vert \,\Big({\cal B}_{n\perp}^{a\mu}(x){\cal B}_{\bar n\perp\mu}^{a}(x)\Big) \Big({\cal B}_{n\perp}^{b\nu}(0){\cal B}_{\bar n\perp\nu}^{b}(0)\Big)\vert p_{n} p_{\bar{n}} \rangle+O(\lambda) \, .
\end{multline}
The interaction is non-local only along the light cone and the transverse directions as  there are no hard partons crossing the cut\footnote{We would see this 
non-locality in the transverse direction by transforming to momentum space which would place a momentum
conserving delta function in the transverse plane between the operators.}.
The  hard matching  can depend only upon $m_h$ and  is given by
\beq H(m_h)=4m_h^4 |C(m_h)|^2   \, ,  \eeq 
where $C(m_h)$ is the matching coefficient from current matching\footnote{Formally, the matching coefficient $C$ is a function of SCET label operators but they only appear in a Lorentz invariant combination reducing to $m_h^2$.}. 
Furthermore
\beq
B_{n\perp}^{a\mu}(x)=\frac{2}{g}{\rm Tr} \left[ T^a \left[ W_n^\dagger(x)i { D}^\mu_{n\perp}W_n(x)\right] \right]
\eeq
and
\beq
{\cal B}_{n\perp}^{a\mu}(x)=S_{n}^{a a'}(x)B_{n\perp}^{a'\mu}(x).
\eeq
$W_n$ is a collinear Wilson line in the fundamental representation defined in $x$-space by
\beq
W_n(x)=P \exp \left(\int_{-\infty}^x \bar n \cdot A_n(\bar n \lambda)d\lambda \right).
\eeq
$S_{n}^{a a'}(x)$ is a soft Wilson line in the adjoint representation
\beq
S_{n}^{a a'}(x)= P \exp \left(\int_{-\infty}^x n \cdot A_s(n \lambda)d\lambda \right)^{a a'}.
\eeq
 Now we factorize the matrix element:
\begin{align}
\eqref{eq:GGtoBB-current-matching}&=H(m_h)\langle p_n|B_{n\perp}^{a'\mu}(x)B_{n\perp}^{c'\nu}(0)|p_n\rangle\langle p_{\bar n}|B_{\bar n\perp}^{b'\mu}(x)B_{\bar n\perp}^{d'\nu}(0)|p_{\bar n}\rangle\nonumber\\
&\,\,\,\,\,\,\,\,\,\, \langle0|S_{n}^{aa'}(x)S_{\bar n}^{ab'}(x)S_{n}^{bc'}(0)S_{\bar n}^{bd'}(0)|0\rangle\\
&=\frac{H(m_h)}{(N_c^2-1)^2}\langle p_n|{\rm tr}[B_{n\perp}^{\mu}(x)B_{n\perp}^{\nu}(0)]|p_n\rangle\langle p_{\bar n}|{\rm tr}[B_{\bar n\perp}^{\mu}(x)B_{\bar n\perp}^{\nu}(0)]|p_{\bar n}\rangle\nonumber\\
&\,\,\,\,\,\,\,\,\,\, \langle0|S_{n}^{ac}(x)S_{\bar n}^{ad}(x)S_{n}^{bc}(0)S_{\bar n}^{bd}(0)|0\rangle
\end{align}
The hadronic states fix the sum over collinear directions to be along the protons' initial states, and we have made use of the color singlet constraint on the hadronic matrix elements. Finally, we have assumed that the
so-called Glauber mode does not contribute to the physical cross section. Proofs that these modes
don't contribute have been given in the more traditional approach to factorization \cite{Bodwin:1984hc,Collins:1988ig}
but within the EFT, where it is known that Glaubers may contribute at the level of amplitudes  \cite{Idilbi:2008vm,Bauer:2010cc}, a proof is still lacking.

\subsubsection{Factorization}
Given the factorized matrix element we now multipole expand it to generate an expression for  the cross-section which scales
homogeneously in the power counting  parameter 
\begin{multline}
\frac{d\sigma}{d p_{\perp}^2 dy}= \frac{C_t^2}{8v^2S(N_c^2-1)}\int \frac{d^4p_h}{(2\pi)^4}(2\pi)\delta^{+}(p_h^2-m_h^2) \delta\!\!\left(y-\half \ln\frac{p_h^+}{p_h^-}\right) \delta(p_{\perp}^2 - \vert \vec p_{h\perp}\vert^2)\\
4(2\pi)^8\int d^4x e^{-i x\cdot p_h}H(m_h)f_{\perp\, g/P}^{\mu\nu}(0,x^+,\vec{x}_{\perp})f_{\perp\, g/P\,\mu\nu}(x^-,0,\vec{x}_{\perp}){\cal S}(0,0,\vec{x}_{\perp})
\end{multline}
which is valid at leading order in $\lambda$.
We have defined the functions, with spin averaging implicit\footnote{In what follows, we will denote both the function and its Fourier transform by the same symbol.}:
\begin{align}
\mathcal{S}(0,0,\vec{x}_{\perp})&=\frac{1}{(2\pi)^2(N_c^2-1)} \langle0|S_{n}^{ac}(x)S_{\bar n}^{ad}(x)S_{n}^{bc}(0)S_{\bar n}^{bd}(0)|0\rangle \, , \nn\\
f_{\perp\, g/P}^{\mu\nu}(0,x^+,\vec{x}_{\perp})&=\frac{1}{2(2\pi)^3}\langle p_n|[B_{n\perp}^{A\mu}(x^{+},\vec{x}_{\perp})B_{n\perp}^{A\nu}(0)]|p_n\rangle \, , \\
f_{\perp\, g/P}^{\mu\nu}(x^-,0,\vec{x}_{\perp})&=\frac{1}{2(2\pi)^3}\langle p_{\bar n}|[B_{\bar n\perp}^{A \mu}(x^{-},\vec{x}_{\perp})B_{\bar n\perp}^{A\nu}(0)]|p_{\bar n}\rangle \nn
\end{align}
We Fourier transform now to express the factorization theorem directly in transverse momentum space:
\begin{align}
f_{\perp\, g/P}^{\mu\nu}(x^+,\vec{x}_{\perp})&=\int \frac{dz}{4\pi}e^{\frac{i}{2}z(x^{+}p_{n}^{-})}\int \frac{d^2\vec{p}_\perp}{(2\pi)^2} e^{i\vec{x}_\perp.\vec{p}_\perp}f_{\perp\, g/P}^{\mu\nu}(z,\vec{p}_{\perp})\\
f_{\perp\, g/P}^{\mu\nu}(z,\vec{p}_{\perp})&=(\bar n \cdot p_n) \langle p_n|[B_{n\perp}^{A\mu}(0)\delta(p_n z-\overline{\mathcal{P}}_n)\delta^{(2)}(\vec{p}_{\perp}-\vec{\mathcal{P}}_{\perp})B_{n\perp}^{A\nu}(0)]|p_n\rangle \, , \\
\mathcal{S}(0,0,\vec{p}_{\perp})&=\frac{1}{(N_c^2-1)} \langle0|S_{n}^{ac}(0)S_{\bar n}^{ad}(0) \delta^2(p_\perp - {\cal P}_\perp)S_{n}^{bc}(0)S_{\bar n}^{bd}(0)|0\rangle \, ,
\end{align}
where ${\cal P}$ is the SCET label-momentum operator.
Using the on-shell constraint for the hogs and the rapidity delta function, we may set $p_{h}^{\pm}=m_h e^{\pm y}$, then, in terms of the momentum space TMDPDF\footnote{TMDPDF with analogous definitions has been discussed intensely in various contents\cite{Cherednikov:2007tw, Cherednikov:2008ua, Cherednikov:2009wk, Collins:2003fm,Ji:2004wu,Hautmann:2007uw,Meissner:2008xs,Pasquini:2008ax}.}, we get:
\begin{multline}
\label{higgs_pt_final_factorization}\frac{d\sigma}{d p_{\perp}^2 dy}= \frac{\pi C_t^2H(m_h)}{2v^2S^2(N_c^2-1)}\int d^2\vec{p}_{1\perp}\int d^2\vec{p}_{2\perp}\int d^2\vec{p}_{s\perp}\delta(p_{\perp}^2-|\vec{p}_{1\perp}+\vec{p}_{2\perp}+\vec{p}_{s\perp}|^2)\\
f_{\perp\, g/P}^{\mu\nu}\Big(\frac{m_h}{\sqrt{S}}e^{-y},\vec{p}_{1\perp}\Big)f_{\perp\, g/P\,\mu\nu}\Big(\frac{m_h}{\sqrt{S}}e^{y},\vec{p}_{2\perp}\Big)\mathcal{S}(\vec{p}_{s\perp}).
\end{multline}

\subsection{Renormalization of Higgs $p_{t}$ Distribution}
Note that to renormalize the soft function and the TMDPDF in transverse momentum space, one must adopt a scheme like 't Hooft-Veltmann or CDR$_{2}$ \cite{Jain:2011iu}. This ensures that the bare operator has an integral number of mass dimensions. In 't Hooft-Veltmann, any observed degree of freedom is in four space-time dimensions. Any loop momenta, spin averages or sum, or internal polarization sums in loops are performed in $d$ space-time dimensions. This implies for the TMDPDF $f_{\perp\, g/P}^{\mu\nu}(z,\vec{p}_{\perp})$, the polarizations $\mu$ and $\nu$ are in four space-time dimensions, and the transverse momentum $\vec{p}_{\perp}$ is in 2-dimensions. The proton is spin averaged, and thus this is performed in $d$-dimensions. In CDR$_{2}$, one allows the polarizations $\mu$ and $\nu$ to be continued to $d$ dimensions, while keeping the observed transverse momentum in two dimensions. 
\subsubsection{Renormalization of the TMDPDF}\label{sec:TMDPDF-RGE}
In perturbation theory, the bare TMDPDF suffers from infra-red, ultra-violet, and rapidity divergences. We can renormalize the ultra-violet and rapidity divergences, while the infra-red divergence is part of the matrix element. Non-perturbatively, this infra-red divergence is cutoff in the hadronic matrix element. For perturbative values of transverse momentum, which we will focus on in this paper,  we can match the TMDPDF onto traditional PDF's and other higher twist hadronic matrix elements (c.f., \eqref{TMDPDF_PDF_matching}). In the matching procedure, the infra-red divergence is canceled, leaving a finite matching coefficient. 

The relation between the bare and renormalized TMDPDF is given as:
\begin{align}
\label{TMDPDF_renorm_cond}f_{\perp\, g/P}^{B\,\mu\nu}(z,\vec{p}_{\perp})&=Z^{f_{\perp}}(\mu,\omega/\nu,\vec p_\perp)\otimes_\perp f_{\perp\, g/P}^{R\,\mu\nu}(z,\vec{p}_{\perp},
\mu,\omega/\nu),
\end{align}
where $\omega$ is the large momentum component carried by the struck parton, the superscripts $B$ and $R$ mean bare and renormalized respectively, and we make use of the notation:
\begin{align}
g\otimes_\perp f(\vec{p})=\int\frac{d^2\vec{q}_{\perp}}{(2\pi)^2}g(\vec{p}_{\perp}-\vec{q}_{\perp})f(\vec{q}_{\perp}).
\end{align}
In this space we normalize the identity operator as follows
\begin{align}\label{eq:Z-consistency}
{\mathbb I} &\equiv (2\pi)^2 \delta^{(2)}(\vec k)   =  \int \frac{d^2\vec k^\prime}{(2\pi)^2} Z_{f_\perp}^{-1}(\vec k - \vec k^\prime) Z_{ f_\perp}( \vec k^{\prime }) .
\end{align}
The anomalous dimensions of the TMDPDF are then given by
\begin{align}
 \gamma_{\nu}^{f_{\perp}}(\vec{p}_{\perp},\mu)&=-(Z^{f^{\perp}})^{-1}\otimes\nu\frac{d}{d\nu} Z^{f_{\perp}}(\vec{p}_{\perp},\mu,\omega/\nu) \, , \nn\\
{\mathbb I} \, \gamma_{\mu}^{f_{\perp}}(\vec{p}_{\perp},\mu,\omega/\nu) &=-(Z^{f^{\perp}})^{-1}\mu\frac{d}{d\mu} Z^{f_{\perp}}(\vec{p}_{\perp},\mu,\omega/\nu).
\end{align}

Notice that $\gamma_\mu$ must necessarily be proportional to $\delta^{(2)}(\vec p_\perp)$ since
hard anomalous dimension must be diagonal in $\vec p_\perp$ space and the sum of
the anomalous dimensions must vanish. In principle there could be plus function dependence
on $\vec p_\perp$ in the TMDPDF $\mu$-anomalous dimension which could cancel with the
soft function contribution, but given that the TMDPDF and the soft function both
are renormalized at the same $\mu$ scale, were such contributions to the anomalous
dimensions to appear they would not contribute to any running. Thus from here on we will
drop the implied $p_\perp$ dependence in  $\gamma_\mu$ for both the TMDPDF and the soft function.

The renormalized function then satisfies the RG and RRG equations:
\begin{align}
\nu\frac{d}{d\nu} f_{\perp\, g/P}^{R\,\mu\nu}(z,\vec{p}_{\perp}, \mu,\omega/\nu)&=\gamma_{\nu}^{f_{\perp}}(\vec{p}_{\perp},\mu)\otimes_{\perp} f_{\perp\, g/P}^{R\,\mu\nu}(z,\vec{p}_{\perp}, \mu,\omega/\nu) \, , \nn\\
\mu\frac{d}{d\mu} f_{\perp\, g/P}^{R\,\mu\nu}(z,\vec{p}_{\perp}, \mu,\omega/\nu)&=\gamma_{\mu}^{f_{\perp}}(\mu,\omega/\nu) f_{\perp\, g/P}^{R\,\mu\nu}(z,\vec{p}_{\perp}, \mu,\omega/\nu) \, .
\end{align}


\subsubsection{Renormalization of the Soft and Hard Functions}
The treatment of the  bare soft and hard functions follows in the same way,
\begin{align}
H^{B}(m_h)&=Z^{H}(\mu,m_h^2)H^{R}(m_h,\mu) \, ,\\
\mathcal{S}_{i}^{B}(\vec{p}_{\perp})&=Z^{S}(\mu,\mu/\nu)\otimes_\perp \mathcal{S}^{R}(\vec{p}_{\perp},\mu,\mu/\nu) \, .
\end{align}
The anomalous dimensions of the soft function are
\begin{align}
\gamma_{\nu}^{S}(\vec{p}_{\perp},\mu)&=-(Z^{S})^{-1}\otimes_\perp \nu\frac{d}{d\nu} Z^{S}(\vec{p}_{\perp},\mu,\omega/\nu) \, , \nn\\
\gamma_{\mu}^{S}(\mu,\mu/\nu){\mathbb I}_{\cal S}  &=-(Z^{S})^{-1}\otimes_\perp \mu\frac{d}{d\mu} Z^{S}(\vec{p}_{\perp},\mu,\omega/\nu) .
\end{align} 
The renormalized soft function then satisfies the RG and RRG equations:
\begin{align}
\label{soft_evo_eqn_nu_higgs}\nu\frac{d}{d\nu} \mathcal{S}^{R}(\vec{p}_{\perp},\mu,\mu/\nu)&=\gamma_{\nu}^{S}(\vec{p}_{\perp},\mu)\otimes_{\perp} \mathcal{S}^{R}(\vec{p}_{\perp},\mu,\mu/\nu) \, , \\
\mu\frac{d}{d\mu} \mathcal{S}^{R}(\vec{p}_{\perp},\mu,\mu/\nu)&=\gamma_{\mu}^{S}(\mu,\omega/\nu)  \mathcal{S}^{R}(\vec{p}_{\perp},\mu,\mu/\nu) \, ,
\end{align}
where again the $\mu$ running can only change the large momentum component.
The hard function has anomalous dimension:
\begin{align}
\gamma_{\mu}^{H}(m_h,\mu)&=-(Z^{H})^{-1}\mu\frac{d}{d\mu} Z^{H}(\mu,m_h) \, ,
\end{align}
and satisfies the RG equation:
\begin{align}
\mu\frac{d}{d\mu}H^{R}(m_h,\mu)=\gamma_{\mu}^{H}(m_h,\mu)H^{R}(m_h,\mu) \, .
\end{align}

As in the case of the Sudakov form factor we have a set of constraints which the anomalous dimensions must obey.
The independence of the physical cross section from $\mu$ and $\nu$ gives
\begin{align}
0&=\gamma_{\mu}^{H}(m_h,\mu)+\gamma_{\mu}^{S}(\nu,\mu)+2\gamma_{\mu}^{f_{\perp}}(\omega/\nu,\mu)\\
0&=\gamma_{\nu}^{S}(\mu,\vec{p}_{\perp})+2\gamma_{\nu}^{f_{\perp}}(\mu,\vec{p}_{\perp}).
\end{align}
This provides an important consistency check on the calculations of each sector. 
Furthermore, we also have the commutativity of the $\mu$ and $\nu$ running, leading to
\begin{eqnarray}\label{eq:mu-nu-constraint-Higgs}
\mu\frac{ \, d}{d\mu} \, \gamma_\nu^{f_{\perp}} ={\mathbb I}   \, \nu \frac{ d}{d\nu}  \gamma_\mu^{f_{\perp}} =  \Gamma_{\rm cusp} \, {\mathbb I} \, ,
\end{eqnarray}
\begin{eqnarray}
\mu \frac{\, d}{d\mu} \, \gamma_\nu^{\cal S} = {\mathbb I} \,  \nu\frac{ \, d}{d\nu} \, \gamma_\mu^{\cal S} = - 2 \Gamma_{\rm cusp} \, {\mathbb I} . \nn
\end{eqnarray}
Here we have used the linearity of $\mu$-anomalous dimensions in its logarithmic term and its relationship to the cusp anomalous dimension.

\subsection{TMDPDF}
In calculating the transverse momentum dependent PDF, it is useful to consider its matching onto the PDF. This will allow us to separate the  ultra-violet from infra-red divergences. So as long as $p_\perp >\Lambda_{QCD}$, we can perform this matching so that the non-perturbatively effects lie in the PDF and its power corrections. The matching  onto 
the PDF is similar to the matching of the so-called beam function in \cite{Stewart:2009yx}\begin{multline}
\label{TMDPDF_PDF_matching}f_{\perp g/P}^{R\,\mu\nu}(z,\vec{p}_{\perp})=\sum_{k}\frac{1}{z}\int_{z}^{1}\frac{dz'}{z'}\Big\{\frac{g_{\perp}^{\mu\nu}}{2}I_{\perp 1\, g/k}(z/z',\vec{p}_{\perp}^2)\\
+\Big(\frac{\vec{p}_{\perp}^{\,\mu}\vec{p}_{\perp}^{\,\nu}}{\vec{p}_{\perp}^{\,2}}+\frac{g_{\perp}^{\mu\nu}}{2}\Big)I_{\perp 2\, g/k}(z/z',\vec{p}_{\perp}^2)\Big\}f_{k/P}^{R}(z')+O\Big(\frac{\Lambda_{QCD}}{|\vec{p}_{\perp}|}\Big),
\end{multline}
where the sum is on species of partons, and the gluon PDF\footnote{Quark mixing is irrelevant for the purposes of this paper.} is defined as
\beq
f_{g/P}(z)=  - z\,  \bar n \cdot p_n \theta(z) \, g_{\perp\mu\nu}\langle p_n\mid
\left[ B^{c \mu}_{n \perp}(0)
\delta(z \, \bar n \cdot p_n -{ \cal {\bar P}} ) B_{n \perp}^{c \nu}(0)\right] \mid p_n \rangle.
\eeq
We adopt the mostly minus metric such that conventions that $\vec{p}_{\perp}^{\,\alpha} \vec{p}_{\perp}^{\,\beta} g_{\perp\alpha\beta}=-\vec{p}_{\perp}^{\,2}$. We make use of the 't Hooft-Veltmann scheme, so the external transverse momenta remains in 2 dimensions, as do the external polarizations on the operator (the free Lorentz induces). The scheme choice is advantageous, as it allows one to renormalize the operator directly in $\vec{p}_{\perp}$ space. At tree level in perturbation theory we have for the TMDPDF and its matching coefficient to the PDF:
\begin{align}
\label{TMDPDF_tree}f_{\perp\, g/g}^{(0)\alpha\beta}(z,\vec{p}_{\perp})&=\delta(1-z)\delta^{(2)}(\vec{p}_{\perp})\frac{g^{\alpha\beta}_{\perp}}{2}\\
\label{TMDPDF_tree_matching_1}I_{\perp 1\, g/g}^{(0)}(z,\vec{p}_{\perp})&=\delta(1-z)\delta^{(2)}(\vec{p}_{\perp})\\
\label{TMDPDF_tree_matching_2}I_{\perp 2\, g/g}^{(0)}(z,\vec{p}_{\perp})&=0
\end{align}

\subsubsection{One-Loop Calculation}
  \begin{figure}
  \label{TMPDF1loop}
    \centering
    \includegraphics[width=14 cm]{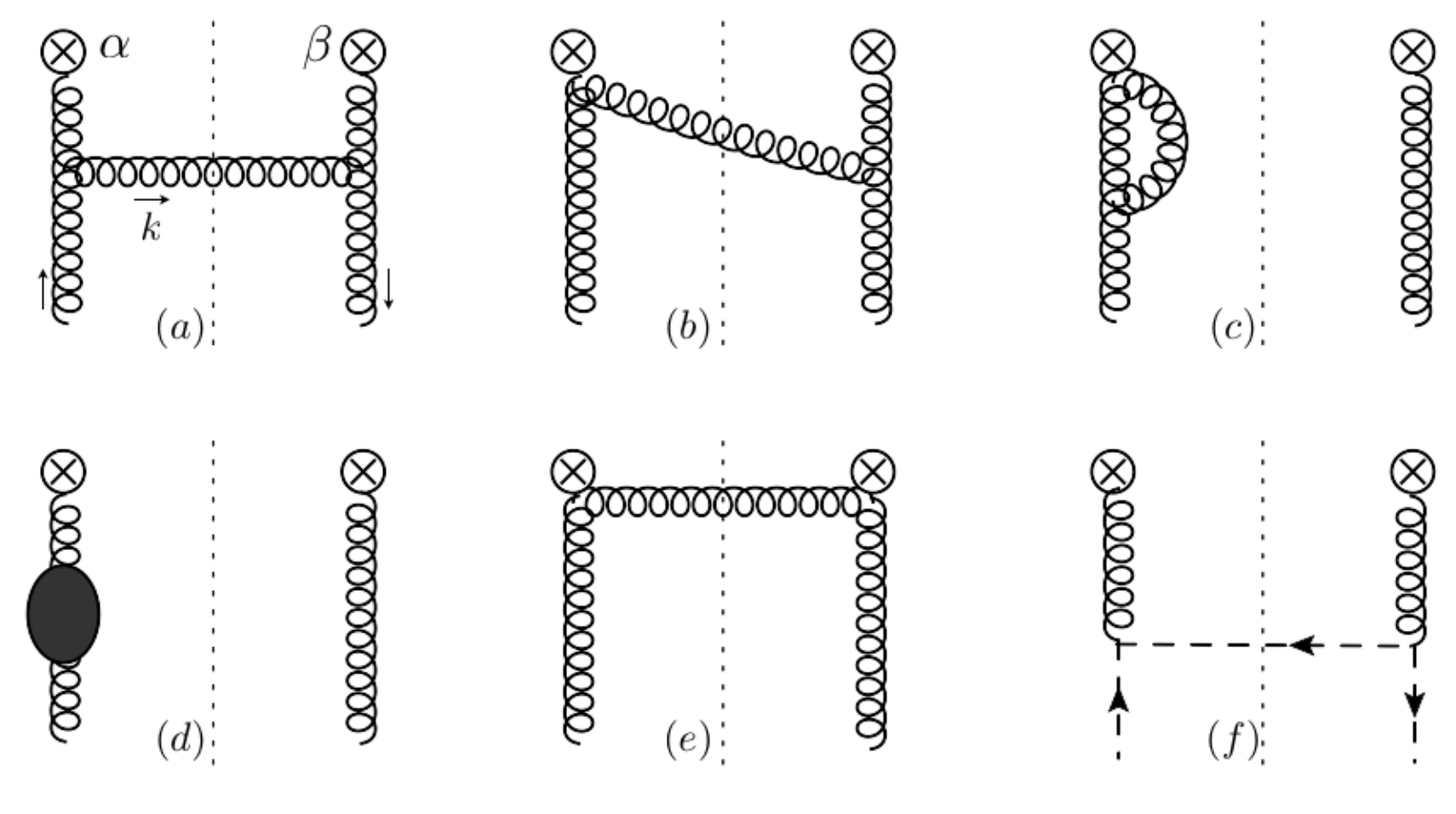}
\caption{Diagrams which contribute to the TMDPDF at one loop. Diagram (f) corresponds to
quark mixing that does not contribute to the one loop running. Diagram (e) vanishes in Feynman gauge.
}
\label{TMPDF1loop}
\end{figure}
At one loop we have for the sum of diagrams in fig. \ref{TMPDF1loop}(a)-(e),
\begin{multline}
f_{\perp\, g/g}^{(1)\alpha\beta}(z,\vec{p}_{\perp})= \frac{g^2C_A\mu^{2\epsilon}}{2}\int \frac{d ^d k}{(2\pi)^{d}} 
\frac{\delta \big(\omega_-(1-z)-k^-\big)}{\omega_-}\delta^{(2)}(\vec{p}_{\perp}-\vec{k}_{\perp})\frac{\delta^{(+)}(k^2)}{(k^+)^2}\\
\times\left[g_{t}^{\alpha\beta}\left(\frac{\omega_- k^+}{2}+\vec{k}_{\perp}^{\,2}+\nu^{\eta}\frac{\omega_- k^+(2\omega_--k^-)}{2(k^-)^{1+\eta}}\right)+(1-\epsilon)\frac{2\omega_-^2\vec{k}_{\perp}^{\,\alpha} \vec{k}_{\perp}^{\,^\beta}}{(\omega_--k^-)^2}\right].
\end{multline}
Note that $\vec{p}_{\perp}$ is strictly two-dimensional, and $\vec{k}_{\perp}$ is in $2-2\epsilon$ dimensions. The momentum conservation delta function constrains only the components of $\vec{k}_{\perp}$ that overlap physical space.  $\omega_-$ is the large component of the light-cone momenta of the incoming parton. Performing the integrals we get the bare TMDPDF at one-loop:
\begin{multline}
f_{\perp\, g/g}^{(1)\alpha\beta}(z,\vec{p}_{\perp})= \frac{g^2C_A\pi^{-\epsilon}}{(2\pi)^{3-2\epsilon}}\frac{\Gamma(1+\epsilon)}{2}\frac{\mu^{2\epsilon}}{(\vec{p}_{\perp}^2)^{(1+\epsilon)}}\\
 \times  \left[\left(-2z+3+\left(\frac{\nu}{\omega_-}\right)^\eta\frac{1+z}{(1-z)^{1+\eta}}\right)g_\perp^{\alpha\beta}-4(1-\epsilon^2)\frac{1-z}{z^2}\frac{\vec{p}_{\perp}^{\,\alpha} \vec{p}_{\perp}^{\,\beta}}{\vec{p}_{\perp}^{\,2}}\right].
\end{multline}
The virtual diagrams vanished in dimensional regularization. 
Their effects however are non-vanishing and will be accounted for when we match onto
the PDF below.
Plus function expanding in $\eta$ and inserting the $\overline{MS}$ factor gives:
\begin{multline}
f_{\perp\, g/g}^{(1)\alpha\beta}(z,\vec{p}_{\perp})= \frac{g^2C_A\pi^{-\epsilon}}{(2\pi)^{3-2\epsilon}}\frac{\Gamma(1+\epsilon)}{2}\frac{\mu^{2\epsilon}e^{\epsilon \gamma_E}}{(\vec{p}_{\perp}^{\,2})^{(1+\epsilon)}}
 \left[ \Big(-2 \frac{\delta(1-z)}{\eta}\big(\frac{\nu}{\omega_-}\big)^\eta + \frac{p_{gg*}}{z}(z)-2\epsilon^2\frac{(1-z)}{z^2}\Big)g_{\perp}^{\alpha\beta}\right.\\
 \left.
-4\frac{(1-z)}{z^2}(1-\epsilon^2) \left(\frac{\vec{p}_{\perp}^{\,\alpha} \vec{p}_{\perp}^{\,\beta}}{\vec{p}_{\perp}^{\,2}}+\frac{g_{\perp}^{\alpha\beta}}{2}\right)\right] \, .
\end{multline}
Where we have written the expression in terms of the gluon splitting function:
\begin{align}
p_{gg*}(z)  &= \frac{2z}{[1-z]_{+}}+ 2\,\theta(1-z)\Bigl[\frac{1-z}{z} +  z(1-z)\Bigr].
\end{align}

\subsubsection{Matching and Renormalization}
%
As we used dimensional regularization to regulate both the UV and IR of the TMDPDF, we match onto the PDF which allows us  to distinguish the $\frac{1}{\epsilon_{UV}}$ and $\frac{1}{\epsilon_{IR}}$ poles. Since the effective theory containing the PDFs (where the scale set by the transverse momentum has been integrated out) must have the same infra-red divergence as the effective theory containing the TMDPDFs, any IR poles will cancel in the matching procedure. Any poles left over must be UV in origin, and are removed by the $Z$-factor in the high scale  theory. The effects of the virtuals, which vanish in dimensional regularization, will arise via the conversion of an IR pole into a UV pole. To determine the matching and $Z$-factor to one loop, we derive a relation by expanding equations \eqref{TMDPDF_PDF_matching} and \eqref{TMDPDF_renorm_cond} to order $\alpha_s$:
\begin{multline}
\label{One_Loop_Matching_Relation_TMDPDF}-Z^{(1)}_{f_{\perp}}\otimes_{\perp} f_{\perp\, g/g}^{(0)\alpha\beta}(z,\vec{p}_{\perp})+ f_{\perp\, g/g}^{(1B)\,\alpha\beta}(z,\vec{p}_{\perp})\\
=\frac{1}{z} \int_{z}^{1} \frac{dz^{'}}{z^{'}}\frac{g_{\perp}^{\alpha\beta}}{2}\left(I^{(1)}_{\perp 1\,g/g}(\frac{z}{z^{'}},\vec{p}_{\perp})f^{(0)}_{g/g}(z^{'})+I^{(0)}_{\perp 1\,g/g}(\frac{z}{z^{'}},\vec{p}_{\perp})f^{(1R)\,}_{g/g}(z^{'})\right) \, .
\end{multline}
We split the bare TMDPDF into divergent and finite pieces, $f_\perp^{(1B)} = f_\perp^{\rm (1div)} + f_\perp^{\rm (1fin)}$,
\begin{align}
\label{TMDPDF_divergences_pt}f_{\perp\, g/g}^{\rm (1div)\alpha\beta}(z,\vec{p}_{\perp})&= -\frac{\alpha_sC_A}{2\pi^2}\Gamma(1+\epsilon)\frac{e^{\gamma_E\epsilon}\mu^{2\epsilon}}{(\vec{p}_{\perp}^{\,2})^{(1+\epsilon)}}\frac{\delta(1-z)}{\eta}g_{\perp}^{\alpha\beta} \nn \\
&- \frac{\alpha_sC_A}{4\pi}\frac{\delta^{(2)}(\vec{p}_{\perp})}{\e}g_{\perp}^{\alpha\beta} \left(\frac{p_{gg*}(z)}{ z}-\delta(1-z)\ln \frac{\nu^2}{\omega_-^2} \right)\\ 
\label{TMDPDF_finite_pt}f_{\perp\, g/g}^{(1\rm fin)\alpha\beta}(z,\vec{p}_{\perp})&=\frac{\alpha_sC_A}{\pi}{\cal L}_0\Big(\mu,\frac{\vec p_\perp}{\mu}\Big)\frac{ g_\perp^{\alpha\beta}}{2}\Bigg(-\ln\Big(\frac{\nu^2}{\omega_-^2}\Big)\delta(1-z)+\frac1{z}p_{gg*}(z)\Bigg)\nonumber\\
&-2\frac{\alpha_sC_A}{\pi}\frac{1-z}{z} {\cal L}_0\Big(\mu,\frac{\vec p_\perp}{\mu}\Big)\left(\frac{\vec{p}_{\perp}^{\,\alpha} \vec{p}_{\perp}^{\,\beta}}{\vec{p}_{\perp}^{\,2}}+\frac{g_{\perp}^{\alpha\beta}}{2}\right) \,.
\end{align}
where we have written the expression in terms of plus distribution ${\cal L}_n=\frac{1}{2\pi\mu^2}\Big[\frac{\mu^2}{\vec{p}^{\,2}}\ln^n\Big(\frac{\mu^2}{\vec{p}^2}\Big)\Big]_{+}^{1}$, whose definition and properties are collected in the appendix \eqref{plus_distr_appendix}. Given that the renormalized PDF at one-loop only contains IR divergences,
\begin{align}
f^{(1R)}_{g/g}(z)&=-\frac1 {\epsilon}\frac{\alpha_s}{2\pi} \left[C_A p_{gg*}(z)+\frac12 \beta_0 \delta (1-z)\right] \, ,
\end{align}
we compare $\epsilon$ divergences in eqn. (\ref{One_Loop_Matching_Relation_TMDPDF}) to obtain the renormalization constant $Z_{f_\perp}$ at one loop
\begin{align}
\label{TMDPDF_Z_pt}Z^{(1)}_{f_{\perp}}(z,\vec{p}_{\perp})&=(2\pi)^2\delta^{(2)}(\vec{p}_{\perp})-4\alpha_sC_A\left(w^2\Gamma(1+\epsilon)\frac{e^{\gamma_E\epsilon}\mu^{2\epsilon}}{(\vec{p}_{\perp}^{\,2})^{(1+\epsilon)}}\frac{1}{\eta} 
- \pi \frac{\delta^{(2)}(\vec{p}_{\perp})}{\e}\Big( \frac{1}{2}\ln \frac {\nu^2}{\omega_-^2}+\frac{1}{4 C_A} \beta_0\Big) \right) \, .
\end{align}
Comparing finite pieces in (\ref{One_Loop_Matching_Relation_TMDPDF}), we get the matching coefficients
\begin{align}
\label{One_Loop_Matching_TMDPDF}I^{(1)}_{\perp 1\, g/g}(z,\vec{p}_{\perp})&=\frac{\alpha_sC_A}{\pi}{\cal L}_{0}\left(\mu,\vec{p}_{\perp}\right)\Bigg(-\ln\Big(\frac{\nu^2}{\omega_-^2}\Big)\delta(1-z)+p_{gg*}(z)\Bigg) \, , \nn\\
I^{(1)}_{\perp 2\, g/g}(z,\vec{p}_{\perp})&=-2\frac{\alpha_sC_A}{\pi}\frac{1-z}{z} {\cal L}_0(\mu,\vec p_\perp) \, .
\end{align}
Note that there is no real singularity as $p_\perp \to 0$ in $I_{\perp 2}$ since the traceless tensor itself vanishes in that limit and hence the plus prescription in $I^{(1)}_{\perp 2\, g/g}$ may be dropped.
There  are finite contributions to the matching from the quark PDF's. These contributions do not effect the running of the TMDPDF and  hence for the purposes of this paper, we neglect these contributions.

The one loop anomalous dimensions can be calculated from $Z_{f_\perp}^{(1)}$,
\begin{align}
\label{TMPDFgammas}
\gamma_{\mu}^{f_{\perp}}(\nu) &= \frac{\alpha_sC_A}{\pi}\ln\Big(\frac{\nu^2}{\omega_-^2}\Big)+ \frac{\alpha_s \beta_0}{2\pi} \, ,\\
\gamma_{\nu}^{f_{\perp}}\Big(\mu,\frac{\vec{p}_{\perp}}{\mu}\Big)&=-8\pi \alpha_sC_A {\cal L}_{0}\Big(\mu,\frac{\vec{p}_{\perp}}{\mu}\Big).
\end{align}
Notice that as long as the scale $\mu$ is taken at the low scale this anomalous dimensions captures all
the physics at NLO. That is, we need not integrate the $\mu$ anomalous dimensions ($\gamma_\mu$) as in
(\ref{eq:full-log}) to calculate $\gamma_\nu$ since the difference involves no large logarithms.
Using the identity
\beq
\mu \frac{d}{d\mu}{\cal L}_0\Big(\mu,\frac{\vec{p}_{\perp}}{\mu}\Big)= -\delta^{(2)}(\vec{p}_\perp)
\eeq
we see that the results for the anomalous dimensions are consistent with RRG  commutativity (\ref{eq:mu-nu-constraint-Higgs}).


\subsection{The $p_T$ Dependent Soft Function}
The tree level soft function is simply:
\begin{align}
\mathcal{S}^{(0)}(\vec{p}_{\perp})=\delta^{(2)}(\vec{p}_{\perp}) \, .
\end{align}
The calculation of the one-loop soft function proceeds as:
\begin{equation}
\mathcal{S}^{(1)}(\vec{p}_{\perp})=4C_A g^2 \mu^{2\epsilon}\nu^{\eta}\int \frac{\dd^dk}{(2\pi)^{d}}\left|2k^3\right|^{-\eta}\frac{\delta^{(+)}(k^2)\delta^{(2)}(\vec{p}_{\perp}-\vec{k}_{\perp})}{k^{-}k^{+}}.
\end{equation}
Using the delta functions and performing the transverse momentum integrals gives 
\begin{equation}
\mathcal{S}^{(1)}(\vec{p}_{\perp})=\frac{2C_A g^2(\pi)^{-\epsilon}}{(2\pi)^{3-2\epsilon}}\frac{\Gamma(1+\epsilon+\eta/2)}{\Gamma(1+\eta/2)}\frac{\nu^\eta\mu^{2\epsilon}}{(\vec{p}_\perp^{\,2})^{1+\epsilon}}\int_{0}^{\infty} \frac{\dd k^-}{k^-}\left|k^--\frac{\vec{p}_{\perp\,}^2}{k^-}\right|^{-\eta} \, .
\end{equation}
The last integral contains the rapidity divergences coming from both the large $k^-$ and $k^+=\frac{\vec{p}_{\perp\,}^2}{k^-}$ limits. This is critical for the soft function in order to cancel the rapidity divergences found in both of the jet sectors, as well as consistent with the isotropic nature of soft radiation.
Finally for the bare soft function we have (including the $\overline{MS}$ factor):
\begin{equation}
\mathcal{S}^{(1)}(\vec{p}_{\perp})=\frac{2C_A g^2}{(2\pi)^3} \frac{e^{\epsilon\gamma_E}\nu^\eta\mu^{2\epsilon}}{(\vec{p}_{\perp}^{\,2})^{1+\epsilon+\frac{\eta}{2}}}\frac{\Gamma(1+\epsilon+\frac{\eta}{2})}{\Gamma\left(1+\frac{\eta}{2}\right)}\frac{2^{-\eta}\Gamma(\frac{1}{2}-\frac{\eta}{2})\Gamma(\frac{\eta}{2})}{\sqrt{\pi}} \, .
\end{equation}
Since the soft function is IR safe, expanding the $p_{\perp}$-space soft function, we have for the divergences and renormalized part:
\begin{align}
Z_{\cal S}^{(1)}(\vec{p}_{\perp})&=(2\pi)^2\delta^{(2)}(\vec{p}_{\perp})+ 4\alpha_sC_A \left[ 2w^2\Gamma(1+\epsilon)\frac{1}{\eta}\frac{e^{\epsilon\gamma_E}\mu^{2\epsilon}}{(\vec{p}_{\perp}^{\,2})^{1+\epsilon}}+\pi\delta^{(2)}(\vec{p}_{\perp})\left( \frac{1}{\epsilon^2} -\frac{ \ln \frac{\nu^2}{\mu^2}}{\epsilon}\right)\right] \, , \nn \\
S^{(1{\rm R})}(\vec{p}_{\perp})&=2\frac{\alpha_sC_A}{\pi}\Bigg(-\frac{\pi}{24}\delta^{(2)}(\vec{p}_{\perp})+\LN\Big(\frac{\nu^2}{\mu^2}\Big){\cal L}_{0}\Big(\mu,\frac{\vec{p}_{\perp}}{\mu}\Big)+{\cal L}_{1}\Big(\mu,\frac{\vec{p}_{\perp}}{\mu}\Big)\Bigg) \, ,
\end{align}
where $w$ is the book-keeping parameter that tracks the number of eikonal vertices, see section (\ref{RRD}). The plus distribution ${\cal L}_n=\frac{1}{2\pi\mu^2}\Big[\frac{\mu^2}{\vec{p}^{\,2}}\ln^n\Big(\frac{\mu^2}{\vec{p}^2}\Big)\Big]_{+}^{1}$ is defined in appendix (\ref{plus_distr_appendix}).

Then the anomalous dimensions at one loop order are:
\begin{align}
\label{softanom}
\gamma_\mu^{\cal S} &= -2 \frac{\alpha_sC_A}{\pi}\ln\Big(\frac{\nu^2}{\mu^2}\Big) \, , \nn\\
\gamma_\nu^{\cal S}(\vec{p}_{\perp},\mu)&=16 \pi \alpha_sC_A {\cal L}_{0}\Big(\mu,\frac{\vec{p}_{\perp}}{\mu}\Big) \, .
\end{align}
This verifies the constraint (\ref{eq:mu-nu-constraint-Higgs}) for the soft anomalous dimensions, and on comparing with (\ref{TMPDFgammas}) we see that we correctly reproduce the constraints
\beq
\gamma_\nu^S+2 \gamma_\nu^{f\perp}=0 \, .
\eeq
Using the result for the hard anomalous dimension,
\beq
\gamma^H = - \frac{\alpha_s}{\pi}\left (2C_A \ln \frac{\mu^2}{Q^2}+\beta_0 \right) \, ,
\eeq
in conjunction with (\ref{TMPDFgammas}) (and its partner with $\omega_-\!\!\rightarrow \!\omega_+$) and (\ref{softanom}) we find that the constraint
\beq
\gamma^H+\gamma^{\cal S}_\mu+2\gamma_\mu^{f_\perp}=0
\eeq
is also satisfied after making the identification $\omega_- \omega_+=Q^2$.


\subsection{Resummation of Rapidity Logarithms and Handling Undesired Singularities} \label{sec:Higgs-resum}
To calculate the resummed cross-section we must evolve the  soft function  in $\nu$ up to the jet scale $\nu_J \sim \omega$ as  shown in fig.~\ref{fig:strategy}. For this purpose we need to solve the $\nu$-RGE and obtain $V_{\cal S}$ at NLL.  First we solve in $b$-space, and transform back to present
the $p_\perp$ space solution. We will show that a naive solution will lead to a well-known unwanted singularity \cite{Frixione:1998dw} of the classic CSS result \cite{Collins:1984kg}. Then we will discuss a careful solution to avoid the undesired singularity.

Solving \eqref{soft_evo_eqn_nu_higgs} in impact-parameter space gives:
\begin{align}
\tilde{S}(b,\mu,\nu)&=\text{Exp}\Bigg[-\frac{2\alpha_s C_A}{\pi}\ln\Big(\frac{\mu^2b^2e^{2\gamma_E}}{4}\Big)\ln\Big(\frac{\nu}{\nu_0}\Big)\Bigg]\tilde{S}(b,\mu,\nu_0) \nn\\
&=\Big(\frac{\mu^2b^2e^{2\gamma_E}}{4}\Big)^{-\omega_s}\tilde{S}(b,\mu,\nu_0) \, , 
\end{align}
where
\begin{align}
\label{eq:omega-Higgs}
\omega_s(\mu, \nu/\nu_0) &= \frac{2\alpha_s C_A}{\pi}\ln\Big(\frac{\nu}{\nu_0}\Big) \, .
\end{align}
Performing the inverse transform gives the resummed soft function in $p_\perp$ space,
\begin{align}
S(\vec{p}_{\perp},\mu,\nu)&=\int \frac{d^2\vec{p}\,'_{\perp}}{(2\pi)^2}V_{S}(\vec{p}\,'_{\perp},\mu,\nu,\nu_0)S(\vec{p}_{\perp}-\vec{p}\,'_{\perp},\mu,\nu_0) \, ,
\end{align}
where,
\begin{align}
\label{eq:V_S-FT}
V_{S}(\vec{p}_{\perp},\mu,\nu,\nu_0)&=  2\pi \int_0^\infty db \, b\, J_0(b \vert \vec p_\perp\vert ) \Big(\frac{\mu^2b^2e^{2\gamma_E}}{4}\Big)^{-\omega_s} \\
\label{eq:V_S-pT}
& = 4\pi e^{-2\omega_s \gamma_E}\frac{\Gamma(1-\omega_s)}{\Gamma(\omega_s)}\frac{1}{\mu^2}\Big[\Big(\frac{\mu^2}{\vec p_{\perp}^2}\Big)^{1- \omega_s}\Big]_{+}^{\infty} \, .
\end{align}
Now, the NLL cross-section for $p_\perp>\Lambda_{QCD}$ is given by
\begin{align}\label{eq:Higgs-NLL-cross-section}
\frac{d\sigma}{d p_\perp^2dy} \bigg\vert_{\vec p_\perp^2 > 0} &=\frac{\pi^2 C_t^2H(m_h)}{2v^2S^2(N_c^2-1)}U_H(m_h^2,\mu^2)\Big(\frac{S}{2m_h^2}\Big) \frac{e^{-2\omega_s\gamma_E}}{\pi}\frac{\Gamma(1- \omega_s)}{\Gamma(\omega_s)}\frac{1}{\mu^2} \Big(\frac{\mu^2}{\vec p_{\perp}^2}\Big)^{1-\omega_s}  \nn \\
&\quad\quad\quad \times f_{g/P}\Big(\frac{m_h}{\sqrt{S}}e^{-y}\Big)f_{g/P}\Big(\frac{m_h}{\sqrt{S}}e^{y}\Big) \, .
\end{align}
Note the singularity at $\omega_s = 1$ in eqn. (\ref{eq:Higgs-NLL-cross-section}) and (\ref{eq:V_S-pT}). This is unavoidable because $\omega_s > 0$ and is typically $\sim 1$. Thus relation (\ref{eq:Higgs-NLL-cross-section}) is not useful for phenomenology, but is useful for generating the fixed order logs at higher orders in perturbation theory.

This singularity arises due to the naive inverse transform performed in eqn. (\ref{eq:V_S-FT}). Note that integral in (\ref{eq:V_S-FT}) gets a singular contribution from $b \ll 1/p_\perp \sim 1/\mu$. In particular when $\omega_s = 1$, the integrand goes like $1/b$ for small $b$ and integral diverges. This shows up as singularity at $\omega_s = 1$ in (\ref{eq:V_S-pT}). This is a UV problem since $b$ is small, and is an unexpected situation because only impact parameters of order $1/p_\perp$ are expected to contribute to the inverse Fourier transform. Therefore, care must be taken at this step to avoid contributions from the region $b \ll 1/p_\perp$.

By making a choice for the scale $\nu_0 = 1/b$ in $\omega_s$, before performing the inverse transform in eqn. (\ref{eq:V_S-FT}), the  $b \ll 1/p_\perp$ region is exponentially suppressed, removing the singularity\footnote{We thank Wouter Waalewijn for making this suggestion.}. One would typically choose $\nu_0 \sim p_\perp$ in the $p_\perp$ space NLL cross-section, to ensure all the large logarithms are resummed. Making a choice $\nu_0 = 1/b$ is an equivalent one up to higher order effects in resummation\footnote{Typically we can only argue $\nu_0 \sim 1/b$ but the arbitrariness is still captured by varying $\nu \sim m_h$ in a reasonable range, thus $\nu_0 = 1/b$ is justified.}. With this choice we have 
\begin{align}
\label{eq:V_S-reviewed}
\hat V_{S}(\vec{p}_{\perp},\mu,\nu)&=  2\pi \int_0^\infty db \, b\, J_0(b \vert \vec p_\perp\vert ) \Big(\frac{\mu^2b^2e^{2\gamma_E}}{4}\Big)^{-\omega_s(\mu,\nu b)} \, ,
\end{align}
and soft function is given by 
\begin{align}
S(\vec{p}_{\perp},\mu,\nu)&=\int \frac{d^2\vec{p}\,'_{\perp}}{(2\pi)^2}\hat V_{S}(\vec{p}\,'_{\perp},\mu,\nu) \hat S(\vec{p}_{\perp}-\vec{p}\,'_{\perp},\mu) \, ,
\end{align}
where
\begin{align}
\hat S(\vec{p}_{\perp},\mu) = 2\pi \int_0^\infty db \, b J_0(b \vert \vec p_\perp \vert ) \tilde S(b, \mu , \nu_0 = 1/b) \, .
\end{align}
It is difficult to obtain a closed form expression of (\ref{eq:V_S-reviewed}), but it is certainly implementable numerically and is free from the undesired singularities. We leave this implementation for a future work. Since the scale $\nu_0$ does not appear in the coupling, the choice of scale setting does not involve any Landau pole, as $\mu$ is left arbitrary in the transforms.

A similar approach was taken in \cite{Becher:2010tm} to cure the singularities, however, there the singularity was interpreted to be some indication of non-perturbative physics. In contrast, it was argued in \cite{Frixione:1998dw}, the problem is completely perturbative: it is solvable by a rearrangement of sub-leading terms of the resummed series, and occurs in regions where $\alpha_s$ is still perturbative. We agree with this and have verified that the problem simply arises from a naive inverse transform that includes contributions from a UV region inappropriate to the soft matrix elements being resummed.

\subsection{Fixed Order Cross-section}

Substituting the tree level matching onto PDFs, and integrating over $p_\perp$ and $y$ we obtain properly normalized integrated cross-section at LO,
\begin{align}
\sigma_{0}&=\frac{\pi C_t^2H(m_h)}{2v^2S^2(N_c^2-1)}\frac{S^2}{2m_h^2}\int dz_1dz_2\delta\Big(m_h^2-z_1z_2S\Big)f_{g/P}(z_1)f_{g/P}(z_2).
\end{align}
This agrees the leading order cross-section given in \cite{Harlander:2001is}.

For the next to  leading order transverse momentum spectrum we get
\begin{align}
\frac{d\sigma}{dp_{\perp}^2dy}\Big\vert_{p_{\perp}>0}&=\frac{\pi C_t^2H(m_h)}{2v^2S^2(N_c^2-1)}\Big(\frac{S}{m_h^2}\Big)\frac{\alpha_s C_A}{4\pi}\frac{1}{p_{\perp}^2}\Bigg\{2\text{ln}\Big(\frac{m_h^2}{p_{\perp}^2}\Big)f_{g/P}\Big(\frac{m_h}{\sqrt{S}}e^{y}\Big)f_{g/P}\Big(\frac{m_h}{\sqrt{S}}e^{-y}\Big)  \\
&+f_{g/P}\Big(\frac{m_h}{\sqrt{S}}e^{-y}\Big)(p_{gg*}\otimes f_{g/P})\Big(\frac{m_h}{\sqrt{S}}e^{y}\Big)+f_{g/P}\Big(\frac{m_h}{\sqrt{S}}e^{y}\Big)(p_{gg*}\otimes f_{g/P})\Big(\frac{m_h}{\sqrt{S}}e^{-y}\Big)\Bigg\} \nn
\end{align}
where $g\otimes h(z)=\int_{z}^{1}\frac{dx}{x}g(x)h(z/x).$
Note that an overall factor of $\pi$ comes from the angular integrations in the transverse momentum convolution variables. This agrees with the result found in \cite{Kauffman:1991jt}.

\subsection{Comparison to Previous Resummation Formalisms}

Working within an EFT formalism, a result nearly identical to \eqref{eq:Higgs-NLL-cross-section} was derived in \cite{Becher:2010tm} at NLL. However, the rapidity logs were exponentiated by hand after summing over sectors\footnote{This sum over sectors is necessary due to the regulator chosen to give meaning to their low-scale matrix elements. Given the regulator implemented in \cite{Becher:2010tm}, each sector is not well-defined and nor renomalizable, but only the combination is.}, without introducing a new scale. Thus  it is not clear how the scale dependence arising at higher orders can be tracked. As explained in detail in the conclusions of sections (\ref{needRRG}) and  (\ref{JB}), there is a fundamental ambiguity in the exponentiation of the rapidity logarithms, since there is freedom in choosing what goes into the exponent one is free to include sub-leading logarithms in the resummation power counting in the exponent or the low-scale matrix elements. In our formalism this ambiguity corresponds to the choice of $\nu$ matching scale.  Varying this scale shuffles sub-leading contributions into or out of the  matrix element. Hence the residual $\nu$ dependence (which is not exponentiated) of the resummed cross-section, can be included in the theoretical error of our prediction by varying $\nu$ in the same way one varies $\mu$ to get a handle on errors form sub-leading term in traditional RG calculations. Such an analysis, at least in the context of Higgs tranverse momentum distributions,  is presently
absent from the literature \footnote{Such an analysis can also be performed in the CSS formalism using Collins most
recent definition of the TMDPDF. For a  discussion see \cite{Aybat:2011zv}.}.

It is also worth noting that our calculation of the transverse momentum distribution is distinct from the work \cite{Becher:2010tm}. Therein, they do not include the soft mode of the effective theory. It is stated that the ``soft mode'' cancels because the typical soft momenta is order $(\lambda^2,\lambda^2,\lambda^2)$. We would call this mode ultra-soft, which is the relevant mode for \sceti. The use of the analytic regulator \footnote{The traditional analytic regulator breaks the eikonal identities needed for exponentiation of the soft function. This complicates the claim made for factorization, though this breaking of eikonal identities has been remedied in a more recent the paper \cite{Becher:2011dz}.}  renders integrals in the actual soft function (having modes $Q(\lambda,\lambda,\lambda)$) scaleless and hence zero.  While technically correct, this method seems to obscure the physics, given that soft radiation clearly plays a role at small $p_\perp$, and thus is must be hidden in other sectors. Finally, the implementation of the analytic regulator in \cite{Becher:2010tm} renders the expanded results in the two  collinear sectors different in form, even though the operators in the effective theory look identical, and thus  one can not define universal TMDPDFs.
  
The classic CSS formalism utilized the so-called Collins-Soper equation to accomplish the rapidity resummation. In particular, it played a key role in establishing the formal $\mu$-independence of the double logarithmic terms in the resummation exponent \cite{Collins:1981uk}. Using this formalism, a resummed formula for transverse momentum distributions was derived \cite{Collins:1984kg}. Nonetheless, the classic CSS formalism suffers from a number of mild deficiencies according to one of the authors\cite{Collins:2011ca}. In particular, the Collins-Soper equation itself suffered from hard to control power corrections, and the hard matching coefficient is ambiguous. These issues stem from the way the rapidity divergences were regulated in the original Collins-Soper approach, where a non-light like axial gauge regulated the divergences. The regulating parameter did not cancel between the low-scale matrix elements, so the choice of the axial gauge vectors that defined the collinear matrix element also affected the hard matching. Nor was the regulating parameter divergences ever removed from the matrix element, so the regulating parameter could not be set to zero. This malady remained true in more modern versions of the Collins-Soper approach adopted in \cite{Collins:1989bt,Dixon:2008gr} where the Wilson lines are deformed off the light-cone.

The authors of \cite{deFlorian:2001zd,Bozzi:2007pn,Catani:2010pd} who have performed the highest order resummations of the transverse momentum spectrum to date have used the resummation formula given in \cite{Collins:1984kg}. The anomalous dimensions and matching needed for resummation were derived by comparison to full QCD calculations in soft and collinear limits, not by direct calculation of the low-scale matrix elements in the Collins-Soper formalism. From this procedure, it is not clear that they can gauge the residual effects of the rapidity resummation, since they do not make use of the Collins-Soper equation directly, nor calculate the resummation from the factorized matrix elements.

Recently Collins has improved further upon the CSS approach in \cite{Collins:2011ca}, fixing the above problems. This new method has many similarities to our approach. In \cite{Collins:2011ca}, the square root of the soft function is included in the collinear sector, and a series of soft-bin subtractions (explicitly represented by inverse soft functions) are carried out to cancel the regulator dependence in the sector. This introduces explicitly a $\ln \frac{p^-}{\mu}$ in the TMDPDF, where $p^-$ is the large  light-cone momentum. The soft factors  remove the rapidity regulator dependence and introduce a $\mu$ dependence in its place. Thus if $p^-$ is large, a single $\mu$ scale cannot eliminate all large logarithms. However, through use of the Collins-Soper equation, the extra large logarithms can be exponentiated by evolving the TMDPDF from low energies, where $p^-\sim p_\perp$,  to the relevant high energies. In the low energy region, all logs can by minimized by a single choice of $\mu$. All logarithms remain minimized when evolving the low-energy TMDPDF via the Collins-Soper equation to the high scale where the experiment takes place. This accomplishes the same effect as our rapidity RG, but we do not need the low-energy TMDPDF as an initial condition (just the PDF). We can take the TMDPDF at fixed energy and transverse momentum to be renormalized at any $\mu,\nu$ point, and evolve to any other point in order to minimize logarithms. Finally, we must mention that there is one important way in which the claims\footnote{We use this terms simply because we have not been able to reconstruct the proof ourselves.} \cite{Collins:2011ca} are stronger then ours. In particular the proof  in \cite{Collins:2011zzd} allows for the possible contribution of Glauber gluons.
In SCET an understanding of Glauber gluons in hard scattering processes \cite{Stewart:SCET2009,Bauer:2010cc} is still lacking.


Another attempt at defining a TMDPDF within the SCET context was presented in \cite{GarciaEchevarria:2011rb}. Using a $\delta$-style regulator, and including a square-root of the inverse soft function in the collinear sector, they were able to eliminate the rapidity divergences within the bare matrix element as in Collins' approach. They do not discuss resummation, nor indicate how, after eliminating the rapidity divergence with no auxiliary parameter, they will accomplish the specific resummation of the rapidity logarithms.  As it stands no choice of $\mu$ will minimize all logarithms in their TMDPDF if the light-cone momentum is large.

Finally, it is worth noting that the approaches taken in \cite{GarciaEchevarria:2011rb} assumes the equivalence of soft-bin subtractions and the inverse soft function. While this is true in many situations for many regulators, this is complicated in the case of a soft function depending on multiple parameters. This is illustrated below, when we factorize the cross-section for left and right broadening. Here the soft function depends on the broadening in both hemispheres, while the jet function only depends on the broadening of a single hemisphere. Thus the soft bin of the jet sector cannot be the the same as the inverse soft function, as the soft-bin will continue to only depend on the broadening of a single hemisphere. Including the square-root of the (inverse) soft function to make the jet function free of rapidity divergences would introduce dependence on the dynamics of the other hemisphere. The fact that the soft function depends on both hemispheres complicates a straightforward definition of a square-root of the soft function, especially since non-global logarithms seem to be a generic feature of multi-region soft functions \cite{Dasgupta:2001sh,Kelley:2011ng,Hornig:2011iu}. Thus it is hard to see how one would generalize Collins' recent approach to the TMDPDF to cases such as jet broadening where  one cannot eliminate the soft function by splitting it up the between the jet functions without inducing dependence on the broadening of both hemispheres in each jet function.

Finally, for other works within SCET transverse momentum factorization \cite{Mantry:2009qz,Gao:2005iu,Idilbi:2005er} only the hard logarithms from running the currents are resummed. The rapidity logarithms are left unsummed.

\section{Jet Broadening }
\label{JB}

Event shapes have  played an important role in precision measurements of the strong coupling $\alpha_s$\cite{Kluth:2006bw}.  A generalized event shape for event $e^-e^+\rightarrow X$ at center of mass energy $\sqrts$,  can be defined \cite{Berger:2003pk} in terms of a parameter $a$ via
\beq
\label{ang}
e(a)= 
\sum_{i\in X}
 \frac{\vert \vec p_{i\perp} \vert }{\sqrts}e^{-\vert \eta_i\vert(1-a)}
\eeq
where $p_{i\perp}$ is the transverse momentum with respect to the thrust axis $\hat t$ of the event, and $\eta_i$ is the rapidity of the $i$'th particle. The thrust axis $\hat t$ is defined by maximizing thrust T \cite{Farhi:1977sg},
\begin{align}
T = \max_{\hat t}  \sum_{i \in X} \frac{\vert  \vec p_i \cdot \hat t \vert}{\sqrts} \, .
\end{align}
$T$ close to 1 corresponds to the special case $a=0$, $e(0)\approx 1-T$ and is also loosely called ``thrust''. Another interesting event shapes is the limit $a=1$ corresponding to ``total jet broadening'' B \cite{Catani:1992jc}, $e(1)=2 B$. The limit $e(a) \ll 1$ isolates events composed of back to back jets.  In the case of thrust, jets are composed of collinear radiation, and the recoil due to soft (ultra-soft in this case) radiation does not affect the jet axis. For jet broadening all radiation with parametrically similar transverse momentum can contribute, so the soft radiation of order $Q(\lambda,\lambda,\lambda)$  recoils the jet off the thrust axis, where $\lambda \sim e(1)$. In both of these cases fixed order perturbation theory will fail when $e(a)$ is small. However, as long as $eQ\gg \Lambda_{QCD}$, we expect non-perturbative effects to be suppressed, though large logarithms of $e$ need to be resummed.

  The pioneering work on jet broadening resummations \cite{Catani:1992jc}
utilized the coherent branching  formalism \cite{Catani:1989ne}. It was later stated \cite{Dokshitzer:1998kz} that
the results in \cite{Catani:1992jc} neglected terms due to recoil of soft gluons.
In this section we will provide a factorization theorem for jet broadening.
 The factorization proofs for  angularity observables (\ref{ang}) in \cite{Hornig:2009vb}
are known to fail as $a$ approaches $1$, since there are  growing power corrections in this limit. The reason for the apparent breakdown of factorization is the fact that in this limit the soft radiation has the same invariant mass as collinear radiation and
one must change the power counting  accordingly to factorize in a consistent fashion. Which is to say that thrust
can be analyzed using \sceti~whereas jet broadening necessitates the use of \scetii.

\subsection{Factorization Theorem}

We start with the expression for differential cross section in QCD for broadening $e$ (strictly speaking, angularity for $a=1$),
\begin{align}\label{eq:QCD-X-sec}
\frac{d\sigma}{de}&=\frac{1}{2 Q^2}   \sum_{i=A,V} \int d^4x L^i_{\mu\nu}(q) e^{i x.q}\langle 0 |j^{\dagger\mu}_i(x)\left(\delta(e-\hat{e})j_i^\nu(0)\right)|0\rangle  \nonumber \\
& = \frac{1}{2 Q^2}  \sum_{i=A,V}L_{\mu\nu}^i(q) \sum_{X} (2\pi)^4 \delta^{(4)}(q-P_X)\langle 0 \vert j_i^{\dagger\mu}(0) \, \delta(e-\hat{e}) \vert X \rangle \langle X \vert  j_i^\nu(0) \vert 0\rangle \, ,
\end{align}
where $\hat{e}$ is the jet broadening operator that first maximizes thrust for a given state $\vert X \rangle$ to determine the thrust axis $\hat t$ and then measures broadening  via, $\hat e \vert X \rangle = \sum_{i\in X} \frac{\vert p_{\!\perp i}\vert }{Q} \vert X \rangle$, where $p_{\!\perp i}$ is transverse momentum measured {\it w.r.t.} $\hat t$. Here $Q$ is center of mass energy and QCD current $j^\mu$ is given by
\begin{align}\label{eq:QCD-current}
j_i^\mu(x)&=\overline{q}(x)\Gamma_i^{\mu}q(x) \, ,
\end{align}
with $\Gamma^\mu = \gamma^\mu$ or $\gamma_5\gamma^\mu$.
$L_{\mu\nu}$ is the leptonic tensor given by
\begin{align}\label{eq:lepton-tensor}
L_{\mu\nu}^V &= - \frac{16\pi^2 \alpha_\text{em}^2}{3Q^2} \bigg( g_{\mu\nu} - \frac{q_\mu q_\nu}{Q^2} \bigg) \biggl[ Q_q^2 + \frac{v_q^2  (v_e^2+a_e^2) - 2 Q_q v_q v_e (1-m_Z^2/Q^2)}
{(1-m_Z^2/Q^2)^2 + \Gamma_Z^2/m_Z^2} \biggr] \, ,\nn \\
L_{\mu\nu}^A &= - \frac{16\pi^2 \alpha_\text{em}^2}{3Q^2}  \bigg( g_{\mu\nu} - \frac{q_\mu q_\nu}{Q^2} \bigg) \biggl[ Q_q^2 + \frac{ a_q^2 (v_e^2+a_e^2) }
{(1-m_Z^2/Q^2)^2 + \Gamma_Z^2/m_Z^2} \biggr] \, ,
\end{align}
where subscript $q$ denotes the (anti)quark flavor, $Q_q$ is the quark charge in units of $\vert{e}\vert$, $v_{q,e}$ and $a_{q,e}$ are the vector and axial couplings of the (anti)quark $q$ and the electron to the $Z$  as e.g. in eq.(A3) of ref.~\cite{Abbate:2010xh}. Here $m_Z$ and $\Gamma_Z$ denote the mass and the width of the $Z$ boson. 

In what follows, by requiring $e\sim \lambda\ll 1$, we will prove a factorization theorem of the form
\begin{align}
\frac{d\sigma}{de}=H J_{n}\otimes J_{\overline{n}}\otimes S \, , 
\end{align}
where each function is a vacuum matrix element of operators that depend on either collinear or soft modes that do not interact.

We begin by first matching the QCD currents onto the \scetii ~currents\footnote{In principle we should include ultra-soft Wilson lines $Y_n$'s as well, but since ultrasoft modes do not contribute to the process the $Y_n$ Wilson lines will cancel. Therefore we drop them already and resort to the phrase ``\scetii~ current''.}  as follows,
\begin{align}
\label{eq:scet2-current-matching}
j^\mu(0)  &=  C_{n_1 n_2}\sum_{n_1,n_2}  \bar{\chi}_{n_1}(0) S^\dagger_{n_1}(0) \,  \Gamma^\mu \,S_{n_2}(0) \chi_{n_2}(0) \, ,
\end{align}
where $C_{n_1 n_2} = C(\bar n_1\!\cdot\! {\cal P}, \bar n_2\!\cdot \!{\cal P}) $, is only a function of large label momentum operators. $\chi_{n_1,n_2}$ are SCET collinear fields while  the $S_n$ are soft Wilson lines extending to infinity
along the $n$ direction. 
Inserting eqn.~(\ref{eq:scet2-current-matching}) into eqn.~(\ref{eq:QCD-X-sec}) we have
\begin{align}\label{eq:intermediate-X-sec}
\frac{d\sigma}{de}
& = \frac{1}{2 Q^2}  L^i_{\mu\nu}  \sum_{n_1,n_2} \sum_{\tilde n_1,\tilde n_2} C_{n_1 n_2} C^*_{\tilde n_1 \tilde n_2} \sum_{X} (2\pi)^4\delta^{(4)}(q-P_X) \\
& \quad \quad \langle 0 \vert  \bar{\chi}_{n_1} S^\dagger_{n_1} \,  \Gamma_i^\mu \,S_{n_2} \, \chi_{n_2} \, \delta(e-\hat{e}) \vert X \rangle \langle X \vert  \bar{\chi}_{\tilde n_2} S^\dagger_{\tilde n_2} \,  \Gamma_i^\nu \,S_{\tilde n_1} \, \chi_{\tilde n_1} \vert 0\rangle \, , \nn
\end{align}
where it is implicitly understood that all fields are evaluated at space-time coordinate $x=0$. Now we decompose the complete set of states as $\sum_{X} \vert X \rangle \langle X \vert = \sum_{\hat n} \sum_{\Xn} \vert \Xn \rangle \langle \Xn \vert $ where $\Xn$ are states with thrust axis along $\hat n$. Here, the states that may have an ambiguity in the choice of the thrust axis, without loss of generality, can be associated with either $\hat n$. For $q^\mu = (Q,\vec 0)$ we have
\begin{align}\label{eq:intermediate-X-sec-1}
\frac{d\sigma}{de}
& = \frac{(2\pi)^4}{2 Q^2}  L^i_{\mu\nu}  \sum_{n_1,n_2} \sum_{\tilde n_1,\tilde n_2} \sum_{\hat n}  C_{n_1 n_2} C^*_{\tilde n_1 \tilde n_2} \\
&\langle 0 \vert  \bar{\chi}_{n_1} S^\dagger_{n_1} \,  \Gamma_i^\mu \,S_{n_2} \, \chi_{n_2} \, \left \{ \sum_{\Xn} \delta(E_X -  Q) \, \delta^{(3)}({\bf P}_X) \delta(e-\hat{e}) \vert \Xn \rangle \langle \Xn \vert  \right \} \, \, \bar{\chi}_{\tilde n_2} S^{\dagger}_{\tilde n_2} \,  \Gamma_i^\nu \,S_{\tilde n_1} \, \chi_{\tilde n_1} \vert 0\rangle \, \nonumber \\
& = \frac{(2\pi)^4}{2 Q^2}  L^i_{\mu\nu}  \sum_{n_1,n_2} \sum_{\tilde n_1,\tilde n_2} \sum_{\hat n} C_{n_1 n_2} C^*_{\tilde n_1 \tilde n_2}\, \langle 0 \vert \, \,  \bar{\chi}_{n_1} S^{\dagger}_{n_1} \,  \Gamma_i^\mu \,S_{n_2} \, \chi_{n_2} \,  \,
\oXn \,\,
\bar{\chi}_{\tilde n_2} S^{\dagger}_{\tilde n_2} \,  \Gamma_i^\nu \,S_{\tilde n_1} \, \chi_{\tilde n_1} \, \, \vert 0\rangle \, , \nn
\end{align}
where we identify the term in braces as the broadening projector $\oXn$, which simplifies as
\begin{align}\label{eq:oXn2}
\oXn & =  \sum_{X_{\hat n}} \delta(\hat E - Q) \,\, \delta^{(3)}(\hat {\bf P}\, ) \,\, \delta(e-\hat e_{\hat n}) \, \, \vert \Xn \rangle \langle \Xn \vert \,  \nonumber \\
 & =  \sum_{X_{\hat n}} \delta(\hat E - Q) \,\, \delta^{(3)}(\hat {\bf P}\, ) \,\, \delta(e-\hat e_{\hat n}) \,  \delta_{\oPp , 0}  \,  \delta_{\oPpb , 0} \, \, \vert \Xn \rangle \langle \Xn \vert \nonumber \\
 & =  \sum_{X} \delta(\hat E - Q) \,\, \delta^{(3)}(\hat {\bf P}\, ) \,\, \delta(e-\hat e_{\hat n}) \,  \delta_{\oPp , 0}  \,  \delta_{\oPpb , 0} \, \, \vert X \rangle \langle X \vert \nonumber \\ 
 & =  \delta\left (\half\bar n \cdot \hat P \!+\! \half n \cdot \hat P - Q \right) \,\,\delta\left (\half\bar n \cdot \hat P \! - \! \half n \cdot \hat P \right ) \,\, \delta^{(2)}(\hat {\bf P}_{\!\!\perp}\, ) \,\, \delta(e-\hat e_{\hat n}) \,  \delta_{\oPp , 0}  \,  \delta_{\oPpb , 0} \, , \nn \\
 & = 2 \, \delta(\bar n \cdot \hat P - Q) \,\,\delta( n \cdot \hat P - Q )\,\, \delta_{\hat {\bf P}_{\!\!\perp} , 0} \,\, \delta(e-\hat e_{\hat n}) \,  \delta^{(2)}(\oPp)  \,  \delta^{(2)}(\oPpb) \, \left (\int d^2k_{r\!\perp}\right ) \nonumber \\
& = 2 \, \delta(\bar n \cdot \hat P - Q) \,\,\delta( n \cdot \hat P - Q ) \,\,  \delta(e-\hat e_{\hat n}) \,  \delta^{(2)}(\oPp)  \,  \delta^{(2)}(\oPpb) \, \left ( \int d^2k_{r\!\perp} \right ) \, ,
\end{align}
where $\oPp$ and $\oPpb$ are defined as 
\begin{align}\label{eq:oPp}
\oPp \vert X \rangle &= \sum_{j \in X} \theta(\hat n\cdot \, \vec p_{\! j}) \vec p_{\!j\!\perp} \, \vert X \rangle \nn \\ 
\oPpb \vert X \rangle &= \sum_{j \in X} \theta(-\hat n\cdot \vec p_{\! j}) \, \vec p_{\!j\!\perp} \, \vert X \rangle \, . 
\end{align}
In the second line we have simply used $\delta_{\oPp , 0} \vert \Xn \rangle = \vert \Xn \rangle$ and $\delta_{\oPpb , 0} \vert \Xn \rangle = \vert \Xn \rangle$ owing to the property of the thrust axis that total transverse momentum in each hemisphere defined by the plane perpendicular to the thrust axis is zero. This form of the broadening projector implements the kinematic constraints imposed by the choice of thrust axis, since zero-transverse momentum flow in each hemisphere defined by $\hat n$ along with small broadening {\it w.r.t.} $\hat n$ uniquely fixes the thrust axis to be $\hat n$. Hence in the third line we have promoted the state $\vert X_{\hat n} \rangle$ to a generic state $\vert X \rangle$. In the fourth line of eqn. (\ref{eq:oXn2}) we have summed over the complete set of states.
For convenience in factorization with continuous labels, we turn the Kronecker-$\delta$s to continuous Dirac-$\delta$s, with the general relation:
\begin{align}\label{eq:perp-identity}
\delta_{\hat {\bf P}_{\!\!\perp} , 0}  = \delta^{(2)}(\hat {\bf P}_{\!\!\perp}) \,\int d^2k_{r\!\perp} \,  ,
\end{align}
where $\int d^2k_{r\!\perp}$ is simply the area of the label-transverse momentum unit cell\footnote{The Kronecker-$\delta$ assures that the discrete-label momentum is zero while the Dirac-$\delta$ ensures that continuous-label momentum is zero up to order $\lambda^2$, i.e. all momenta belonging to the unit cell are considered to be zero, hence the area of the unit cell appears as the proportionality constant in the identity~(\ref{eq:perp-identity}).}; $k_{r}$ refers to the residual momentum. In the final step we have used  $\delta_{\hat {\bf P}_{\!\!\perp} , 0} = \delta_{\oPp + \oPpb , 0} = \delta_{0,0}  = 1$.
For $e\ll 1$, constraints put in by all the $\delta$-functions ensure that the broadening projector selects the dijet states with the thrust axis $\hat n$. 

Given the properties of the broadening projector for small broadening, the light-cone directions of the collinear fields must be within a small cone (of radius $\lambda$) about the $\hat n$-collinear directions. Fields whose directions are an order $1$ or more displaced from $\hat n$ must produce too great a broadening, and so are excluded. Therefore by making a parametrization transformation \cite{Manohar:2006nz}, we can set the directions of the collinear fields to be exactly  parallel or anti-parallel to $\hat n$.
With further constraints from quark number conservation we only have two choices: the quark jet is either along the $\hat n$ or $-\hat n$ direction. Since the observable is symmetric under charge conjugation, the two are equivalent, and we get
\begin{align}\label{eq:intermediate-X-sec-2}
\frac{d\sigma}{de}
& = \frac{(2\pi)^4}{2 Q^2}  L^i_{\mu\nu}   \,\, 2 \sum_{\hat n} C_{n \bar n} C^*_{n \bar n}\, \langle 0 \vert \, \,  \bar{\chi}_{\bar n} S^{\dagger}_{\bar n} \,  \Gamma_i^\mu \,S_{n} \, \chi_{n} \,  \,
\oXn \,\,
\bar{\chi}_{n} S^{\dagger}_{n} \,  \Gamma_i^\nu \,S_{\bar n} \, \chi_{\bar n} \, \, \vert 0\rangle \\
& = \frac{(2\pi)^4L^i_{\mu\nu}}{Q^2}\,\, 2 \sum_{\hat n}  \vert C_{n\bar n}\vert^2  \,  
\langle 0 \vert \, \,  \bar{\chi}_{\bar n} S^{\dagger}_{\bar n} \,  \Gamma_i^\mu \,S_{n} \, \chi_{n} \,  \,
 \delta(\bar n \cdot \hat P - Q) \,\,\delta( n \cdot \hat P - Q ) \,\, \nonumber \\
& \hspace{5cm} \times  \delta(e-\hat e_{\hat n}) \,  \delta(\oPp)  \,  \delta(\oPpb)  \,\,\,
\bar{\chi}_{n} S^{\dagger}_{n} \,  \Gamma_i^\nu \,S_{\bar n} \, \chi_{\bar n} \, \, \vert 0\rangle \, \left ( \int d^2k_{r\!\perp} \right ) \, ,\nonumber
\end{align}
where we have used eqn.~(\ref{eq:oXn2}). The matrix element in the last equation does \emph{not} depend upon the direction $\hat n$ but only on the large labels $Q$ and broadening $e$. The choice of $n$ inside the matrix element is only representative of a light cone vector necessary for calculation, but these calculations would yield identical results for different $n$. Also, $C_{n\bar n}$ is only a function of boost invariant $n\!\cdot\! P \, \bar n \! \cdot\! P = s = Q^2$, so we define the hard function independent of the light cone direction, $H(Q^2,\mu) = \vert C_{n\bar n}\vert^2$. Therefore we can safely factor out the matrix element out of the sum over $\hat n$ and can write
\begin{align}\label{eq:intermediate-X-sec-3}
\frac{d\sigma}{de}
& = \frac{(2\pi)^4 L^i_{\mu\nu}}{Q^2} \, H(Q^2,\mu) \,  \langle 0 \vert \, \,  \bar{\chi}_{\bar n} S^{\dagger}_{\bar n} \,  \Gamma_i^\mu \,S_{n} \, \chi_{n} \,  \,
 \delta(\bar n \cdot \hat P - Q) \,\,\delta( n \cdot \hat P - Q ) \,\,  \\
 & \hspace{4 cm} \times  \delta(e-\hat e_{\hat n}) \,  \delta(\oPp)  \,  \delta(\oPpb)  \,\,\,
\bar{\chi}_{n} S^{\dagger}_{n} \,  \Gamma_i^\nu \,S_{\bar n} \, \chi_{\bar n} \, \, \vert 0\rangle \, \left (  2 \sum_{\hat n^\prime} \int d^2k_{r\!\perp} \right ) \, , \nn
\end{align}
where $\hat n$ from here on is a fixed vector, say $\hat z$, {\it i.e.} $n^\mu = (1,0,0,1)$ and $\bar n^\mu = (1,0,0,-1)$. Now we use \cite{Fleming:2007qr}
\begin{align}
\int d^2k_{\perp} \sum_{\rm cones} = 2 \sum_{\hat n} \int d^2k_{r\!\perp} = \frac{Q^2}{4} \int d\Omega = \pi Q^2 \, ,
\end{align}
where $k_\perp$ is the label-momentum and $k_{r\!\perp}$ is the residual momentum. The cones subtend an angular area of order 1, while $\hat n$ directions label cones of order $\lambda$. Then we achieve:
\begin{align}\label{eq:intermediate-X-sec-4}
\frac{d\sigma}{de}
& = (2\pi)^4 \pi \, H(Q^2,\mu)L^i_{\mu\nu}(q) \, \\
& \langle 0 \vert  \bar{\chi}_{\bar n} S^{\dagger}_{\bar n} \,  \Gamma_i^\mu \,S_{n} \, \chi_{n} \,  \,
 \delta(\bar n \cdot \hat P - Q) \,\,\delta( n \cdot \hat P - Q ) \,\,  \delta(e-\hat e_{\hat n}) \,  \delta^{(2)}(\oPp)  \,  \delta^{(2)}(\oPpb)  
\bar{\chi}_{n} S^{\dagger}_{n} \,  \Gamma_i^\nu \,S_{\bar n} \, \chi_{\bar n}  \vert 0\rangle . \nonumber
\end{align}
We have two choices for lepton tensor which correspond to $\Gamma_i^\mu = \gamma^\mu$ or $\gamma_5\gamma^\mu$ in the hadron matrix element. We can simplify the lepton tensor by noting that $\bar \chi_{\bar n} \, ( q \slashed \, , \, \gamma_5 q\slashed ) \chi_n = 0$ for $q_\perp = 0$, thus we can safely replace $L^{i}_{\mu\nu}(q)$ with $L^{i}(Q^2) \, g_{\mu\nu}$. We now have
\begin{align}\label{eq:intermediate-X-sec-5}
\frac{d\sigma}{de}
& = (2\pi)^4 \pi \, H(Q^2,\mu)\, L^i(Q^2) \,  \\ 
& \langle 0 \vert \, \,  \bar{\chi}_{\bar n} S^{\dagger}_{\bar n} \,  \Gamma_i^\mu \,S_{n} \, \chi_{n} 
 \delta(\bar n \cdot \hat P - Q) \,\,\delta( n \cdot \hat P - Q ) \,\,  \delta(e-\hat e_{\hat n}) \,  \delta^{(2)}(\oPp)  \,  \delta^{(2)}(\oPpb)  
\bar{\chi}_{n} S^{\dagger}_{n} \,  \Gamma_\mu^i \,S_{\bar n} \, \chi_{\bar n} \, \, \vert 0\rangle . \nonumber
\end{align}
We are almost ready to factorize. The operators $\hat e_{\hat n}$, $\oPp$ and $\oPpb$ still mix $n$-collinear, $\bar n$-collinear and soft sectors. We take care of this by considering following identity operators
\begin{align}\label{eq:perp-identity-op}
1 = \int d^2 k_{1\!\perp} \, d^2 k_{2\!\perp} \, d^2 k_{1\!\perp}^\prime \, d^2 k_{2\!\perp}^\prime \, \delta^{(2)}(\oPp^{(c)} - k_{1\!\perp}) \, \, \delta^{(2)}(\oPpb^{(\bar c)} - k_{2\!\perp}) \, \delta^{(2)}(\oPp^{(s)} - k_{1\!\perp}^\prime) \, \delta^{(2)}(\oPpb^{(s)} - k_{1\!\perp}^\prime)
\end{align}
and 
\begin{align}\label{eq:en-identity}
1 = \int de_n \, de_{\bar n} \, de_s \,  \delta(e_n - \hat e_{\hat n}^{(c)}) \, \delta(e_{\bar n} - \hat e_{\hat n}^{(\bar c)}) \, \delta(e_s - \hat e_{\hat n}^{(s)})  \, ,
\end{align}
where operators with superscript $(c)$ has the same action as their parent operator on $n$-collinear particles and fields but they give zero for all else. Similarly $(\bar c )$ for $\bar n$-collinear and $(s)$ for soft. We now insert (\ref{eq:perp-identity-op}) and (\ref{eq:en-identity}) in the expression~(\ref{eq:intermediate-X-sec-5}). Since there are only $n$-collinear, $\bar n$-collinear and soft sectors contributing to ~(\ref{eq:intermediate-X-sec-4}) we must have $\oPp^{(c)} + \oPp^{(s)} = \oPp$, ~$\oPpb^{(\bar c)} + \oPpb^{(s)} = \oPpb$ and $\hat e_{\hat n} = e_{\hat n}^{(c)} + e_{\hat n}^{(\bar c)} + e_{\hat n}^{(s)}$. Using this and integrating over $k_{1\!\perp}^\prime$ and $k_{2\!\perp}^\prime$, we have
\begin{align}\label{eq:intermediate-X-sec-6}
\frac{d\sigma}{de}
& = (2\pi)^4 \pi \, H(Q^2,\mu) L^{i}(Q^2) \, \int de_n \, de_{\bar n} \, de_s \, \int d^2 k_{1\!\perp} \, d^2 k_{2\!\perp} ~\langle 0 \vert \, \,  \bar{\chi}_{\bar n} S^{\dagger}_{\bar n} \,  \bar \Gamma_i^\mu \,S_{n} \, \chi_{n} \,  \, \nn \\
 &  \quad \times  \delta(\bar n \cdot \hat P - Q) \,\,\delta( n \cdot \hat P - Q ) \,\, 
 \delta(e - e_n - e_{\bar n} -e_s) 
 \delta(e_n-\hat e_{\hat n}^{(c)}) \delta(e_{\bar n}-\hat e_{\hat n}^{(\bar c)}) \delta(e_s-\hat e_{\hat n}^{(s)})  \nn \\
 & \quad \times  \delta^{(2)}(\oPp^{(c)} - k_{1\!\perp}) \delta^{(2)}(\oPpb^{(\bar c)} - k_{2\!\perp}) \delta^{(2)}(\oPp^{(s)} + k_{1\!\perp})  \delta^{(2)}(\oPpb^{(s)} + k_{2\!\perp})  \,\,\,
\bar{\chi}_{n} S^{\dagger}_{n} \,  \Gamma^i_\mu \,S_{\bar n} \, \chi_{\bar n} \, \, \vert 0\rangle \, . 
\end{align}
The hard work is done, now we can simply factorize because each operator acts only on either sector (note that $n \cdot {\hat P}$ and $\bar n \cdot \hat P$ get leading contribution only from $\bar n$-collinear and $n$-collinear sectors respectively; in other words multipole expansion ensures that we can safely assume $n \cdot {\hat P}$ acts only on $\chi_{\bar n}$ field and so on). After using color conservation for the collinear matrix elements and the Fierz transformations, $\gamma^\mu_{ab} \gamma_{\mu\,cd} \to -(\frac{\nslashinline}{2})_{ad}(\frac{\nbarslashinline}{2})_{cb}$ and $(\gamma^5\gamma^\mu)_{ab} (\gamma^5\gamma_\mu)_{cd} \to -(\frac{\nslashinline}{2})_{ad}(\frac{\nbarslashinline}{2})_{cb}$~, we have,
\begin{align}\label{eq:factorized-X-sec-2}
\frac{d\sigma}{de}
& =  N_c \left ( \frac{-L^A(Q^2) - L^V(Q^2) }{4\pi}\right ) H(Q^2,\mu)\, \int de_n \, de_{\bar n} \, de_s \,\, \delta(e - e_n - e_{\bar n} -e_s) \, \int d^2 k_{1\!\perp} \, d^2 k_{2\!\perp} \,\,\,  \nonumber \\
 & \quad \quad \quad \quad \frac{(2\pi)^3}{N_c} \,  \langle 0 \vert \, \,   \bar{\chi}_{\bar n}   \,\delta( n \cdot \hat P - Q ) \,\, \delta(e_{\bar n} -\hat e_{\hat n})    \,\, \delta^{(2)}(\hat P_{\!\!\perp} - k_{2\!\perp})  \, \frac{\nslash}{2} \, \chi_{\bar n}   \, \, \vert 0\rangle \nonumber \\
  & \quad \quad \quad \quad \frac{(2\pi)^3}{N_c}{\rm tr} \,  \langle 0 \vert \, \,  \frac{\nbarslash}{2} \, \,  \chi_{n}   \,\delta( \bar n \cdot \hat P - Q ) \,\, \delta(e_n -\hat e_{\hat n})    \,\, \delta^{(2)}(\hat P_{\!\!\perp} - k_{1\!\perp})  \, \,  \bar{\chi}_{n}    \, \, \vert 0\rangle \nonumber \\
& \quad \quad \quad \quad \frac{1}{N_c}{\rm tr} \, \langle 0 \vert \, \,     S^{\dagger}_{\bar n} \, S_{n} \, \, \delta^{(2)}(\oPp + k_{1\!\perp}) \, \,  \delta^{(2)}(\oPpb + k_{2\!\perp})  \,\, \delta(e_s-\hat e_{\hat n}) \,\, 
  S^{\dagger}_{n} \, S_{\bar n}  \, \, \vert 0\rangle \, , 
\end{align}
where we have also reduced transverse momentum and broadening operators to their respective parent operators (with a further reduction of transverse momentum operator in the collinear sectors to label momentum operators). Trace is over both color and Dirac indices. In eqn. (\ref{eq:factorized-X-sec-2}), we have a factorization theorem in which the third and the fourth lines represent the broadening jet function composed only of collinear fields, and in the last line we have a matrix element composed of the soft Wilson lines only. Written compactly, factorization theorem is
\begin{align}\label{eq:factorized-X-sec-3}
\frac{d\sigma}{de}
& =  \sigma_0 \, H(Q^2,\mu) \, \int de_n \, de_{\bar n} \, de_s \,\, \delta(e - e_n - e_{\bar n} -e_s) \, \int d^2 k_{1\!\perp} \, d^2 k_{2\!\perp} \nn \\
& \quad \quad \quad {\cal J}_{n}(Q,e_n,\vec k_{1\perp}^2) \,  {\cal J}_{\bar n}(Q,e_{\bar n},\vec k_{2\perp}^2) \, {\cal S}(e_s,\vec k_{1\perp},\vec k_{2\perp})
\end{align} 
where $\sigma_0$ is the Born cross-section given by (see, for example, appendix A of ref.~\cite{Abbate:2010xh})
\begin{equation} \label{eq:si0}
  \sigma_0^q = \frac{4\pi \alpha_\text{em}^2 N_c}{3Q^2} \biggl[ Q_q^2 + \frac{(v_q^2 + a_q^2) (v_e^2+a_e^2) - 2 Q_q v_q v_e (1-m_Z^2/Q^2)}
{(1-m_Z^2/Q^2)^2 + \Gamma_Z^2/m_Z^2} \biggr]
\,.
\end{equation}
A straightforward generalization of this result gives the factorization theorem for the left and right broadening \cite{Rakow:1981qn}
\begin{align}\label{eq:factorized-X-sec-LR}
\frac{d\sigma}{de_L de_R}
& =  \sigma_0 \, H(Q^2,\mu) \, \int de_n \, de_{\bar n} \, de_s^L\, de_s^R \,\, \delta(e_R - e_n  - e_s^R) \delta(e_L - e_{\bar n} - e_s^L) \,\int d \vec k^2_{1\!\perp} \, d \vec k^2_{2\!\perp} \nn \\
& \quad \quad \quad {\cal J}_{n}(Q,e_n,\vec k_{1\perp}^2) \,  {\cal J}_{\bar n}(Q,e_{\bar n},\vec k_{2\perp}^2) \, {\cal S}(e_s^R, e_s^L, \vec k_{1\perp}^2,\vec k_{2\perp}^2) \, ,
\end{align} 
where only the soft function changes
\begin{align}
{\cal S}(e_s^R, e_s^L, \vec k_{1\perp}^2,\vec k_{2\perp}^2) &= \frac{\pi^2}{N_c}{\rm tr} \, \langle 0 \vert \, \,     S^{\dagger}_{\bar n} \, S_{n} \, \, \delta^{(2)}(\oPp + k_{1\!\perp}) \, \,   \delta(e_s^R-\hat e_{\hat n}^{R}) \nn \\
& \hspace{3cm}\delta^{(2)}(\oPpb + k_{2\!\perp})  \,\, \delta(e_s^L-\hat e_{\hat n}^{L}) \,\, 
  S^{\dagger}_{n} \, S_{\bar n}  \, \, \vert 0\rangle \, .
\end{align}
Here $\hat e_{\hat n}^{R,L}$ are defined as $\hat e_{\hat n}^{R,L} \vert X \rangle = (\sum_{i \in X} \theta(\pm p_{i3})\vert p_{i\perp} \vert/Q ) \vert X \rangle$. This simply comes about by changing the identity insertion (\ref{eq:en-identity}) appropriately. Note that in this form each function only depends on the magnitude of the transverse momenta which is particularly convenient, thus we have changed the overall integration measure appropriately. We present the bare definition of the jet and the soft functions in the next section where we discuss their renormalization.



\subsection{Broadening jet and soft functions: definition and renormalization}

The naive definitions of the jet and the soft function obtained in the previous section contain unregulated rapidity and UV divergences. We will regulate the UV divergences in dimensional regularization as usual and for the rapidity divergences we will adopt the regulator prescribed in this work. For fermion free abelian theory we can put the rapidity regulator into the collinear and soft Wilson lines and for the non-abelian case we follow sec. \ref{sec:non-abelian-regulator}. Bare quark jet function is given by
\begin{align}\label{eq:broadening-bare-jet}
{\cal J}^{\rm bare}_{n}(Q,e_n,\vec k_{\perp}^2) =  \frac{(2\pi)^{3-2\epsilon}}{N_c}{\rm tr} \, \langle 0 \vert   \frac{\nbarslash}{2}   \chi_{n}(0) \delta( \bar n \cdot \hat {\cal P} - Q ) \,\, \delta(e_n -\hat e_{\hat n}) \delta^{(2-2\epsilon)}(\hat {\cal P}_{\!\!\perp} - k_{\!\perp})  \bar{\chi}_{n}(0) \vert 0\rangle \, ,
\end{align}
where $\hat {\cal P}$ is the standard SCET label operator and here we are working with continuous labels. There is analogous equation for bare anti-quark jet function but its functional dependence is the same as the quark jet function so it is simply obtained by replacing $e_n \to e_{\bar n}$. The bare soft function is given by
\begin{align}\label{eq:broadening-bare-soft}
{\cal S}_{\rm bare}(e_s^R, e_s^L, \vec k_{1\perp}^2,\vec k_{2\perp}^2) &=   \frac{\pi^{2-2\epsilon}(\vec k_{1\perp}^2)^{-\epsilon}(\vec k_{2\perp}^2)^{-\epsilon}}{N_c\, \Gamma^2(1-\epsilon)}{\rm tr} \, \langle 0 \vert \, \,     S^{\dagger}_{\bar n}(0) \, S_{n}(0) \, \, \delta^{(2-2\epsilon)}(\oPp + k_{1\!\perp}) \, \,   \delta(e_s^R-\hat e_{\hat n}^{R}) \nn \\
& \hspace{2.5cm}\delta^{(2-2\epsilon)}(\oPpb + k_{2\!\perp})  \,\, \delta(e_s^L-\hat e_{\hat n}^{L}) \,\, 
  S^{\dagger}_{n}(0) \, S_{\bar n}(0)  \, \, \vert 0\rangle \, ,
\end{align}
where $\oPp$ and $\oPpb$ were defined in eqn.~(\ref{eq:oPp}).
For both the jet and soft functions all fields are evaluated at $x=0$. Note that after accounting for the dimensions of the bare fields, bare jet and soft functions are integer dimensional objects, as required for operator renormalization.

In our formalism rapidity divergences appear as counter terms just like the UV divergences and therefore can be renormalized away via renormalization constants. Renormalized quantities can be calculated as usual
\begin{align}\label{eq:broadening-ren}
 {\cal J}^{\rm ren}(e,\vec k^{\,2};\mu,\nu/Q) &= \int d e^\prime \int \frac{d^2\vec k^\prime}{(2\pi)^2} Z_{\cal J}^{-1}(e - e^\prime, (\vec k - \vec k^\prime)^2;\mu,\nu/Q) {\cal J}^{\rm bare}(e^\prime , \vec k^{\prime 2},Q) \nn \\
{\cal S}^{\rm ren}(e_R,e_L,\vec p^{\, 2},\vec q^{\,2};\mu,\nu)  &= \int d e_R^\prime d e_L^\prime \int \frac{d^2\vec p^{\, \prime}}{(2\pi)^2}\frac{d^2\vec q^{\, \prime}}{(2\pi)^2} \,  {\cal S}^{\rm bare}(e_R^\prime,e_L^\prime , \vec p^{\, \prime 2},\vec q^{\, \prime 2}) \\
& \hspace{2 cm} Z_{\cal S}^{-1}(e_R - e_R^\prime,e_L - e_L^\prime, (\vec p - \vec p^{\, \prime})^2,(\vec q - \vec q^{\, \prime})^2;\mu,\nu) \, , \nn 
\end{align}
where $Z^{-1}_{\cal J\!,S}$ only contain terms that are divergent in $\eta$ and $\epsilon$. Note that after expanding in $\eta$ and $\epsilon$ all vectors are Euclidean 2-vectors, hence we have a 2-dimensional vector convolution only. Henceforth, in this section, we drop the subscript $\perp$ on $2$-vectors. $Z_{\cal J\!,S}$ follow the standard constraints
\begin{align}\label{eq:Z-consistency}
{\mathbb I}_{\cal J}(e,\vec k) &\equiv (2\pi)^2 \delta^{(2)}(\vec k) \delta(e)  = \int d e^\prime \int \frac{d^2\vec k^\prime}{(2\pi)^2} Z_{\cal J}^{-1}(e - e^\prime, (\vec k - \vec k^\prime)^2) Z_{\cal J}(e^\prime , \vec k^{\prime 2}) \\
{\mathbb I}_{\cal S}(e_L,e_R,\vec p, \vec q) &\equiv (2\pi)^4 \delta^{(2)}(\vec p)\delta^{(2)}(\vec q) \delta(e_R)\delta(e_L)  = \int d e_R^\prime d e_L^\prime \int \frac{d^2\vec p^{\, \prime}}{(2\pi)^2}\frac{d^2\vec q^{\, \prime}}{(2\pi)^2} \,  Z_{\cal S}(e_R^\prime,e_L^\prime , \vec p^{\, \prime 2},\vec q^{\, \prime 2}) \nn \\
& \hspace{6 cm} Z_{\cal S}^{-1}(e_R - e_R^\prime,e_L - e_L^\prime, (\vec p - \vec p^{\, \prime})^2,(\vec q - \vec q^{\, \prime})^2) \, . \nn
\end{align}

Using the consistency condition and the fact that the bare functions do not depend upon the renormalization scales $\mu$ and $\nu$ we get the RG equations in $\mu$ and $\nu$. The renormalization group equations for the jet functions are given by
\begin{align}
\frac{\mu \, d {\cal J}^{\rm ren}}{d\mu} &= \gamma _{\mu }^{\cal J}(\mu, \nu/Q)  \, {\cal J}^{\rm ren}(e,\vec k^{\,2};\mu,\nu/Q) \\
\frac{\nu \,  d {\cal J}^{\rm ren}}{d\nu} &= \int d e^\prime \int \frac{d^2\vec k^\prime}{(2\pi)^2} \gamma _{\nu }^{\cal J}(e - e^\prime, (\vec k - \vec k^\prime)^2;\mu) \, {\cal J}^{\rm ren}(e^\prime,\vec k^{\,\prime 2};\mu,\nu/Q) \, , \nn
\end{align}
where the anomalous dimensions are obtained via
\begin{align}\label{eq:jet-AD-eqn}
{\mathbb I}_{\cal J} \,\times  \gamma _{\mu }^{\cal J}(\mu, \nu/Q)  &= - \int d e^\prime \int \frac{d^2\vec k^\prime}{(2\pi)^2} Z_{\cal J}^{-1}(e - e^\prime, (\vec k - \vec k^\prime)^2) \frac{d}{d \ln \mu}Z_{\cal J}(e^\prime , \vec k^{\prime 2};\mu,\nu/Q) \\
\gamma _{\nu }^{\cal J}(e,\vec k^{\,2};\mu)  &= - \int d e^\prime \int \frac{d^2\vec k^\prime}{(2\pi)^2} Z_{\cal J}^{-1}(e - e^\prime, (\vec k - \vec k^\prime)^2) \frac{d}{d \ln \nu} Z_{\cal J}(e^\prime , \vec k^{\prime 2};\mu,\nu/Q)  \, . \nn
\end{align}
For the soft function, RG equations are similar with more variable dependencies, therefore for brevity we only show relevant variables and represent convolutions via $\otimes$. We have for RG equations
\begin{align}
\frac{\mu \, d {\cal S}^{\rm Rena}}{d\mu} &= \gamma _{\mu }^{\cal S}(\mu, \nu/\mu)  \, {\cal S}^{\rm ren}(e_R,e_L,\vec p^{\,2},\vec q^{\,2};\mu,\nu/\mu) \\
\frac{\nu \,  d {\cal S}^{\rm ren}}{d\nu} &=  \gamma _{\nu }^{\cal S}(e_R,e_L,\vec p^{\,2},\vec q^{\,2};\mu) \otimes {\cal S}^{\rm ren}(\ldots;\mu,\nu/\mu) \, , \nn
\end{align}
and for anomalous dimensions
\begin{align}\label{eq:soft-AD-eqn}
{\mathbb I}_{\cal S} \,\times  \gamma _{\mu }^{\cal S}(\mu, \nu/Q)  &= - Z_{\cal S}^{-1} \otimes \frac{d}{d \ln \mu} Z_{\cal S}(\ldots; \mu, \nu/\mu) \\
\gamma _{\nu }^{\cal S}(e_R,e_L,\vec p^{\,2},\vec q^{\,2};\mu)  &= - Z_{\cal S}^{-1} \otimes \frac{d}{d \ln \nu} Z_{\cal S}(\ldots;\mu,\nu/\mu)  \, . \nn
\end{align}
We emphasize that $\mu$-RG equations do not involve convolutions and $\gamma_\mu$ does not have any kinematical dependence. Reasons for this were made clear in section \ref{sec:TMDPDF-RGE}.

We have a few consistency conditions on the anomalous dimensions. Firstly, sum of the $\mu$-anomalous dimensions in the IR sectors (jets and soft) should add up to the negative of the hard anomalous dimension, that is
\begin{align}\label{eq:sum-o-mu}
\gamma_\mu^{\cal J}(\mu,\ln \frac{\nu}{Q^-}) + \gamma_\mu^{\cal \bar J}(\mu,\ln \frac{\nu}{Q^+}) + \gamma_\mu^{\cal S}(\mu,\ln \frac{\nu}{\mu}) + \gamma_\mu^{H}(\mu,\ln \frac{\mu^2}{Q^-Q^+}) = 0 \, ,
\end{align}
where we have {\it explicitly} shown the logarithmic dependence in the anomalous dimensions which can only be linear. Note that in c.o.m. frame $Q^+ = Q^- = Q$.
Secondly, we have a consistency condition for $\nu$-anomalous dimensions analogous to eqn. (\ref{eq:nu-consistency}) 
\begin{align}\label{eq:sum-o-nu}
&  \quad \quad \quad  \nu\frac{ \, d }{d \nu} \left[\left (Z_{\cal J} Z_{\cal \bar J} \right ) \otimes Z_{\cal S}\right] = 0 \nn \\
& \Rightarrow  \quad \, {\mathbb I}_{\cal \bar J}(e^\prime, \vec q^2) \gamma_\nu^{\cal J}(e,\vec p^2)    + {\mathbb I}_{\cal J}(e, \vec p^2) \,  \gamma_\nu^{\cal \bar J}(e^\prime,\vec q^2)  + \gamma_\nu^{\cal  S}(e,e^\prime,\vec p^2,\vec q^2) = 0 \, .
\end{align}Lastly, the independence of renormalization scales, $\mu$ and $\nu$ implies that the UV and rapidity RGs commute, which then gives the constraints
\begin{align}\label{eq:mu-nu-constraint}
\mu\frac{ \, d}{d\mu} \, \gamma_\nu^{\cal J} &= {\mathbb I}_{\cal J} \,  \nu \frac{\, d}{d\nu} \, \gamma_\mu^{\cal J} =  \Gamma_{\rm cusp}^{q} \,  {\mathbb I}_{\cal J}   , \nn \\
\mu \frac{\, d}{d\mu} \, \gamma_\nu^{\cal S} &=  {\mathbb I}_{\cal S} \, \nu\frac{ \, d}{d\nu} \, \gamma_\mu^{\cal S} = - 2 \Gamma_{\rm cusp}^{q}  \, {\mathbb I}_{\cal S}.
\end{align}
Here we have used the linearity of $\mu$-anomalous dimensions in its logarithmic term and its relationship to the cusp anomalous dimension in $\gamma_\mu^H$. This concludes the formal discussion on renormalization and running. We present the calculation and results in rest of this section.


\subsection{Jet  Function Calculation up-to NLO}
Bare quark jet function was defined in eqn.~(\ref{eq:broadening-bare-jet}), which is what we will calculate here. Anti-quark jet function is obtained by $e_R \to e_L$ and $\vec p^{\, 2} \to \vec q^{\, 2}$. For tree level jet function we have
\bea\label{eq:tree-jet-JB}
{\cal J}^{(0)}(e_R,\vec{p}^{\, 2})&=&\delta(e_R - |\vec{p}\, |/Q ) \, ,
\eea
which is all what we need for the NLL cross-section.
For NLO-singular cross-section, we only need to consider the one-loop jet function at $\vec{p}=0$. 
Non-zero transverse momentum implies presence of at least a soft radiation, which then gives contribution to the cross-section at NNLO.
We will use $\eta$-regulator as prescribed in section \ref{RRD} to regulate rapidity divergences. To regulate IR and UV divergences we will use dimensional regularization with $d=4-2\epsilon$. Virtual diagrams all vanish as they will be scaleless. So we need to calculate only the real diagrams and interpret $1/x^{1+a}$ as the distribution $[\frac{\theta(x)}{x^{1+a}}]_+^{^{\infty}}$\footnote{this is the plus distribution with boundary at $+\infty$ {\it i.e.}, $\int_{-\infty}^\infty dx [\frac{\theta(x)}{x^{1+a}}]_+^{^\infty} = 0$.}. We use Feynman gauge for this calculation. At one-loop, the jet function is simply the sum of all real diagrams\footnote{We have verified by using gluon mass as an explicit IR regulator that IR divergences cancel between the real and virtual diagrams and that we obtain the same results as presented here.}. We get, for $\vec p = 0$,
\begin{figure}[h]
    \includegraphics[width=15cm]{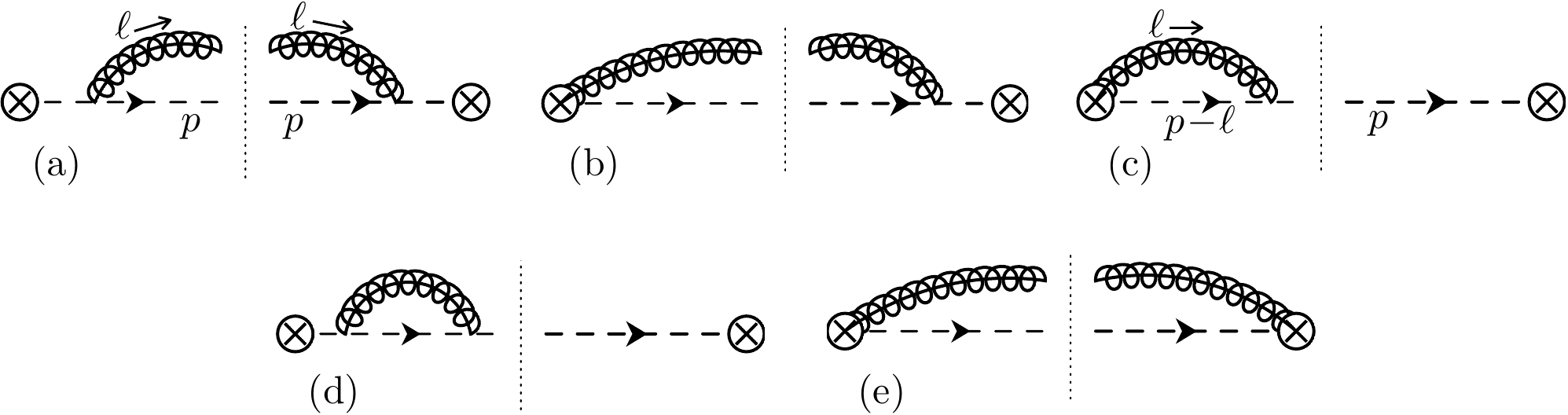}
   \centering
\caption{\label{JBjet}Diagrams contribution to broadening jet function. Diagram (c) is zero in Feynman gauge and virtual diagrams (d) and (e) are zero in dim.-reg.}
\end{figure}
\begin{align}\label{J_pT}
  {\cal J}^{(1)}_{\rm bare}(e_R,0) &= \Big(\frac{e^{\gamma_E} \mu^2}{4\pi}\Big)^\e\, \frac{1}{2N_c} 
  \int\! {\dd}^d p\,  \delta^{+}(p^2) \int \frac{{\dd}^d \ell}{(2\pi)^{d-1}}  \delta^{+}(\ell^2)
   \\ 
   & \quad \times 
  \delta(k^- - \ell^- - p^-) \delta^{d-2}(\ell_\perp + p_\perp)\, \delta( e_R -|\vec{\ell}_{\perp}|/Q- |\vec{p}_{\perp} |/Q ) \,
  \nn \\ & \quad \times\left (
  {\rm{tr}}\Big[ \frac{\nbarslash}{2}\, \frac{ i (\ell\slashed+ p\slashed)}{(\ell+p)^2}\,i g \gamma^\mu T^a\,  p\slashed (-g_{\mu\nu})\, i g \gamma^\nu T^a \frac{ i(\ell\slashed + p\slashed)}{(\ell+p)^2} \Big] 
  \right.\nn\\
  &\quad\quad\quad\left.  
  +2 w^2 \nu^{\eta} \,{\rm{tr}}\Big[ \frac{\nbarslash}{2}\, \frac{ i (\ell\slashed+ p\slashed)}{(\ell+p)^2}\, i g \gamma^\mu T^a\,  p\slashed (-g_{\mu\nu})\frac{  gT^a \nbar^{\nu}\nu^{\eta}  }{(\ell^-)^{1+\eta}}\Big] 
  \right)\nn\\
 &=\frac{\alpha_s C_F}{2\pi} \frac{e^{\epsilon \gamma_E}}{\Gamma(1-\epsilon)}\, \Big(\frac{2\mu}{Q}\Big)^{2\epsilon} \, \infplus{1}{e_R^{1+2\epsilon}}\, \left[   (1-\epsilon)-w^2\left(\frac{\nu}{Q}\right)^{\eta}  \frac{4}{\eta(1-\eta)}\right] \, . \nn 
\end{align}
Expanding in $\eta$ and then in $\epsilon$ we get,
\begin{align}\label{eq:jet-bare-JB}
  {\cal J}_{\rm bare}^{(1)}(e_R,0)  &= \frac{\alpha_s(\mu) C_F}{\pi} \Bigg [ -\frac{2w^2e^{\epsilon \gamma_E}}{\eta \, \Gamma(1-\epsilon)} \Big(\frac{2\mu}{Q}\Big)^{2\epsilon} \, \infplus{1}{e_R^{1+2\epsilon}}\,  +\frac{1}{\epsilon} \ln \frac{\nu}{Q} \delta(e_R)  + \frac{3}{4\epsilon}\delta(e_R) \nn \\
  & \quad\quad\quad\quad \quad \quad  - \frac{3Q}{4\mu} \plusf{2\mu}{Q e_R} - \frac{Q}{\mu}\plusf{2\mu}{Q e_R} \ln\frac{\nu}{Q}  + \frac{1}{4}\delta(e_R) \Bigg ] \, ,
\end{align}
where we have expressed the finite parts in terms of the standard plus distributions which are related to those with the infinity boundary via
\begin{align}
\infplus{1}{x^{1+a}} = -\frac{1}{a} \delta(x) + \plusf{1}{x} - a \plusf{\ln x}{x} + \ldots
\end{align}
and obey
\begin{align}
\int_0^1 \plusf{\ln^n x}{x} = 0.
\end{align}
Since we did not calculate $\vec p^{\, 2}$ dependence we only give the renormalized part of the one loop jet function which is what we need for the NLO cross-section,
\begin{align}
  {\cal J}_{\rm ren}(e_R,0)  &=\delta(e_R - |\vec{p}\, |/Q )+ \frac{\alpha_s(\mu) C_F}{\pi} \Bigg [ - \frac{3Q}{4\mu} \plusf{2\mu}{Q e_R} \!\!- \frac{Q}{\mu}\plusf{2\mu}{Q e_R} \!\!\ln\frac{\nu}{Q}  + \frac{1}{4}\delta(e_R) \Bigg ] \, .
\end{align}


\subsection{Soft Function Calculation up-to NLO}

The bare soft function was defined in eqn.~(\ref{eq:broadening-bare-soft}).
 The  tree-level soft-function is given by
\begin{align}
{\cal S}^{(0)} = \delta(e_R) \delta(e_L) \delta(\vec p^{\, 2})  \delta(\vec q^{\, 2}) \, .
\end{align}
\begin{figure}[h]
    \includegraphics[width=17cm]{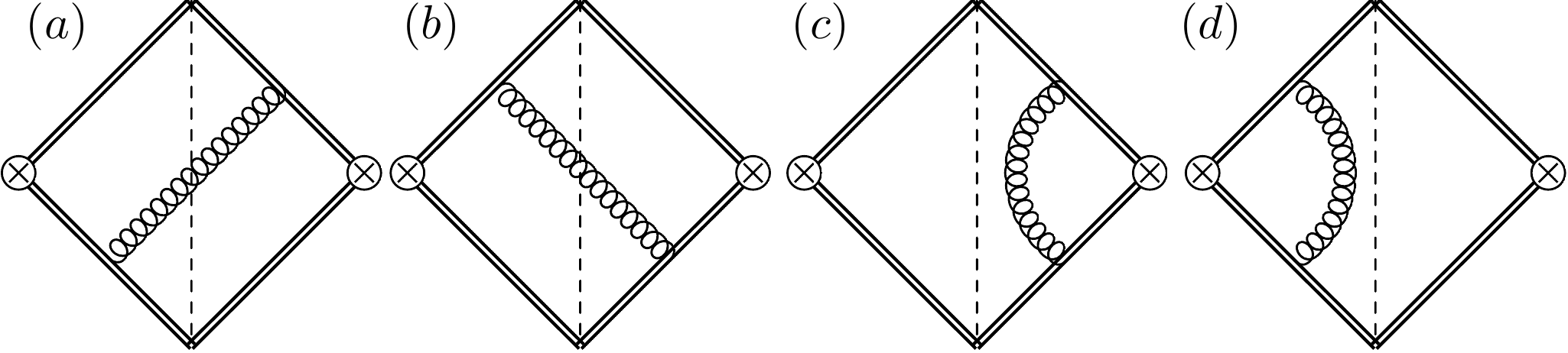}
   \centering
\caption{\label{JBSoft}Diagrams contribution to broadening soft function. The virtual diagrams (c) and (d) are zero in pure dim.-reg.}
\end{figure}
At one-loop  we regulate rapidity divergences with $\eta$ and use dimensional regularization for UV and IR divergences, just as in the previous section\footnote{We have verified that regulating IR divergences with gluon mass at one-loop yields the same results as presented here.}. In Feynman gauge only two real radiation diagrams contribute to the one loop soft function which are identical to each other leaving
\begin{align} \label{S_pT}
&{\cal S}^{(1)}(e_R,e_L, \vec p^{\,2}, \vec q^{\,2})= 4 g^2 w^2 C_F\mu^{2\epsilon}\nu^{\eta}  \frac{\pi^{2-2\epsilon}(\vec p^{\, 2})^{-\epsilon} (\vec q^{\, 2})^{-\epsilon}}{\Gamma^2(1-\epsilon)} \int  \frac{d^dk}{(2\pi)^{d-1}}\delta^{(+)}(k^2)\frac{|2k^3|^{-\eta}}{\overline{n}.k \; n.k} \\
 &\!\!\!\!\! \quad \times \!\! \left[ \theta(k^3)\delta(e_R- |\vec k_t |/Q)\delta^{d-2}(\vec{k}_t-\vec p \, )\delta(e_L)\delta^{d-2}(\vec q\,)+\theta(-k^3) \delta(e_L- |\vec k_t |/Q)\delta^{d-2}(\vec{k}_t-\vec q \, ) \delta\left(e_R\right) \delta^{d-2}(\vec p\,) \right ] \nn \\
& = \frac{\alpha_s C_F w^2}{ \pi}   \frac{e^{\epsilon \gamma_E}\Gamma\left(\frac12 -\frac{\eta}{2}\right)\Gamma\left(\frac{\eta}2\right)}{2^{\eta}\sqrt{\pi}\Gamma(1-\epsilon)} \frac{\nu^\eta}{Q^\eta}\frac{\mu^{2\epsilon}}{Q^{2\epsilon}}  
 \left[ \infplus{1}{e_R^{1+2\epsilon+\eta}} \!\! \delta(Q^2 e_R^2 -\vec p^{\, 2}) \delta(e_L) \delta(\vec q^{\, 2}) 
   + \bigg ( \!\!
  \begin {array} {c} 
  L \leftrightarrow R \\ 
  \vec p \leftrightarrow \vec q 
  \end {array} \!\! \bigg)
 \right] \, .\nn
\end{align}
Expanding in $\eta$, then in $\epsilon$ and combining with the  tree-level result we have the bare soft function at NLO,
\begin{align}\label{eq:NLO-soft-bare-JB}
& {\cal S}^{\rm bare}(e_R,e_L, \vec p^{\,2}, \vec q^{\,2}) = 
\delta(e_R) \delta(e_L) \delta(\vec p^{\, 2})  \delta(\vec q^{\, 2}) 
+ \frac{\alpha _s(\mu )C_F}{\pi } \Bigg [  \Bigg ( \frac{2 w^2 e^{\epsilon  \gamma _E} }{\eta  \, \Gamma (1-\epsilon )}\frac{\mu^{2\epsilon} }{ Q^{2\epsilon } } \infplus{1}{e_R^{1+2\epsilon}} \nn \\
& + \delta(e_R)\left(\frac{1}{2\epsilon ^2}-\frac{1}{2\epsilon }\ln \frac{\nu ^2}{\mu ^2}\right) - \frac{2Q}{\mu} \plusf{\mu \ln(Q e_R/\mu)}{Q e_R} +  \frac{Q}{\mu} \plusf{\mu}{Q e_R} \ln \frac{\nu^2}{\mu^2} -\frac{\pi^2}{24}\delta(e_R) \Bigg ) \nn \\
& \times  \delta(Q^2 e_R^2 - \vec p^{\, 2}) \delta(e_L)   \delta(\vec q^{\, 2})  
+ \bigg ( \!\!
  \begin {array} {c} 
  L \leftrightarrow R \\ 
  \vec p \leftrightarrow \vec q 
  \end {array} \!\! \bigg) \,\,
\Bigg ] \, .
\end{align}
As in  the jet case, we have written the  rapidity divergences in terms of plus-distributions with boundary at $\infty$ and the finite corrections in terms of the standard plus-distributions. 

Using eqn.~(\ref{eq:broadening-ren}) we extract the renormalization constant for the soft function,
\begin{align}
&Z_{\cal S}(e_R,e_L, \vec p^{\,2}, \vec q^{\,2}) = 16 \pi^2 \delta(e_R) \delta(e_L) \delta(\vec p^{\, 2})  \delta(\vec q^{\, 2}) 
+ (16 \pi^2)  \frac{\alpha _s(\mu )C_F}{\pi } \Bigg [  \Bigg \{ \frac{2 w^2 e^{\epsilon  \gamma _E} }{\eta  \, \Gamma (1-\epsilon )}\frac{\mu^{2\epsilon} }{ Q^{2\epsilon } } \infplus{1}{e_R^{1+2\epsilon}} \nn \\
& \hspace{2cm} + \delta(e_R)\left(\frac{1}{2\epsilon ^2}-\frac{1}{2\epsilon }\ln \frac{\nu ^2}{\mu ^2}\right)  \Bigg \}
  \delta(Q^2 e_R^2 - \vec p^{\, 2}) \delta(e_L)   \delta(\vec q^{\, 2})  
+ \bigg ( \!\!
  \begin {array} {c} 
  L \leftrightarrow R \\ 
  \vec p \leftrightarrow \vec q 
  \end {array} \!\! \bigg) \,\,
\Bigg ] \, ,
\end{align}
leaving the renormalized soft function
\begin{align}
& {\cal S}^{\rm ren}(e_R,e_L, \vec p^{\,2}, \vec q^{\,2}) = 
\delta(e_R) \delta(e_L) \delta(\vec p^{\, 2})  \delta(\vec q^{\, 2}) 
+ \frac{\alpha _s(\mu )C_F}{\pi } \Bigg [  \Bigg \{ 
- \frac{2Q}{\mu} \plusf{\mu \ln(Q e_R/\mu)}{Q e_R} \nn \\
& \hspace{2cm} +  \frac{Q}{\mu} \plusf{\mu}{Q e_R} \ln \frac{\nu^2}{\mu^2} -\frac{\pi^2}{24}\delta(e_R) \Bigg \} 
  \delta(Q^2 e_R^2 - \vec p^{\, 2}) \delta(e_L)   \delta(\vec q^{\, 2})  
+ \bigg ( \!\!
  \begin {array} {c} 
  L \leftrightarrow R \\ 
  \vec p \leftrightarrow \vec q 
  \end {array} \!\! \bigg) \,\,
\Bigg ] \, .
\end{align}
The anomalous dimensions can be calculated using eqn.~(\ref{eq:soft-AD-eqn}). At one-loop we have,
\begin{align}\label{eq:soft-AD-JB}
\gamma_\mu^{\cal S}(\mu, \nu/\mu) &=  - \frac{2\alpha _s(\mu )C_F}{\pi }\ln \frac{\nu ^2}{\mu ^2}\\
\gamma_\nu^{\cal S}(e_R,e_L, \vec p^{\,2}, \vec q^{\,2})  &=  \frac{2\alpha _s(\mu ) C_F}{\pi} (16\pi^2) \frac{Q}{\mu} \plusf{\mu}{Q e_R} \delta(Q^2 e_R^2 - \vec p^{\, 2}) \delta(e_L)   \delta(\vec q^{\, 2})  
+ \bigg ( \!\!
  \begin {array} {c} 
  L \leftrightarrow R \\ 
  \vec p \leftrightarrow \vec q 
  \end {array} \!\! \bigg)   \, . \nn
\end{align}
It is a straightforward exercise to check that constraint (\ref{eq:mu-nu-constraint}) for soft anomalous dimensions are satisfied up to order $\alpha_s$. We cannot do a direct check on constraints (\ref{eq:sum-o-mu}) and (\ref{eq:sum-o-nu}) but an indirect check comes from calculating the cross-section at order $\alpha_s$ with the bare matrix elements. We will perform this check in sec. \ref{sec:NLO-JB}.


\subsection{NLL Soft Function}

 To calculate the NLL cross-section for jet broadening we must evolve the  soft function  in $\nu$ up to the jet scale ($Q$) as  shown in fig.~\ref{fig:strategy}. For this purpose we need to solve the $\nu$-RGE and obtain $V_{\cal S}$ at NLL. Its easiest to solve this equation in conjugate space where we make Fourier transform w.r.t. $\vec p, ~\vec q$ and make Laplace transform w.r.t. $e_L, ~ e_R$. In the conjugate space $\nu$-RGE reads
\begin{align}
\frac{  d}{d \ln(\nu)}  \tilde {\cal S}(\tau_R,\tau_L,b_R,b_L;\mu,\nu/\mu) = \tilde \gamma_\nu^{\cal S}(\tau_R,\tau_L,b_R,b_L;\mu)  \, \tilde {\cal S} \, ,
\end{align}
where $\vec b_R~(\vec b_L)$ is Fourier conjugate to $\vec p~(\vec q)$ and $\tau_{R,L}$ is Laplace conjugate to $e_{R,L}$. Since $\gamma_\nu$ has no explicit dependence on $\nu$ solution of this RGE is simply
\begin{align}
\tilde {\cal S}(\ldots;\mu, \nu/\mu) = \exp\! \left (  \tilde\gamma_\nu^{\cal S}(\ldots;\mu) \, \ln\frac{\nu}{\nu_0} \right ) \, \tilde{\cal S}(\ldots;\mu,\nu_0/\mu) \, .
\end{align}
Therefore,
\begin{align}
V_{\cal S} = LF^{-1}\left [\exp\! \left (  \tilde\gamma_\nu^{\cal S}(\tau_R,\tau_L,b_R,b_L;\mu) \, \ln\frac{\nu}{\nu_0} \right )\right ] \, ,
\end{align}
where $LF^{-1}$ refers to the inverse Laplace and inverse Fourier transform on $\tau_{R,L}$ and $b_{R,L}$, respectively. From Laplace-Fourier transform of the result in eqn.~(\ref{eq:soft-AD-JB}) we find that
\begin{align}
 \tilde\gamma_\nu^{\cal S} = - 2\frac{\alpha _s(\mu ) C_F}{\pi } \left[\ln  \left(\sqrt{b_R^2Q^2+\tau _R^2}+\tau _R\right)+\ln  \left(\sqrt{b_L^2Q^2+\tau _L^2}+\tau _L\right)+\ln \frac{\mu ^2 e^{2\gamma _E}}{4Q^2}\right] \,
\end{align}
which gives
\begin{align}\label{eq:VS-LF-space}
V_{\cal S}^{\rm NLL} = LF^{-1}\left [ \left(\frac{\mu }{2Q}\right)^{-2\omega_s } e^{-2\omega_s  \gamma _E} \bigg(\sqrt{b_R^2Q^2+\tau _R^2}+\tau _R\bigg)^{-\omega_s } \bigg(\sqrt{b_L^2Q^2+\tau _L^2}+\tau _L\bigg)^{-\omega_s }\right ] \, ,
\end{align}
where
\begin{align}\label{eq:omega-JB}
\omega_s \equiv \omega_{\rm NLL}(\alpha_s(\mu) , \nu/\nu_0) =  \frac{2 \alpha_s(\mu) C_F}{\pi} \, \ln\frac{\nu}{\nu_0} .
\end{align}
Using eqns. (\ref{eq:inv-L-transforms}) and (\ref{eq:inv-F-transform}), we evaluate inverse Laplace-Fourier transforms in eqn.~(\ref{eq:VS-LF-space}), to obtain the $\nu$-evolution factor $V_{\cal S}$ at NLL in the physical space,
\begin{align}
V_{\cal S}^{\rm NLL} =(16\pi ^2)(\mu  Q)^{-2\omega_s }  \frac{\omega_s ^2e^{-2\omega_s  \gamma _E} }{\Gamma ^2(\omega_s ) } \infplus{\theta(e_R)}{e_R^{1+\omega_s}} \infplus{\theta(e_L)}{e_L^{1+\omega_s}} \frac{\theta(Q e_R-|\vec p\,|)}{(Q^2 e_R^2-\vec p^{\, 2})^{1-\omega_s}} \frac{\theta(Q e_L-|\vec q\,|)}{(Q^2 e_L^2-\vec q^{\, 2})^{1-\omega_s}} \, .
\end{align}
Note that $\omega_s$ is always positive for us ($\nu>\nu_0$), so there is no non-integrable singularity in the last two fractions and other terms are properly plussed. Convolving this with the tree-level soft function gives us the NLL soft function,
\begin{align}\label{eq:NLL-soft-JB}
{\cal S}_{\rm NLL} = (\mu  Q)^{-2\omega_s }  \frac{\omega_s ^2e^{-2\omega_s  \gamma _E} }{\Gamma ^2(\omega_s ) } \infplus{\theta(e_R)}{e_R^{1+\omega_s}} \infplus{\theta(e_L)}{e_L^{1+\omega_s}} \frac{\theta(Q e_R-|\vec p\,|)}{(Q^2 e_R^2-\vec p^{\, 2})^{1-\omega_s}} \frac{\theta(Q e_L-|\vec q\,|)}{(Q^2 e_L^2-\vec q^{\, 2})^{1-\omega_s}} \, .
\end{align}
This very interesting formula illustrates how the soft function changes when we take into account infinite gluon emissions from a quark between the rapidities of order $|\ln(\nu_0/\mu)|$ and $|\ln(\nu/\mu)|$. Resummation of all the large rapidity logarithms require that we choose $\mu \sim \nu_0 \sim Q \, e_{L,R}$ and $\nu \sim Q$. This would mean that ${\cal S}_{\rm NLL}$ accounts for recoil from  gluon emissions of rapidities, $|y| \lesssim | \ln(e_{L,R}) |$. Note that the unwanted singularities that arose in section \ref{sec:Higgs-resum} do not arise here. Reason being, there is an absolute cutoff on the transverse momentum governed by the measured $e_{L,R}$. Thus no unwanted UV contributions can arise in the inverse transform. Alternately, there is no singularity from $b \to 0$ in eqn. (\ref{eq:VS-LF-space}) for any value of $\omega_s$.


\subsection{Jet Broadening Spectrum at NLL}\label{sec:NLL-JB}

It is a straight forward exercise to calculate NLL differential cross-section now. NLL formula is given by
\begin{align}
\frac{d\sigma^{\rm NLL}}{de_L de_R}
& =  \sigma_0 \, H_{\rm tree}(Q^2,\mu_H) U_H^{\rm NLL} (Q^2,\mu_H,\mu) \, \int de_n \, de_{\bar n} \, de_s^L\, de_s^R \,\, \delta(e_R - e_n  - e_s^R) \delta(e_L - e_{\bar n} - e_s^L) \nn \\
& \quad \quad \quad \,\int d \vec p^{\,2} \, d \vec q^{\,2}  {\cal J}_{\rm tree}(Q,e_n,\vec p^{\,2}) \,  {\cal J}_{\rm tree}(Q,e_{\bar n},\vec q^{\,2}) \, {\cal S}_{\rm NLL}(e_s^R, e_s^L,  \vec q^2,\vec p^2) \, ,
\end{align} 
where $H_{\rm tree} = 1$ and $U_H^{\rm NLL}$ is the evolution factor for the hard running. It is the same as for other angularities (for example thrust) and can be found in, for example App. {\bf C.2} of ref. \cite{Jain:2011xz}. Tree-level jet functions were given in eqn. (\ref{eq:tree-jet-JB}) and are simply $\delta$-functions. Taking the NLL soft function from eqn. (\ref{eq:NLL-soft-JB}) and performing straight forward $\delta$-function integrals we get
\begin{align}
\frac{d\sigma^{\rm NLL}}{de_L de_R} = \sigma_0 U_H^{\rm NLL}(Q^2,\mu_H,\mu) \frac{\mu^{-2\omega_s}}{Q^{-2\omega_s}} \frac{\omega_s^2  e^{-2\omega_s  \gamma _E}}{4\, \Gamma ^2(\omega_s ) }\frac{1}{e_L^{1-\omega_s }e_R^{1-\omega_s }} 
\Bigg [ \int _0^1dx \frac{x }{\left(1-\frac{x}{2}\right)^{1+\omega_s} }\frac{1}{(1-x)^{1-\omega_s}} \Bigg]^2
\end{align} 
where we have non-dimensionalized the last remaining integral and dropped the plus prescription since the integrals are well defined. Performing this integral yields the master formula for NLL cross-section
\begin{align}
\frac{d\sigma^{\rm NLL}}{de_L de_R} = \sigma_0 U_H^{\rm NLL}(Q^2,\mu_H,\mu) \frac{\mu^{-2\omega_s}}{Q^{-2\omega_s}}  \frac{e^{-2\omega_s  \gamma _E}}{\Gamma ^2(\omega_s ) }\frac{1}{e_L^{1-\omega_s }e_R^{1-\omega_s }}
\Big [ 1- \frac{\omega_s}{2^{-\omega_s }}B_{\frac12}(1+\omega_s ,0)  \Big]^2
\end{align}
where $B_z(a,b) = \int_0^z d x\,  x^{a-1}(1-x)^{b-1}$, is the incomplete beta function and $\omega_s$ was defined in eqn.~(\ref{eq:omega-JB}). Using this result we obtain the NLL cross-section for wide jet broadening $B_W$ \cite{Catani:1992jc},
\begin{align}
\frac{d\sigma ^{\text{NLL}}}{dB_W}= \sigma_0 U_H^{\rm NLL}(Q^2,\mu_H,\mu) \frac{\mu^{-2\omega_s}}{Q^{-2\omega_s}}  \frac{e^{-2\omega_s  \gamma _E}}{\Gamma (1+\omega_s )\Gamma (\omega_s ) }\frac{2^{1+2\omega_s }}{B_W^{1-2\omega_s }} 
\Big [ 1- \frac{\omega_s}{2^{-\omega_s }}B_{\frac12}(1+\omega_s ,0)  \Big]^2
\end{align}
and total jet broadening \cite{Catani:1992jc}
\begin{align}\label{eq:BT-NLL}
\frac{d\sigma ^{\text{NLL}}}{dB_T}=\sigma _0  U_H^{\rm NLL}(Q^2,\mu_H,\mu) \frac{\mu^{-2\omega_s}}{Q^{-2\omega_s}} \frac{e^{-2\omega_s  \gamma _E}4^{\omega_s }}{\Gamma (2\omega_s ) }\frac{1}{B_T^{1-2\omega_s }} 
\Big [ 1- \frac{\omega_s}{2^{-\omega_s }}B_{\frac12}(1+\omega_s ,0)  \Big]^2 \, .
\end{align}


\subsection{Jet Broadening Spectrum at LO}\label{sec:NLO-JB}
 We will compute the spectrum at LO using the bare matrix elements to elucidate the cancellation of rapidity divergences and corresponding scale $\nu$. This distribution accounting for one real radiation is given by 
 \begin{align}
\frac1{\sigma_0}\frac{\dd \sigma^{\rm LO}}{\dd e_L \dd e_R}=& ~ \delta (e_L)\delta (e_R) \left(1+H_{\rm bare}^{(1)} (Q^2)\right) + \delta(e_L)  J_{\rm bare}^{(1)}(e_R, \vec p^{\, 2} = 0)+ \delta(e_L)  J_{\rm bare}^{(1)}(e_L, \vec q^{\, 2} = 0) \nn \\
 & + 4 Q^4 \int \dd e_n \dd e_{\bar n} \, e_n \, e_{\bar n} \, {\cal S}^{(1)}_{\rm bare} (e_R-e_n,e_L-e_{\bar n} , Q^2 e_n^2, Q^2 e_{\bar n}^2 )
\end{align}
which can be computed by using (\ref{eq:jet-bare-JB}) and (\ref{eq:NLO-soft-bare-JB}) along with corresponding tree level results. We obtain
\begin{align}\label{eq:LO-LR-cross-section}
\frac1{\sigma_0}\frac{\dd \sigma^{\rm LO}}{\dd e_L \dd e_R}  =&~ \delta (e_R)\delta(e_L)\left(1+H_{\text{bare}}^{(1)}(Q,\mu )\right) \\
& +\frac{\alpha _s(\mu )C_F}{\pi }\delta (e_L)\delta (e_R)\left(\frac{1}{\epsilon ^2}+\frac{1}{\epsilon }\ln \frac{\mu ^2}{Q^2}+\frac{3}{2\epsilon }+\frac{1}{2}-\frac{\pi ^2}{12}\right) 
 + \Bigg \{ \frac{\alpha _s(\mu )C_F}{\pi }\delta(e_L) \nn \\
& \left(-\frac{Q}{\mu} \plusf{2\mu  \ln(Q e_R/(2\mu ))}{Q e_R} - \frac{3Q}{4\mu } \plusf{2\mu }{Q e_R} -\frac{Q}{2\mu }\plusf{2\mu }{Q e_R} \ln \frac{\mu ^2}{Q^2} \right )+(L\leftrightarrow R)\Bigg \} \, . \nn
\end{align}
As expected both the rapidity divergences and $\nu$-dependence cancels out.  The sum of the UV divergences in the IR sector add up to give the expected form consistent with the hard anomalous dimension, i.e they cancel with the  UV divergences in the bare hard function \cite{Manohar:2003vb,Bauer:2003di},
\begin{align}
H_{\rm bare}^{(1)} =\frac{\alpha _s(\mu )C_F}{\pi }\left( -\frac{1}{\epsilon^2} -\frac{1}{\epsilon}\ln\frac{\mu^2}{Q^2} -\frac{3}{2\epsilon}-\frac{1}{2}\ln ^2\frac{\mu^2}{Q^2}-\frac{3}{2}\ln \frac{\mu^2}{Q^2}-4+\frac{7\pi ^2}{12}\right) .
\end{align}
Using eqn.~(\ref{eq:LO-LR-cross-section}) we calculate the singular contribution to the LO total broadening distribution
\begin{align}
\frac{1}{\sigma _0}\frac{d\sigma ^{\text{LO}}}{dB_T}=- \frac{\alpha _s(\mu )C_F}{\pi \,  B_T}(4 \ln B_T+3) \, ,
\end{align}
which agrees with \cite{Catani:1992jc}.

\subsection{Numerics}
\begin{figure}[h]
    \includegraphics[width=10cm]{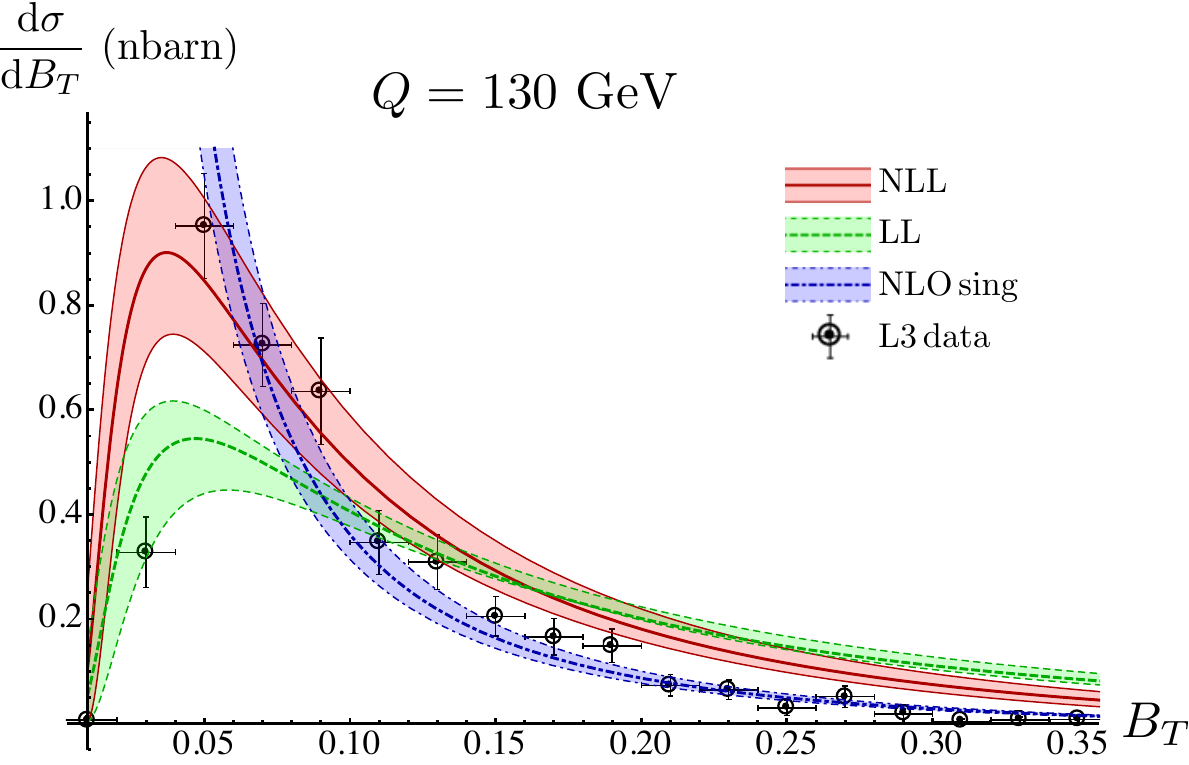}
   \centering
\caption{\label{L3}Total Jet Broadening at 130 GeV. }
\end{figure}

In fig.~(\ref{L3}) we have plotted the theory cross section and the data \cite{Achard:2004sv}. The resummed error bands are the geometric mean of the $\nu$-variation and $\mu$-variation.
We see that given the large experimental error bars the agreement with the data is reasonable.
Complete jet function calculation up to NLO, although will not change the NLL resummed spectrum, is expected to bring down the scale uncertainty significantly. In addition, the NNLL calculation will reduce the theory errors further. It is worth noting that by including both $\nu$- and $\mu$-variations, we gain a well controlled theoretical uncertainty estimation, while the uncertainty analysis for resummation using traditional methods could be ambiguous and may under or over estimate the uncertainties as we will discuss in the next section.
Here we have not included the theory errors due to power corrections. In the small $B_T$ region these are non-perturbative and scale as $\Lambda_{QCD}/(B_TQ)$ and can be expected to be of order $20$-$30$\%. 


\subsection{Comparison to Previous Results}

In our previous work \cite{Chiu:2011qc} where we introduced the rapidity renormalization group, we presented results for NLL total broadening distribution with an unnecessary approximation that oversimplified the structure of $\gamma_\nu^{\cal S}$. As a consequence, our result presented in \cite{Chiu:2011qc} was not accurate at NLL when power counting the resummation in exponent. The result presented in here in eqn.~(\ref{eq:BT-NLL}) differs from one in \cite{Chiu:2011qc}  by a factor of $\big [ 2^{\omega_s}- \omega_s 4^{\omega_s}B_{\frac12}(1+\omega_s ,0)  \big]^2$. This extra factor agrees with the extra factor mentioned in the ``{\it Note added}'' of ref. \cite{Becher:2011pf} which also considered jet broadening in the context of SCET. Our result in eqn.~(\ref{eq:BT-NLL}) agrees with \cite{Becher:2011pf} up to the distinction between our $\omega_s$ given in eqn.~(\ref{eq:omega-JB}) corresponding to  factor $\eta$ in \cite{Becher:2011pf}. The distinction is conceptual and an important one when gauging scale dependence. We explain this below.

 In   the formalism  developed in  \cite{Chiu:2007dg},  which was also used by   Ref. \cite{Becher:2011pf}, a single logarithm of $Q^2$ appears in the combined result of the logarithm of IR sectors. The coefficient of the logarithm of $Q^2$ is extracted and the logarithm is  exponentiated  when calculating the resummed cross-section. This coefficient is unique, however, the scale associated with $Q^2$ in the logarithm is ambiguous in this formalism. Ref. \cite{Becher:2011pf} effectively make this choice same as the renormalization scale $\mu$, hence $\ln(Q^2/\mu^2)$ show up in $\eta$, the factor corresponding to $\omega_s$. They could have made an alternate choice, for example $\ln(Q^2/(Q B_T)^2)$, since $\mu \sim Q B_T$. The two choices differ in that they give significantly different estimate for $\mu$-variation at NLL, one choice underestimates it while other overestimates. In our formalism such an ambiguity does not arise since  $\nu$  is unrelated to $\mu$ and can be varied  independently to gauge the errors associated with  the choice of this scale. We find that  our combined $\mu$ and $\nu$-variation errors are significantly smaller than the result of ref. \cite{Becher:2011pf}. This becomes important in phenomenological applications of the NLL distribution, for example fitting $\alpha_s$ from the total jet broadening data.

Ref. \cite{Becher:2011pf}  state that our ``analytic regulator" \footnote{Our regulator is not an analytic regulator by any definition that we are aware of, though it has the appearance of an analytic regulator at one loop for real radiation in the collinear sector only.} will not necessarily reproduce full QCD since we regulate in the effective theory. We have proven in this article that the regulator leads to a correct cancellation of $\eta$ divergences, preserves non-Abelian exponentiation, the soft-collinear  gauge invariance  and factorization in SCET. In SCET the analytic regulator leads to gauge dependent collinear and soft functions and breaks non-Abelian exponentiation, though it appears it can be modified to remedy these faults \cite{Becher:2011dz}.


\section{Application to Exclusive Processes and End  Point Singularities}
\label{ex}
 We now discuss the application of RRG to the exclusive processes.
 There is a long standing problem in such processes, upon which, we hope to shed some light.
 In particular, many exclusive processes such as 
 the pion form factor at large $Q^2$ \cite{Geshkenbein:1982zs}, the $\rho - \pi$ form factor \cite{Chernyak:1983ej}, and the $B \rightarrow \pi l \nu$ form factor \cite{Akhoury:1993uw}  
 are plagued by end-point divergences.  In SCET these divergences arise in problems that fall with in the domain of
 \scetii.  Thus to avoid this issue in a sensible fashion one is forced to remain in \sceti~ \cite{Bpipi,BKpi} at the cost
 of loss of predictive power. If we could make sense of the end point singularities in \scetii, this
 would  increase predictive power by allowing for one to write down the rates in terms of
 light-cone wave functions. 
 Overcoming these end point singularities in a systematic fashion is thus highly desirable.
  
End point divergences arise when integrating over the momentum fractions in light cone wave functions.
 Schematically these divergence amplitudes are of the form
 \beq
 A= \int_0^1 dx C(x) \phi_H(x) 
 \eeq
 where $x$ is a momentum fraction $\phi_H$ is a light-cone wave function for hadron $H$, and $C$ is a
 perturbative hard matching coefficient. It is often the case that $C(x)$ is singular at the lower end-point.
 While we cannot calculate $\phi_H$ from first principles, we can say something about its form in the limit
 where the hard scale is taken to infinity \cite{Lepage:1980fj}.  In particular when $x$ approaches zero $\phi_H$ 
 vanishes linearly in $x$, whereas it is often the case that $C(x)\sim 1/x^2$ leading to a divergence.
Ref.  \cite{Manohar:2006nz}  correctly account for this singularity as a rapidity divergence and performs a zero bin subtraction.  However,   in \scetii~ subtracting the double counting region can at best move the boundary between the soft and collinear sectors. The rapidity divergences and their accompanying logarithms still exist after the  subtraction since they are associated with the boundary (see section (\ref{def})).
As such, the  discussion in \cite{Manohar:2006nz} was incomplete in the sense that the method does not allow for a resummation. 
The crucial distinction, to be discussed below,  between their regulator and the one employed here is that they imposed manifest
boost invariance in each sector, whereas we explicitly break it with our regulator and only ask that
the final answer be boost invariant.

 We follow \cite{Beneke:2003pa,Manohar:2006nz} in studying this issue within the context of a toy model of $B$ decays.
 We will provide a proof in principle that  exclusive $B$-decays can be factorized in SCET$_{\rm II}$ and all the logarithms can be resummed. We take the example of $B \to \ell \nu \gamma$ where all fields are taken to be scalars. The physical case of fermions was considered in \cite{Beneke:2003pa} but does not suffer from an endpoint singularity until the subleading order.
 The scalar case on the other hand does, and was originally thought to be non-factorizable due to these divergences  \cite{Becher:2003qh}.
  Manohar and Stewart were able to regulate all the integrals and obtained the correct IR divergences  after employing a zero-bin procedure. However, they were unable to resum the rapidity logarithms that appear in the ratio $\mu^+/\mu^-$, the scales associated with their rapidity regularization. This was due to a mismatch in hard logarithms and hard anomalous dimension. We will use the same set up as ref. \cite{Manohar:2006nz} (see section VII A therein for details) but we employ the regulator introduced earlier in this paper. We will also include a region that was not discussed in the analysis
  of ref. \cite{Manohar:2006nz}  that we find essential for solving the problem in all frames of reference. We will ignore the wave function renormalization in our analysis as it is straightforward to include.
 
  \begin{figure}
  \centering
    \includegraphics[width=14 cm]{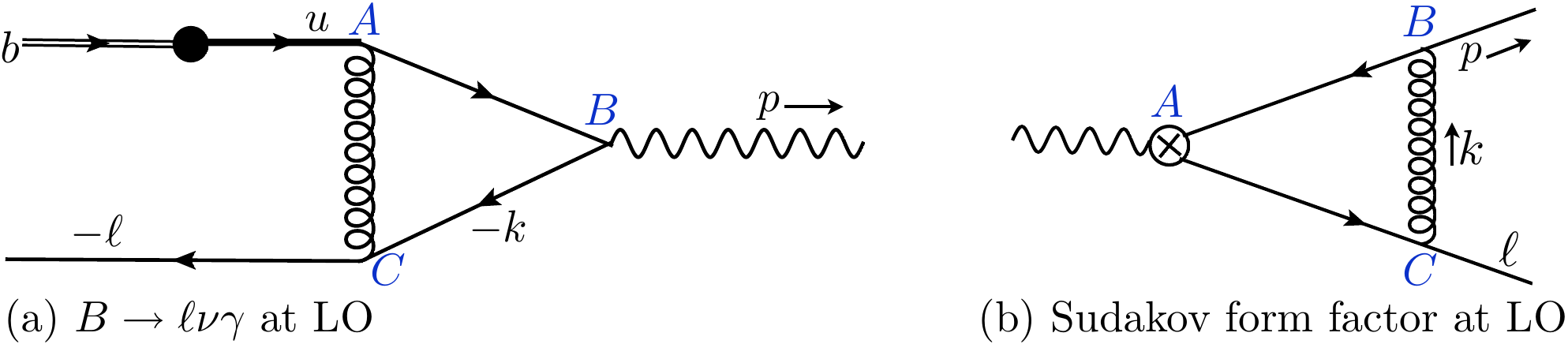}
\caption{ \label{fig:mapping} (a) $W$-boson and its decay is not shown. $b u W$ vertex is represented by \textbullet. $u$ quark represented with the thick line is the hard parton with off-shellness $m_B \Lambda_{\rm QCD}$. Hard interaction takes place at vertex A and external on-shell partons are at vertex B and C. (b) Corresponding Sudakov form factor demonstrating hard interaction at vertex A. On-shell $n$-collinear parton is at vertex B and on-shell $\bar n$-collinear parton is at vertex C.
}
\end{figure}
 
 At next-to leading order and in partonic approximation, the process  $B \to \ell \nu \gamma$ is represented by the Feynman diagram in fig.~\ref{fig:mapping}(a). In this process, observed in $B$-meson rest frame, $p$ is $n$-collinear with $p^- \sim m_B$ and $p^2=0$. While, $\ell$ is soft with $\ell^+ \sim \Lambda_{\rm QCD}$ and $\ell^2=0$. After radiating the vector boson, a hard interaction takes place at vertex A and is only sensitive to the Lorentz invariant combination $p^- \ell^+$. The external on-shell particles (or particles with off-shellness at most $\Lambda_{\rm QCD}^2$) are at vertices B and C.  This full theory integral has IR divergences but is UV finite. To control IR divergences we choose to regulate the (scalar) quark BC with a non-zero mass $m$, as was done it  ref.~\cite{Manohar:2006nz}. We must take $m\sim \Lambda_{\rm QCD}$ to ensure proper scaling of the IR physics. This integral  has a double logarithm, $\ln^2(p^-\ell^+)/m^2$, that is large. A proper factorization should separate the hard scale $p^-\ell^+$ from the non-perturbative scale $m$ and provide a method for resummation.  We will now show that this problem is identical to the Sudakov form factor at LO as far as the momentum flow is concerned \footnote{A similar argument was used in 
 \cite{Becher:2003qh}, however these authors worked in \sceti ~power counting with off-shellness regulator which lead to the erroneous conclusion that the 
 amplitude  did not factor due to the existence of so-called messenger modes which are just the boosted ultra-soft modes.}.
  
We make a boost to a frame where $\ell^+ = p^- \sim \sqrt{m_B \Lambda_{\rm QCD}}$. We will refer to this as the {\it symmetric} frame. In this frame, diagram in fig.~\ref{fig:mapping}(a) immediately maps to the familiar picture of the massive Sudakov form factor at one loop shown in fig.~\ref{fig:mapping}(b), where  the hard interaction is at vertex A and (anti-)collinear on-shell quarks are at vertices  B and C with $p^- = \ell^+ = Q$. The hard scale here is $p^-\ell^+ = Q^2$ and IR scale is set by the mass $m$ of the vector boson exchanged between two quarks. The full theory integral is the same in the two situations when all particles are replaced by the scalars as in the toy example considered here. Previously we factorized the  Sudakov form factor  into ($\bar n$)$n$-collinear and soft regions corresponding to the situation when the loop momentum $k$ becomes ($\bar n$)$n$-collinear or soft.  The same factorization applies to  $B$-decays in the symmetric frame with the identification, $Q^2 \sim m_B \Lambda_{\rm QCD}$, which then yields $\lambda^2 = m^2/Q^2 \sim \Lambda_{\rm QCD}/m_B$.

\begin{figure}
  \centering
    \includegraphics[width=15.5 cm]{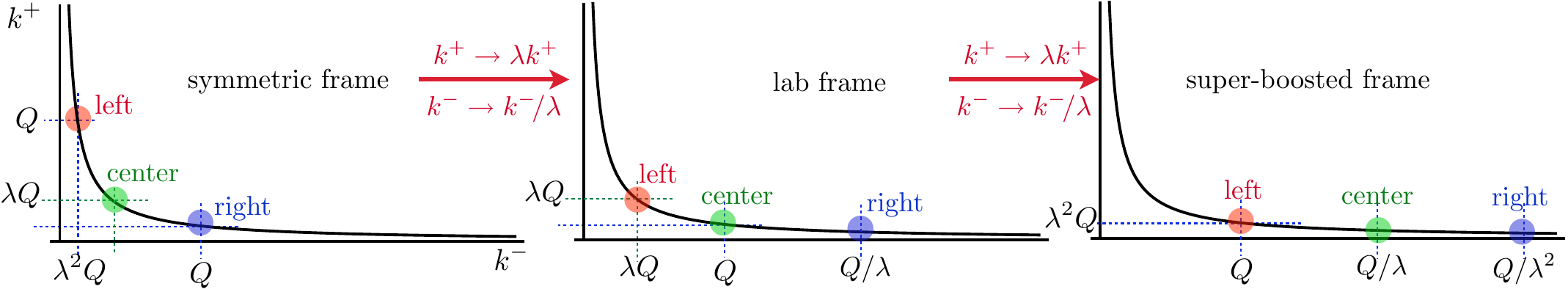}
\caption{ \label{fig:frames} Modes and Frames: relationship between symmetric, lab and super-boosted frame is shown along with the modes in each frame with their location on the hyperbola.
}
\end{figure}

Ref.~\cite{Manohar:2006nz}, analyzed the problem in the $B$-meson rest frame (or lab frame), but did not include the region that corresponds to the soft region of the symmetric frame in their analysis.
This  region looks like a collinear mode in the lab frame but with a lower rapidity compared to the collinear modes that have the same scaling as the photon's momenta. In fig.~\ref{fig:frames}, we show the three IR modes required for this problem in three very different frames of reference. To avoid confusion, we will refer to the modes as left, center and right modes corresponding to their location in the mode diagram. The three frames of reference considered are the {\it symmetric} frame, the {\it lab} frame in which $B$-meson is at rest and the {\it super-boosted} frame where $B$-meson itself is $n$-collinear. In the super-boosted frame both the initial parton and the final photon are $n$-collinear with different rapidity hierarchy. We will now demonstrate a factorization in the symmetric frame which essentially carries over to all other frames. The calculation of the operators changes between frames along with corresponding rapidity logarithms but the end result after the resummation stays the same. To get the same result in all frames it is crucial that our rapidity regulator be employed
in a consistent fashion.

For brevity, we will not present the details of the operators and factorization here and refer the reader to ref.~\cite{Manohar:2006nz}\footnote{The integral corresponding to the missing region and operator in this reference will be referred here as $I_{\rm center}$.}. After suppressing an overall factor of $i eg^2 G /(p^-\ell^+)$, we have the full theory integral for the diagram shown in fig. \ref{fig:mapping}(a),
\begin{align}\label{eq:full-theory}
I_{\text{full}} &= \frac{e^{\epsilon  \gamma _E}\mu ^{2\epsilon }}{(4\pi )^{\epsilon }}\int \frac{d^dk}{(2\pi )^d}\frac{1}{\left[k^2-\ell ^+ k^- + i 0\right ]\left[k^2-m^2+ i 0\right]\left[k^2-p^-k^+ + i 0\right]} \\
&= \frac{-i}{16 \pi ^2\left(p^-\ell ^+\right)}\left[\text{  }\frac{1}{2}\ln ^2\left(\frac{p^-\ell ^+}{m^2}\right)+\frac{\pi ^2}{3}\right] \nn 
\end{align}
where we see the large double logarithm that requires resummation. Note that there is no UV divergence in this full theory result, and that all the IR divergence are regulated by $m^2$.
In the symmetric frame, just like the Sudakov problem, the factorization is accounted for by the following integrals, each corresponding to a different region and operator,
 \begin{figure}[t]
  \centering
    \includegraphics[width=12 cm]{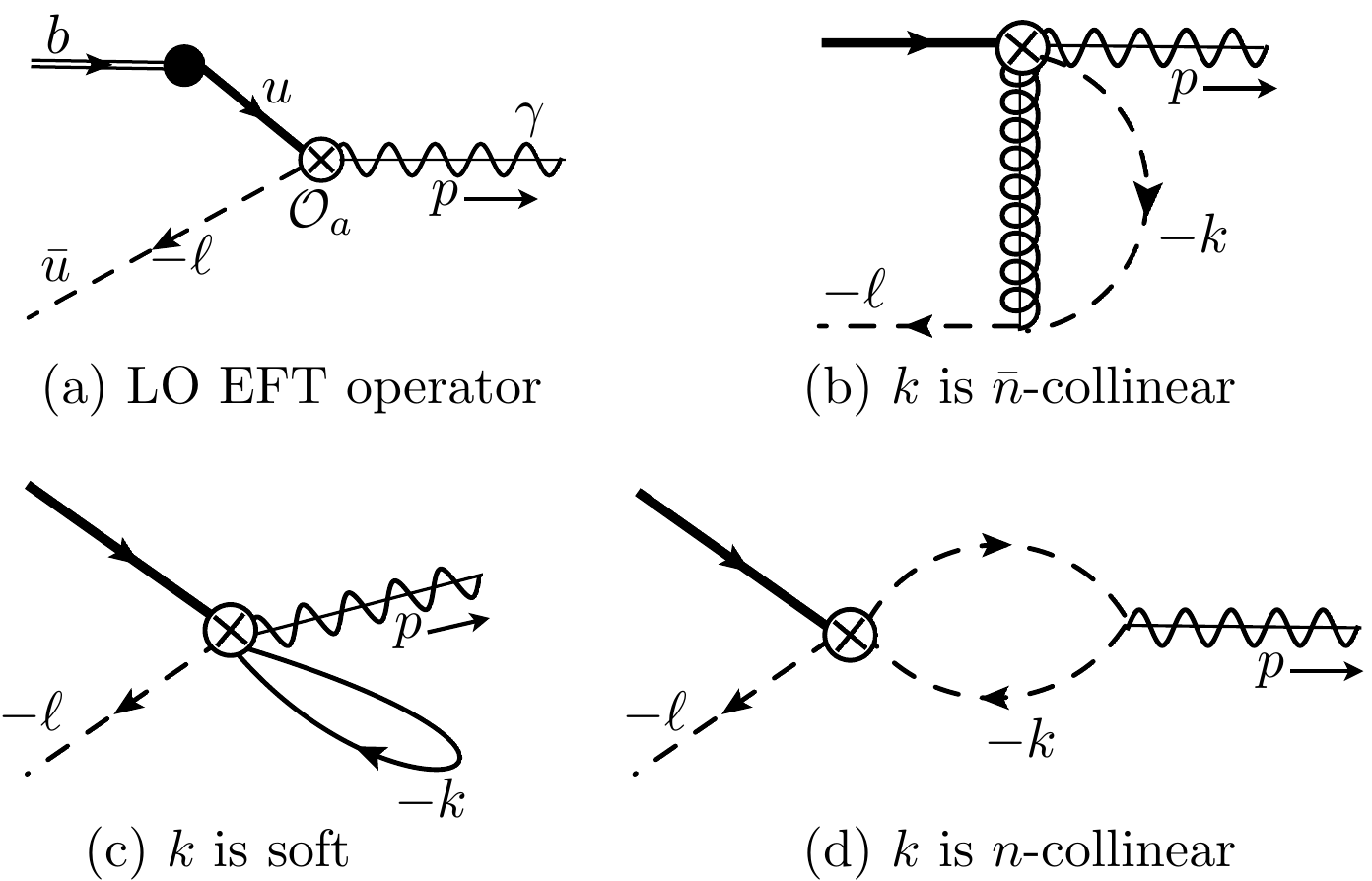}
\caption{ \label{fig:exclusive-loops} Diagrams in the effective theory contributing to $B$-decay in the symmetric frame. Each $\otimes$ corresponds to a different operator.
}
\end{figure}
\begin{align}
 I_{\text{left}} &=\frac{e^{\epsilon  \gamma _E}\mu ^{2\epsilon }}{(4\pi )^{\epsilon }}\int \frac{d^dk}{(2\pi )^d}\frac{f_\eta^{\rm (left)}(k)}{\left[k^2-\ell ^+ k^-  + i 0\right]\left[k^2-m^2 + i 0 \right]\left[-p^-k^+ + i 0 \right]} \, , \nn \\
I_{\text{center}} &=\frac{e^{\epsilon  \gamma _E}\mu ^{2\epsilon }}{(4\pi )^{\epsilon }}\int \frac{d^dk}{(2\pi )^d}\frac{f_\eta^{\rm (center)}(k)}{\left[-\ell ^+ k^- + i 0 \right]\left[k^2-m^2 + i 0 \right]\left[-p^-k^+ + i 0 \right]} \, ,\nn \\
I_{\text{right}} &= \frac{e^{\epsilon  \gamma _E}\mu ^{2\epsilon }}{(4\pi )^{\epsilon }}\int \frac{d^dk}{(2\pi )^d}\frac{f_\eta^{(\rm right)}(k)}{\left[-\ell ^+ k^-  + i 0\right]\left[k^2-m^2  + i 0\right]\left[k^2-p^-k^+  + i 0\right]} \, .
\end{align}
The diagrams corresponding to each of these integrals are shown in figs.~\ref{fig:exclusive-loops}(b), (c) and (d) respectively. They collectively provide renormalization to operator ${\cal O}_a$ shown in fig.~\ref{fig:exclusive-loops}(a). Here $f_\eta(k)$ (shown in Table 1) corresponds to the additional rapidity regularization required to evaluate the integrals which arises from minimally regulating the corresponding operator by inserting an appropriate factor of $f_\eta$ therein. The presence of the regulator breaks the boost-invariance in the otherwise invariant integrands. Therefore, in all frames generated by boosts along $\hat n$ the same integrands  arise but the regulator takes different forms in each frame. Our regulator is simply the correct limit of $w^2 \nu^\eta |2k^3|^{-\eta}$ in each sector, where $w$ is the bookkeeping parameter discussed earlier. In Table~\ref{tab:regulator} we show the regulator in each frame for each mode. 
We will now discuss the calculation and renormalization in each frame.

\begin{table}[b]
\begin{equation*}
\begin{array}{|l|c|c|c|} \hline
 \text{modes} & \text{~~symmetric frame~~} & \text{~~lab or B-meson rest frame~~} & \text{~~super-boosted frame~~} \\ \hline\hline
 \text{left} & w^2\nu ^{\eta }\left|k^+\right|^{-\eta } & w^2\nu ^{\eta }\left|2k^3\right|^{-\eta } & w^2\nu ^{\eta }\left|k^-\right|^{-\eta } \\ \hline
 \text{center} & w^2\nu ^{\eta }\left|2k^3\right|^{-\eta } & w^2\nu ^{\eta }\left|k^-\right|^{-\eta } & w^2\nu ^{\eta }\left|k^-\right|^{-\eta } \\ \hline
 \text{right} & w^2\nu ^{\eta }\left|k^-\right|^{-\eta } & w^2\nu ^{\eta }\left|k^-\right|^{-\eta } & w^2\nu ^{\eta }\left|k^-\right|^{-\eta } \\ \hline
\end{array}
\end{equation*}
\caption{\label{tab:regulator} $f_\eta$ for each mode in each frame.}
\end{table}

\subsection{The symmetric frame}
Using the rapidity regulator as shown in Table~\ref{tab:regulator} for the symmetric frame and set $w=1$ we get,
\begin{align}\label{eq:sym-integrals}
I_{\text{right}} &=\frac{-i}{16 \pi ^2\left( p^-\ell ^+\right)}\left[-\frac{e^{\gamma  \epsilon }\text{  }\Gamma (\epsilon )}{\eta }\left(\frac{\mu ^2}{m^2}\right)^{\epsilon }-\frac{1}{\epsilon }\ln \frac{\nu }{p^-}-\ln \frac{\mu ^2}{m^2} \ln \frac{\nu }{p^-}+\frac{\pi ^2}{6}\right] \, \\
I_{\text{center}} & =\frac{-i}{16\pi ^2p^-\ell ^+}\left[\frac{2 e^{\gamma  \epsilon }\text{  }\Gamma (\epsilon )}{\eta }\left(\frac{\mu ^2}{m^2}\right)^{\epsilon }-\frac{1}{\epsilon ^2}+\frac{2 }{\epsilon }\ln \frac{\nu }{\mu }+\frac{1}{2}\ln ^2\left(\frac{\mu ^2}{m^2}\right)+2 \ln \frac{\mu ^2}{m^2} \ln \frac{\nu }{\mu }+\frac{\pi ^2}{12}\right] \nn \, \\
I_{\text{left}} & =\frac{-i}{16 \pi ^2\left(p^-\ell ^+\right)}\left[-\frac{e^{\gamma  \epsilon }\text{  }\Gamma (\epsilon )}{\eta }\left(\frac{\mu ^2}{m^2}\right)^{\epsilon }-\frac{1}{\epsilon }\ln \frac{\nu }{\ell ^+}-\ln \frac{\mu ^2}{m^2} \ln \frac{\nu }{\ell ^+}+\frac{\pi ^2}{6}\right] \nn \, .
\end{align}
We immediately notice that the rapidity divergences cancel out in the sum of the three integrals, giving for the total bare effective theory contribution
\begin{align}\label{eq:bare-EFT}
I_{\text{EFT}}^{\rm (bare)}=\frac{-i}{16 \pi ^2\left(p^-\ell ^+\right)}\left[ -\frac{1}{\epsilon ^2}+\frac{1}{\epsilon }\ln \frac{p^-\ell ^+}{\mu^2}+\frac{1}{2} \ln^2\left(\!\frac{\mu^2}{m^2}\!\right)+\ln \frac{\mu ^2}{m^2} \ln \frac{p^-\ell ^+}{\mu^2}+\frac{5\pi ^2}{12}\right] \, .
\end{align}
The UV divergences in this result should be canceled by the counter term of the LO operator ${\cal O}_a$. Therefore we obtain\footnote{Note that in this toy example couplings $e$ and $g$ are quantities of mass dimension one.}
\begin{align}\label{eq:counter-term}
\delta J_{a}=\frac{e g^2G}{16 \pi ^2\left(p^-\ell ^+\right)}\left(\frac{1}{\epsilon ^2}-\frac{1}{\epsilon }\ln \frac{p^-\ell ^+}{\mu ^2}\right) \, , 
\end{align}
for the counter term of $J_a$, the matching coefficient to ${\cal O}_a$. The renormalized EFT contribution is 
\begin{align}
I_{\text{EFT}}^{(\text{ren})}=\frac{-i}{16 \pi ^2\left(p^-\ell ^+\right)}\left( \frac{1}{2} \ln ^2 \!\left(\!\frac{\mu ^2}{m^2}\!\right)+ \ln \frac{\mu ^2}{m^2} \ln \frac{p^-\ell ^+}{\mu ^2}+\frac{5\pi ^2} {12}\right) \, .\end{align}
Subtracting this from the full theory result of eqn. (\ref{eq:full-theory}), we get the one-loop matching coefficient,
\begin{align}
J_{a}=e G+\frac{e g^2G}{16 \pi ^2\left(p^-\ell ^+\right)}\left(\text{  }\frac{1}{2}\ln ^2\left(\frac{\ell ^+p^-}{\mu ^2}\right)-\frac{\pi ^2}{12}\right) \, .
\end{align}
First thing we note is that the divergences in the counter term (\ref{eq:counter-term}) are consistent with the logarithms in the matching coefficient $J_a$, that is anomalous dimension obtained using eqn. (\ref{eq:counter-term}) can be used to resum logarithms in $J_a$. Secondly, each operator corresponding to the left, right and center modes can be run independently in $\nu$ to resum the rapidity logarithms in the infrared sector. Apart from the technicalities of operator mixing, the running strategy works the same as in the case of the Sudakov form factor discussed earlier. Therefore we have shown in principle that exclusive $B$-decays can be factorized and resummed in \scetii, contrary to the previous claims \cite{Becher:2003qh}.

\subsection{The super-boosted frame}
Using the rapidity regulator for the super-boosted frame as shown in Table~\ref{tab:regulator} we need to calculate $I_{\rm center}$ and $I_{\rm left}$ only, as  $I_{\rm right}$ is the same in all frames.
\begin{align}
I_{\rm center} &=\frac{-i e^{\epsilon  \gamma _E}\Gamma (\epsilon )}{16\pi ^2 p^-\ell ^+} \left(\frac{\mu ^2}{m^2}\right)^{\epsilon }\int _0^{\infty }dk^-\frac{\nu ^{\eta }}{\left(k^-\right)^{1+\eta }} = 0 \, , \\
I_{\rm left} & =\frac{e^{\epsilon  \gamma _E}\mu ^{2\epsilon }}{(4\pi )^{\epsilon }} \nu^\eta \int \frac{d^dk}{(2\pi )^d}\frac{|k^-|^{-\eta}}{\left[k^2-\ell ^+ k^-  + i 0\right]\left[k^2-m^2 + i 0 \right]\left[-p^-k^+ + i 0 \right]}  \nn \\
& =\frac{-i }{16\pi ^2 p^-\ell ^+ }\left[\frac{e^{\gamma  \epsilon } \Gamma (\epsilon )}{\eta }\frac{\mu ^{2\epsilon }}{m^{2\epsilon }}-\frac{1}{\epsilon ^2}+\frac{1}{\epsilon }\ln \frac{\nu  \ell ^+}{\mu ^2}+\frac{1}{2} \ln ^2\left(\frac{\mu ^2}{m^2}\right)+\ln \frac{\mu ^2}{m^2}\ln \frac{\nu  \ell ^+}{\mu ^2}+\frac{\pi ^2}{4}\right] \nn \, ,
\end{align}
where $I_{\rm center}$ vanished because the last integral in $k^-$ was scaleless. Note that the left integral now has exactly the same $\eta$-divergence as it was in the sum of $I_{\rm center}$ and $I_{\rm left}$ in the previous case, so are the structure of $\nu$-logarithms. Therefore, the sum of three sectors, $I_{\rm EFT}^{\rm (bare)}$, still yields the same result as in eqn.~(\ref{eq:bare-EFT}) of the symmetric frame. 

\subsection{The lab frame}\label{sec:lab-frame}
Using the rapidity regulator as shown in Table~\ref{tab:regulator} for the lab frame we note that $I_{\rm right}$ is same as in eqn.~(\ref{eq:sym-integrals}) and $I_{\rm center} = 0$ as was in the super-boosted frame. It only remains to calculate $I_{\rm left}$, whose exact evaluation is cumbersome. Therefore, for this frame we only give the $\eta$-divergence structure of the left integral\footnote{Naively, one may expect that this integral had an overlap with the central region. The overlap integral is obtained by taking $k$ to be collinear ($k\mu\sim Q(\lambda^2,1,\lambda)$) in the integrand which exactly reproduces $I_{\rm center}$ of this frame once the regulator is transformed as $|2 k_3|^{-\eta}\to |k^-|^{-\eta}$. Thus, there is no overlap between the regions.},
\begin{align}\label{eq:left-lab}
I_{\text{left}} & =\frac{e^{\epsilon  \gamma _E}\mu ^{2\epsilon }}{(4\pi )^{\epsilon }}\nu ^{\eta }\int \frac{d^dk}{(2\pi )^d}\frac{\left|2k_3\right|{}^{-\eta }}{\left[k^2-\ell ^+ k^- + i0\right]\left[k^2-m^2+ i0\right]\left[-p^-k^+ + i0\right]} \, \\
& = \frac{-i}{16\pi ^2p^-\ell ^+}\left[\frac{ e^{\epsilon  \gamma _E} \Gamma (\epsilon )}{\eta }\left(\frac{\mu ^2}{m^2}\right)^{\epsilon }+ \eta\text{-finite} \right]\, . \nn
\end{align}
Details of this calculation are shown in app.~\ref{app:B-decay}. This is exactly the divergence required to cancel the rapidity divergence in $I_{\rm right}$.


\section{Conclusion}
In this paper we have presented a formalism which allows one to factorize and resum observables
which are sensitive to soft recoil.  These observables fall within the confines of \scetii\, in which
soft and collinear modes have the same invariant mass scalings. It is because of this equality of scalings
that one runs into rapidity divergences which force us to introduce a new regulator with an associated
scale. We presented a proof that while individual sectors have rapidity divergences when one sums
over sectors these divergences cancel as they must, since they are an artifact of factorization. 
The sectors contain soft and collinear function which are gauge invariant, and process independent.
For transverse momentum distribution we are able to define a gauge invariant and universal transverse
momentum dependent parton distribution function.  Once the regulator is implemented one can sum
the rapidity logarithms by use of the rapidity renormalization group, which corresponds to sliding the cut-off
which separates collinear and soft modes on the mass shell hyperbola.

We demonstrated our formalism by showing how one can sum the logarithms in the massive Sudakov form factor, as well
as in the Higgs transverse momentum distribution and jet broadening. In the case of the Higgs distribution
we give a generalized factorization theorem which goes beyond the classic CSS result in that it allows for
jets in the central region. At leading order in the matching at the hard scale our results reduce to those of 
CSS. We also showed how our formalism can be used to renormalize exclusive processes with end point
singularities, which allows one to complete the original calculations of Manohar and Stewart who showed how 
to sensibly handle the end point divergences using \scetii\, and the zero-bin subtraction method.


\section*{Acknowledgments}
Work supported by \uppercase{DOE}
contracts \uppercase{DOE-ER}-40682-143 and
\uppercase{DEAC02-6CH03000}. D. Neill is supported by NSF LHC Theory Initiative grant PHY-0705682 and a DOE graduate fellowship. The authors
acknowledge the hospitality of the INT at the Univ. of Washington where some of this work was performed. We also thank Iain Stewart, Aneesh Manohar and Wouter Waalewijn for discussions and comments on the manuscript.


\appendix


\section{Gauge Invariance and Rapidity Regulators}
\label{A}
In this section, we will prove the gauge invariance  is not spoiled by regulating rapidity divergences concentrating
on the generalized Wilson line regulator ($\eta$). Generalizing the proof to the delta regulator follows
in a simple fashion. We also show that regulating rapidity divergences  with the $\eta$ or delta regulators  is consistent with non-Abelian
exponentiation.

\subsection{Regularization at Higher Orders in the  Sudakov Form Factor}
Given our operator definition, which we will see will have to be slightly amended  when going to higher orders, it is not clear at all that gauge invariance is retained once the regulator is inserted. 
However, a simple argument shows that, at least within any set of covariant gauges, the regulator will preserve gauge invariance. Consider first the case of one gluon emission.  In this case the gauge dependent piece of the propagator will generate an extra factor of $n \cdot k$ which then eliminates any rapidity divergence arising from that term. As such, for  the integral involving the gauge piece we may set $\eta$ to zero and gauge invariance follows.

At higher orders we must  modify the regulator in order to manifestly preserve both gauge invariance and eikonal exponentiation. We begin by considering the renormalization of the soft function which, as opposed to
the jet function, is  a pure Wilson line.  Let us recall some basic facts about Wilson lines and their renormalization: 1) The anomalous dimensions of the cusped Wilson line is at most linear in logarithms \cite{Korchemsky:1987wg,Manohar:2003vb}. 2) The result exponentiates at the level of the integrands \cite{Gatheral:1983cz,Frenkel:1984pz,Laenen:2008gt}, with each color weight appearing only once. At each order there are a set of graphs which are two eikonal line irreducible (2EPI), i.e. they can not be disconnected by cutting two eikonal lines.  These graphs generate a color weight which does not appear in any lower order graphs. The sum of these graphs is called a CWEB. Thus in the exponent only CWEBs appear. These two facts imply that the sum of integrands which form a CWEB have no higher order power beyond $1/\epsilon^2$ for UV or $1/\eta$ for rapidity. Poles of order $1/\eta^2$ would violate condition 1) since all integrals have  UV divergent transverse momentum integrals that are regulated by dim. reg. thus leading to anomalous dimensions which are not
monomials in logarithms.
 \begin{figure}
   \centering
     \includegraphics[width=14.5 cm]{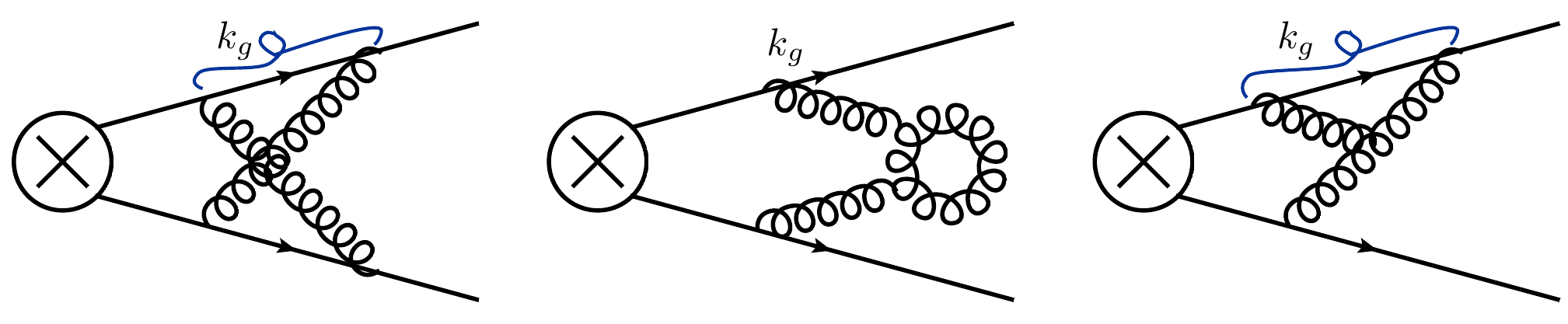}
   \caption{ \label{CWEB}
}
\end{figure}

To prove gauge invariance we should choose our regulator so that we may sum
the diagrams which are part of a CWEB. Doing so allows us to take advantage of the lack of sub-divergences  in this subset of diagrams.
Thus instead of regulating the individual momenta, we will regulate the group momenta $n \cdot k_g$. Where $k_g$ is the total momentum flowing into a CWEB on the  eikonal line . 
So we will re-write our regulated Wilson line as 
\beq
W_n=\sum_{\rm perms}  \exp\left[-\frac{g w}{\nbar \cdot { \cal P}} \frac{\mid \nbar \cdot {\cal P}_g\mid^{-\eta}}{\nu^{-\eta}}\nbar \cdot A_n \right]
\eeq
\beq
S_n=\sum_{\rm perms}  \exp\left[-\frac{gw}{ n \cdot  { \cal P}} \frac{\mid 2 {\cal P}_{3g}\mid^{-\eta/2}}{\nu^{-\eta/2}}n\cdot A_s \right]
\eeq
Note that non 2EPI diagrams will thus be regulated differently then 2EPI diagrams.
Consider the 2EPI diagrams which contribute to the $O(\alpha^2)$ CWEB shown in
Fig. \ref{CWEB}. The  sum of the integrands  contains no sub-divergences.
By regulating only the group momentum  the integrands may still be added.
So the sum of the integrands have the general form
\beq
I_{CWEB}\sim \int d^dk_g d^dkD(k^2,k_g^2, k\cdot k_g) \frac{\mid k_{3g}\mid^{-\eta}}{(\bar n \cdot k_g+i \epsilon )( n \cdot k_g-i\epsilon)}\frac{ N(n \cdot k, \nbar \cdot k , n\cdot k_g, \nbar \cdot k_g)}{D(k^2,k_g^2, k\cdot k_g,n \cdot k, \nbar \cdot k , n\cdot k_g, \nbar \cdot k_g)}\eeq
Here we have explicitly pulled out the first eikonal propagator.
Note that in all contributions to the CWEB the final gluon attaching to the eikonal line will carry the same 
momentum as the eikonal line itself.  
Therefore,  if we now consider a general covariant gauge with polarization sum
\beq
\sum \epsilon_\mu \epsilon_\nu \sim g_{\mu\nu} +(1-\xi) \frac{k_\mu k_\nu}{k^2} \, .
\eeq
The gauge dependent piece will necessarily cancel the final eikonal propagator and
since the CWEB has at most an order $1/\eta$ rapidity divergence the gauge dependent
piece will always be finite. Given that the gauge dependent pieces are finite, we may set $\eta$ to
zero in those contributions, leading to a gauge invariant result.

 \begin{figure}
    \centering
   \includegraphics[width=6cm]{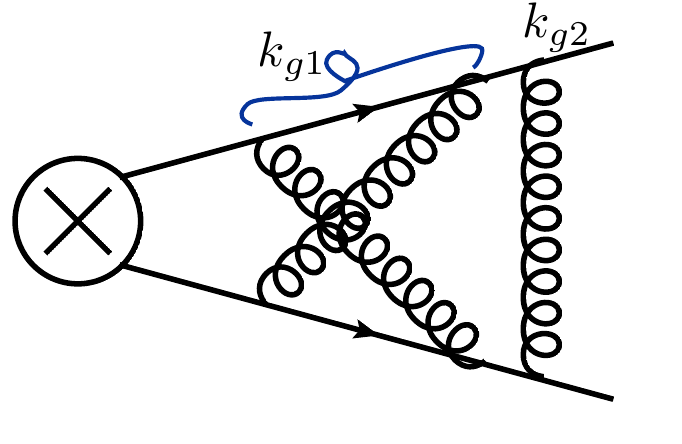}
\caption{ \label{non-2EPI} An example of a diagram composed of two CWEBS. The group momentum for the CWEBs is labelled $k_{go}$.
}
\label{binz}
\end{figure}

Now let us consider the set of diagrams which are not 2EPI such as the one shown in Fig.  \ref{non-2EPI}.
The order $\alpha^2$ CWEB involves  two additional sub-diagrams corresponding
to the vacuum polarization and the Y-graph which are not shown. 
Eikonal exponentiation implies that the sum of this diagrams factorize into a product.
That is, this contribution comes from interference terms in the expansion of the exponent and
will arise with a color factor $C_F^2 C_A$.  The rapidity regulator preserves this
property by regulating the group momentum of the individual CWEB independently.
That is the integrand for this diagrams will be of the form
\beq
\int d^dk_{g1} d^dk_{g2} d^dk \mid k_{3g1}\mid^{-\eta}\mid k_{3g2}\mid^{-\eta}....
\eeq
By defining the regulator in this way we preserve the eikonal identities utilized to show
that the sum of graphs yields the  product of a one and two loop integral.

The renormalization and gauge invariant nature of the jet functions follows by a similar line of
reasoning discussed above.  However, there is a crucial  difference between the jet and soft functions. Since the jet only involves one Wilson line the notion of a 2EPI diagram
is no longer applicable. However, as will be discussed in the next section, it is simple
to define a CWEB for the jet function when noting that every CWEB diagram in the soft sector
has a parent diagram in the full theory which has a collinear limit. There is also the matter of pure
self energy diagrams. Such diagrams on the non-Wilson line part of the jet function, 
obviously have no rapidity divergence since there are no eikonal lines involved, i.e. they are identical
to pure QCD.
All of the eikonal self-energy graphs  vanish after including the proper soft-bin subtraction, since the full theory self energy graphs are reproduced by the purely collinear diagrams. This can be seen from explicit calculation as well.

\subsection{Regularization  of Generalized Soft  and Collinear Functions}\label{sec:non-abelian-regulator}
For generalized observables the soft functions are more complex then the one which arises in  the Sudakov form factor. Typically we are interested in soft matrix elements which arise
from the amplitude squared and where we measure some aspect of the soft radiation that can be written as
\beq
S(p)\equiv \langle 0 \mid S_n
S_{\bar n}^\dagger
 \delta( p-{\cal P}) S_n^\dagger S_{\bar n}  \mid 0 \rangle
\eeq
where $p$ stands for a set of momenta components which scale as some non-zero power of $\lambda$. The soft function will  necessarily be accompanied by at least one collinear function which can be written as
 \beq
J(p) \equiv \langle \psi \mid (\bar O W^\dagger_n) \delta (p-{\cal P}) (W_n O) \mid \psi \rangle.
\eeq
The operator $O$ here is either a quark or a gluon field and $W_n$ here is a collinear Wilson line in either adjoint or fundamental representation depending upon the case. The delta function measures the kinematic quantity of
interest  and $\psi$ is either a vacuum (``Jet Functions'') or
a hadronic state (``Beam functions'', ``TMDPDF's'' or generalizations thereof). 

All of these functions will contain rapidity divergences in \scetii\, upon factorization. It is only the combination
that must be (rapidity) finite.  The finiteness of the total result is predicated on the fact
that the diagrams are regulated in a consistent fashion.  In particular, as touched upon
in the previous section, the regulated EFT diagrams must come as pieces of the asymptotic
expansion of the ``regulated'' full theory diagram. We use quotes here to remind the reader that
the full theory has no rapidity divergences.  It should be clear that
there are many ways in which to insert our rapidity regulator into the full theory
diagram. However, regulating the full theory diagram does not guarantee that the EFT
diagrams will be regulated. One can imagine that the full theory, upon expansion, has
rapidity divergences in sub-diagrams and that a particular choice of regulator
could in general lead to unregulated EFT diagrams.
Furthermore, we should choose a method such that our generalized function are {\it universal}.
That is, they should be process independent, though they will always be scheme dependent just
as with any parton distribution function. 
Finally, we need to ensure that  the regulator preserves eikonal exponentiation.
These criteria are not logically independent so it should not surprise the reader that
its relatively simple to ensure that all are satisfied.  In particular once we ensure 
that we preserve exponentiation, the other criteria are automatically satisfied.

To discern a proper prescription we begin with the soft function where the notion of a CWEB
is clear. As in the case of the Sudakov form factor we regulate the total momentum
emitted in a CWEB.  We will work in the Feynman gauge and the proof of gauge
invariance (covariant gauges) follows by an the identical argument given for the Sudakov form factor.  Note that for the soft function there is no need to ever calculate any diagram which
is not a  CWEB.  At a fixed order a non-CWEB diagram will not contribute to the anomalous dimensions. Furthermore, for the purposes of matching, all non-CWEB diagrams can be
determined by expanding the exponent. i.e. non CWEB diagrams are simply products 
of CWEBS (recall each CWEB is affiliated with a color Casimir). Summing all the CWEB integrands
ensures that the diagrams are marginally divergent. Thus, we regulate the soft function by
inserting a factor of $\mid k_{3g} \mid ^{-\eta}$ where $k_{3g}$ is the total momentum
flowing from the $n$ to $\bar n$ side. The marginal nature of the divergence assures us that
this choice of regularization is sufficient.

We may associate a collinear CWEB  with each soft CWEB. This is true despite the fact
that there is only one eikonal line in the jet function.  While there is no soft contribution to the collinear function, after soft-bin subtraction,   we may consider the soft limit of the diagram for the express purpose of determining whether a diagram is part of a CWEB.
 This is exactly what happens in the soft limit of the parton distribution function \cite{Korchemsky:1992xv}.
Thus given our choice of regulator for the soft function  we regulate the collinear function
by inserting a factor of $\mid k_{g\pm} \mid^{-\eta}$ where $k_g$ is the total momentum flowing
off the Wilson line. Note that the rapidity divergences in the collinear function will
exponentiate (as required by the cancellation of rapidity divergences), but the entire function, including UV divergences, does not.

 \begin{figure}
    \includegraphics[width=13cm]{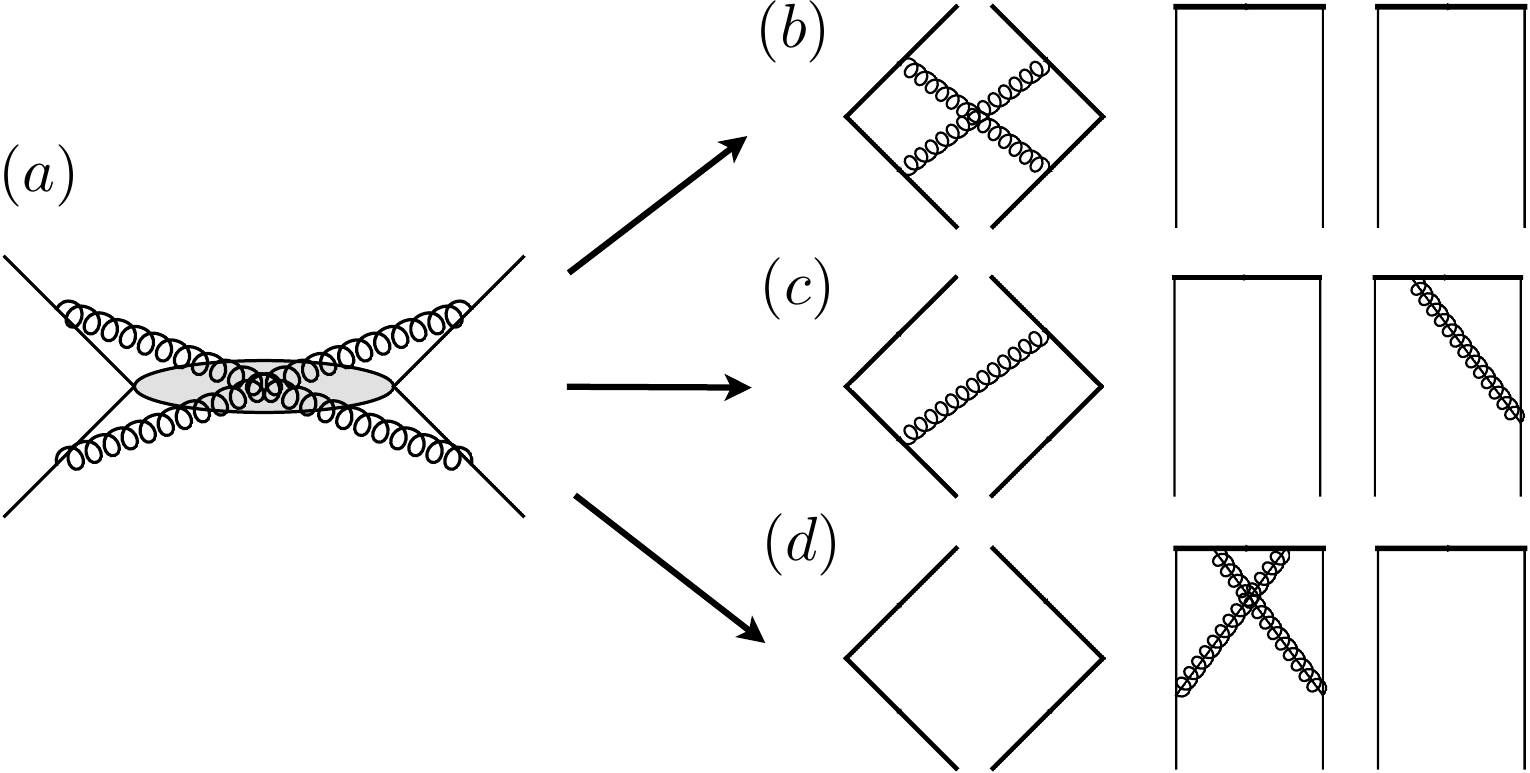}
   \centering
\caption{ \label{mother}The correspondence between the full theory diagram and the diagrams in the effective theory. The first column on the right is the soft function contribution while the next two columns correspond to the collinear functions in the $n$ and $\bar n$ directions respectively.
Not shown are mirror diagrams in which $n \leftrightarrow \bar n$. The blob corresponds to
possible hard pieces or soft non-hadronic (e.g. the Higgs) final states. The darkened lines are eikonalized. The incoming lines in the collinear functions are either quarks or gluons. }
\end{figure}

\begin{figure}
  \includegraphics[width=13cm]{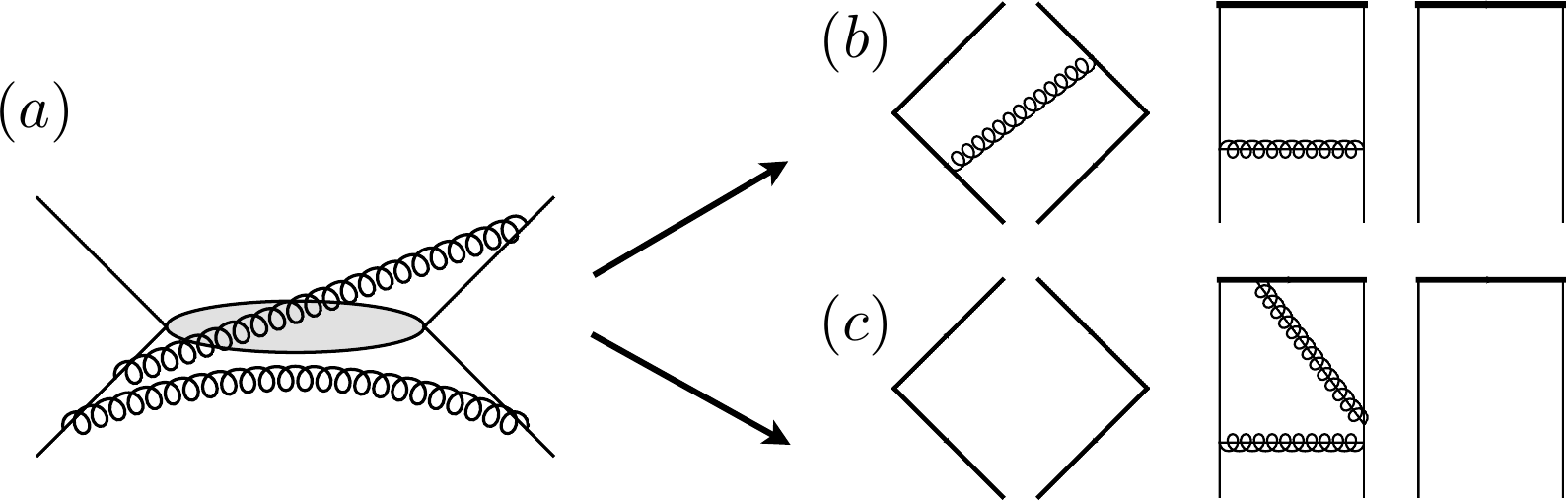}
     \centering
\caption{ \label{MD2}The gluon running straight across the cut is not part of any CWEB and does not
exponentiate.}
\end{figure}

Let us illustrate how this works at two loops.  On the left hand side of Fig. \ref{mother} we 
have a full theory diagram. We will ignore hard gluons since they are handled trivially.
The full theory diagrams can be expanded around three  effective theory
diagrams (plus their mirrors). The possibilities correspond to two softs, one soft one collinear
and two collinears (in each direction). The resulting factorized EFT  diagrams are shown on the right hand side
of the figure. The purely soft diagram corresponds to a contribution to the $O(\alpha_s^2)$ CWEB. Note that at two loops the purely soft (or collinear) diagrams
will not be marginally divergent, only the sum of CWEB diagrams will be. The sum of the diagrams has
no rapidity divergence.
Note that the factorization which occurs in the middle diagram in general will only
 occur once all diagrams have been summed over.
 In Fig. \ref{MD2} we have a case where the collinear function
is not purely eikonal. i.e. the gluon running straight across the cut is not part of any CWEB
and multiplies the exponentiated CWEB.

In the end the regularization is straight forward. In the soft function insert a factor
of $\mid 2k_3 \mid^{-\eta}$ where $ k_3$ is the total momentum in the spatially longitudinal
direction flowing from $n$ to $\nbar$. In the collinear functions (whether they are beam or jet)
insert a factor of $\mid k_\pm \mid^{-\eta}$, where $k$ is the total momentum flow out of the Wilson line. This defines process independent
collinear and soft functions, including TMDPDF's. 
Once any such function is calculated
it can be used in any process as long as all the contributing function are calculated in the
same scheme.  Note that not all schemes will share this property of universality.
For instance, suppose we chose to regulate the momentum flowing across the cut, in which
case purely virtual diagrams would be set to zero. This scheme is attractive for
its relative simplicity of definition. However, while this would give a sensible
beam function or any such generalized PDF, it would not regulate any collinear
function which corresponds to a vacuum matrix element since the total momentum
across the cut will sum to a constant.

Finally we must consider the case of self energy diagrams. As we argued at the end of the
section on the Sudakov form factor, the self energies must be purely collinear since these
graphs are identical to the full theory graphs. All eikonal self energies must therefore vanish
after soft bin subtraction. 

\subsection{The Use of Other Regulators}
As was briefly mentioned previously, it is certainly possible to use other regulators in this
formalism.  We emphasize that the rapidity renormalization and resummation itself is regulator independent formalism,
just as the renormalization and resummation of traditional ultra-violet divergences are independent 
of dimensional regularization, but are often most conveniently performed in dimensional regularization. To
be able to renormalize and resum rapidity divergences is also in the same sense a regulator independent procedure, though to implement in any given instance \emph{some} regulator must be used. Given that,
there are certain properties a regulator should fulfill:
\begin{itemize}
\item Gauge Invariance
\item Preserve Non-Abelian Exponentiation
\item Have a universal definition for generalized soft and jet functions
\item Clearly delineates sectors
\end{itemize}

As an example of a regulator that can be engineered to satisfy all four conditions, we  consider the $\delta$-regulator of \cite{Chiu:2009yx}. At first, the $\delta$-regulator does not manifestly preserve eikonal exponentiation, as can be seen even at the Abelian level where the famous eikonal identity
\beq
\label{eikonal_identity_with_delta}\sum_i \sum_{perms} \frac{1}{n \cdot k_1+\delta}\frac{1}{n \cdot(k_1+k_2)+\delta}... \frac{1}{n \cdot (k_1+.....k_i)+\delta}= \prod_{a=1}^i \frac{1}{n\cdot k_a+\delta}+O(\delta)
\eeq
is no longer obeyed due to the $O(\delta)$ terms. Nonetheless by adding all the elements of a CWEB only the $\delta$ divergences of a single log will appear, and these $O(\delta)$ terms can be safely set to zero. That is to say, only the finite and log divergent pieces survive the $\delta\rightarrow 0$ limit: in a CWEB, higher order terms are never needed. Further, the arguments outlined above illustrating the rapidity integration finiteness of gauge dependent terms work as well with the $\delta$-regulator: the regulator can safely be set to zero in such terms.

The subtlety of the $\delta$ regulator is in the jet sectors. Before one can renormalize the $\delta$ divergence in the jet sector, one must perform a ``soft''-bin subtraction. This subtraction does not remove the rapidity divergence, but rather forces that sector to have the correct rapidity divergence \cite{Chiu:2009yx}. This shifting of \emph{rapidity} divergences in \scetii\, is  analogous to zero-bin subtractions in \sceti. The zero-bin in \sceti\, enforces each sector to have the correct \emph{ultra-violet} divergences, by removing the ultrasoft divergences which should not be attributed to that sector. Any rapidity divergence completely disappears (sector by sector)   in \sceti\, (with the inclusion of the zero-bin) but ultra-violet divergences remain. This accentuates the fundamental distinction between the two theories. Whereas \sceti\, has mode factorization in the invariant mass, \scetii\, has mode factorization in rapidity. The evidence of such factorization in a perturbative calculation is the divergences arising in the integrals of each sector, whose renormalization allows for the resummation of large logarithms.

After the soft-bin subtraction of the jet sector has been performed, one can renormalize  both the rapidity and ultra-violet divergences. In the case of the $\delta$-regulator, this will trade the regulator $\delta$ for an arbitrary parameter analogous to our scale $\nu$. The regulator itself should be formally removed, and and arbitrary scale corresponding to the renormalization point introduced.  At this point one would see that the inclusion of the terms with a positive power of $\delta$ in \eqref{eikonal_identity_with_delta} was unnecessary.
\section{Soft-Bin Subtractions}\label{soft_bin_subtractions}
In SCET the zero bin subtraction plays an important role even when it formally vanishes. In \sceti\,  collinear modes have invariant masses which are parametrically larger then the ultra-soft modes. Thus when we calculate a collinear loop we should not expect an ultra-soft divergence, only a collinear one.  Invariably when calculating loops of the collinear modes one does generate Ultra-soft (non-collinear) divergences which should be attributed to the ultra-soft sector. By doing zero-bin subtraction  (by zero-bin subtraction, we mean  the subtraction of the ultra-soft contribution to an integral, where the loop momenta is taken to scale as $Q(\lambda^2,\lambda^2,\lambda^2)$ in light-cone coordinates) this IR divergence is morphed into a UV divergences. This is sometimes called the ``pull-up" mechanism
\cite{Hoang:2001rr,Manohar:2006nz}.
 
 In \scetii\, since the soft and collinear modes are on the same mass shell hyperbola we do a soft-bin subtraction.  In doing a  ``soft-bin''  subtraction we subtract the soft region, where the loop momenta is taken to scale as $Q(\lambda,\lambda,\lambda)$. The soft-bin of the collinear mode serves the purpose of cutting off the integral at the proper spot on the rapidity hyperbola. This is most clearly seen when one regulates with a $\delta$ regulator  as was shown in \cite{Chiu:2009yx}. In this work the authors introduce a different regulator for each  mode.
The particle $i$ propagator get replaced with
 \beq
 \frac{1}{(p_i+k)^2-m_i^2}\rightarrow \frac{1}{(p_i+k)^2-m_i^2+\Delta_i}.
 \eeq
 For the two collinear modes, we will call the momenta $p_\pm$. $\Delta$ is designed to distinguish (cut-off) the $+$ momenta from the soft region. However, the emission of a $k_+$ gluon of the $p_-$ lines leads to an eikonal propagator $P_-$ of the form
 \beq
 P_-=\frac{1}{k_++\frac{\Delta}{p_-}}.
 \eeq
These emission build up a Wilson line in the $+$ sector, with an apparent violation of factorization, since the collinear $+$ sector now depends upon the wrong cut-off (i.e. the $\pm$ sector is sensitive to $\Delta/p_-$). This cut-off must be pulled-up to the correct sector by the soft-bins and this is exactly what the soft-bin accomplishes \cite{Chiu:2009yx} \footnote{Note that there is no collinear subtraction from the soft  because the soft diagrams are insensitive to the large scales and there can be no overlap. This is not true if there are external particles with soft momenta  as in the B meson decays discussed in \cite{Manohar:2006nz} and  in section (7).}.
  
  When using an $\eta$ regulator the soft-bin is scaleless and is thus vanishing. Its formal subtraction from the collinear integrals has the same physical effect as in the case of the delta regulator.  To see this more clearly we may work with both the $\eta$ and $\delta$  regulators where the limit  $\delta \rightarrow 0$  is take first. In this case the soft-bin subtraction acts to eliminate the ``wrong" delta cut-off and replace it with the RRG scale $\nu$.

Finally, it is important to emphasize that the collinear matrix elements and Lagrangian are well-defined only with the implicit soft or zero-bin subtractions \cite{Manohar:2006nz,Chiu:2009yx}. In particular, the subtractions are necessary to make the matrix element gauge independent \cite{Chiu:2009yx} in each sector. This role of the soft-bin is  unrelated to the issue of rapidity divergences, as we have proven the rapidity divergences to be gauge independent. These soft-bin subtractions can often be related to a matrix element of soft Wilson lines, and often these matrix elements of Wilson lines are the inverse of the soft-function found in factorization. In this guise they appear in the modern CSS formalism \cite{Collins:2011ca}, where the inverse soft factors play a similar role to ensure gauge independence. However, as is the case in section \ref{JB}, the inverse of the soft-function cannot be always identified with the soft-bin subtraction (even if the subtraction is representable as some matrix element of Wilson lines), due to the differing kinematics and phase spaces of the soft function and jet functions.

\section{Higgs $b$ calculation}
\subsubsection{$b$-space TMDPDF}
For completeness we note the impact parameter space results, where the transform is strictly two-dimensional. At tree level, we have:
\begin{align}
\tilde{f}^{(0)\alpha\beta}_{\perp g/g}(z,\vec{b})=\frac{\delta(1-z)}{(2\pi)^4}(2\pi)^2\frac{g_{\perp}^{\alpha\beta}}{2}
\end{align}
At one loop, we get:
\begin{multline}
\tilde{f}^{(1)\alpha\beta}_{\perp g/g}(z,\vec{b})= \frac{1}{(2\pi)^2}\frac{g^2 C_A e^{-\epsilon\gamma_E}}{8\pi^2}\Gamma(1-\epsilon)\frac{1}{\epsilon}\left(\frac{\mu be^{\gamma_E}}{2}\right)^{2\epsilon} \left[ -2 \frac{\delta(1-z)}{\eta}\Big(\frac{\nu}{\omega}\big)^\eta g_{\perp}^{\alpha\beta}+\frac{p_{gg*}(z)}{z}g_\perp^{\alpha\beta}\right.
\\ \left.-2\epsilon^2\frac{(1-z)}{z^2}g_{\perp}^{\alpha\beta}-4\epsilon(1-\epsilon)\frac{(1-z)}{z^2}\Big(\frac{b^\alpha b^\beta}{b^2}+\frac{1}{2}g_{\perp}^{\alpha\beta}\Big)\Big)\right]
\end{multline}
Ignoring the traceless term and expanding in $\eta$ and $\epsilon$ give:
\begin{align}
\nonumber \tilde{f}^{(1div)\alpha\beta}_{\perp g/g}(z,\vec{b})&= -\frac{1}{(2\pi)^2}\frac{g^2 C_A e^{-\epsilon\gamma_E}}{4\pi^2}\Gamma(1-\epsilon)\frac{1}{\epsilon}\left(\frac{\mu be^{\gamma_E}}{2}\right)^{2\epsilon} \frac{\delta(1-z)}{\eta} g_\perp^{\alpha\beta}\\
&\,\,\,\,\,\,\,\,\,+\frac{g^2 C_A}{(2\pi)^4}\frac{1}{2\epsilon}g_{\perp}^{\alpha\beta}\Bigg(\frac{p_{gg*}(z)}{z}-\delta(1-z)\ln\Big(\frac{\nu^2}{\omega^2}\Big)\Bigg)\\
\tilde{f}^{(1 \rm fin)\alpha\beta}_{\perp g/g}(z,\vec{b})&=-\frac{1}{2\pi}\alpha_s C_A g_{\perp}^{\alpha\beta}\ln\Big(\frac{\mu^2b^2e^{2\gamma_E}}{4}\Big)\Bigg(\frac{p_{gg*}(z)}{z}-\delta(1-z)\ln\Big(\frac{\nu^2}{\omega^2}\Big)\Bigg)
\end{align}
We can find the appropriate renormalization factor through the matching procedure to the PDF (see \eqref{One_Loop_Matching_Relation_TMDPDF}):
\begin{align}
Z_{f_\perp}^{(1)}(\nu,\vec{b},\mu)&=1+\frac{\alpha C_A e^{-\epsilon\gamma_E}}{\pi}\Gamma(1-\epsilon)\frac{1}{\epsilon}\left(\frac{\mu be^{\gamma_E}}{2}\right)^{2\epsilon} \frac{1}{\eta}- \frac{\alpha_s C_A}{2\pi\e}\Big( \ln \frac{\nu^2}{\omega^2}+\frac{1}{4 C_A} \beta_0\Big)
\end{align}
The evolution equations satisfied are 
\begin{align}
\mu\frac{d}{d\mu}\tilde{f}^{\alpha\beta}_{\perp g/P}(z,\vec{b},\mu,\nu)&=\gamma^{f_{\perp}}_{\mu}(\nu)\tilde{f}^{\alpha\beta}_{\perp g/P}(z,\vec{b},\mu,\nu)\\
\nu\frac{d}{d\nu}\tilde{f}^{\alpha\beta}_{\perp g/P}(z,\vec{b},\mu,\nu)&=\tilde{\gamma}^{f_{\perp}}_{\nu}(b\mu)\tilde{f}^{\alpha\beta}_{\perp g/P}(z,\vec{b},\mu,\nu)
\end{align}
At one loop we have
\begin{align}
\tilde{\gamma}^{f_{\perp}}_{\mu}(\nu)&=\frac{\alpha_s C_A}{\pi}\left[\ln\Big(\frac{\nu^2}{\omega^2}\Big)+\frac{\beta_0}{4 C_A}\right]\\
\tilde{\gamma}^{f_{\perp}}_{\nu}(b\mu)&=\frac{\alpha_s C_A}{\pi}\ln\Big(\frac{\mu^2b^2e^{2\gamma_E}}{4}\Big)
\end{align}

\subsubsection{$b$-space Soft Function}
In impact parameter space we have for the tree-level result and one-loop correction:
\begin{align}
\tilde{S}^{(0)}(\vec{b})&=\frac{1}{(2\pi)^2}\\
\tilde{S}^{(1)}(\vec{b})&=\frac{1}{(2\pi)^2}\frac{2^{-2\epsilon-\eta}C_A g^2}{4\pi^{2}}e^{\epsilon\gamma_E}\nu^\eta\mu^{2\epsilon}b^{2\epsilon+\eta}\frac{\Gamma(-\epsilon-\frac{\eta}{2})}{\Gamma \left(1+\frac \eta 2 \right)}\frac{2\Gamma(\frac{\eta}{2})\Gamma(-\eta )}{\Gamma\left(-\frac{\eta }{2}\right)}
\end{align}
Separating out the divergent and finite parts gives:
\begin{align}
\tilde{S}^{(1)div}(\vec{b})&=\frac{1}{(2\pi)^2}\frac{g^2C_A}{2\pi^2}\frac{1}{\eta}\Gamma(-\epsilon)e^{-\epsilon\gamma_E}\Big(\frac{b\mu e^{\gamma_E}}{2}\Big)^{2\epsilon}+\frac{g^2C_A}{4\pi^2\epsilon^2}+\frac{g^2C_A}{4\pi^2\epsilon}\ln\Big(\frac{\mu^2}{\nu^2}\Big)\\
Z^{(1)}_{S}(\vec{b})&=1-\frac{\alpha_s C_A}{(2\pi)^4\pi}\left[\frac{2}{\eta}\Gamma(-\epsilon)e^{-\epsilon\gamma_E}\Big(\frac{b\mu e^{\gamma_E}}{2}\Big)^{2\epsilon}+\frac{1}{\epsilon^2}+\frac{1}{\epsilon}\ln\Big(\frac{\mu^2}{\nu^2}\Big)\right]\\
\tilde{S}^{{\rm R}(1)}(\vec{b})&=-\frac{1}{(2\pi)^3}\alpha_s C_A\Bigg(-2\ln\Big(\frac{b^2\mu^2e^{2\gamma_E}}{4}\Big)\ln\Big(\frac{b^2\nu^2e^{2\gamma_E}}{4}\Big)-\ln^2\Big(\frac{b^2\mu^2e^{2\gamma_E}}{4}\Big)-\frac{\pi^2}{6}\Bigg)
\end{align}

The evolution equations satisfied are 
\begin{align}
\mu\frac{d}{d\mu}\tilde{S}(\vec{b},\mu,\nu)&=\gamma_{S}(\mu,\nu)\tilde{S}(\vec{b},\mu,\nu)\\
\label{soft_evo_eqn_higgs_nu}\nu\frac{d}{d\nu}\tilde{S}(\vec{b},\mu,\nu)&=\tilde{\gamma}_{Srap}(b\mu)\tilde{S}(\vec{b},\mu,\nu)
\end{align}
At one loop we have
\begin{align}
\tilde{\gamma}^{S}_{\mu}(\mu,\nu)&=\frac{2\alpha_s C_A}{\pi}\ln\Big(\frac{\mu^2}{\nu^2}\Big)\\
\tilde{\gamma}^{S}_{\nu}(b\mu)&=-\frac{2\alpha_s C_A}{\pi}\ln\Big(\frac{\mu^2b^2e^{2\gamma_E}}{4}\Big)
\end{align}
\section{Jet Broadening Resummation in Fourier-Laplace Space}
For completeness, we will also perform RRG in Fourier-Laplace space, as in comparison to some older literatures. The factorization theorem in b-$\tau$ space can be written as
\begin{align}
\frac{1}{\sigma_0}\frac{\dd^2\sigma}{\dd \tau_R \dd \tau_L} 
& =  \, H(Q^2,\mu) \int d b_{1} \, d b_{2} 
 {J}_{n}(Q,\tau_R,b_{1}) \,  {J}_{\bar n}(Q,\tau_{L}, b_{2}) \, {S}(\tau_R,\tau_L,b_1,b_2),
\end{align}
where
\beq\label{Jbt}
J_n(Q,\tau_R,b_1) \equiv \Omega_{2-2\epsilon} b_1^{1-2\epsilon}\int_{0}^{\infty}de_n e^{-e_n\tau_R} \int\frac{d^{2-2\epsilon}p_1}{(2\pi)^{2-2\epsilon}}e^{-ip_{1\perp} \cdot b_1}{\cal J}_{n}(e_n,Q, \vec{p}_{1\perp}).
\eeq

The soft function in $b-\tau$ space is given by
\begin{align} \label{Sbt}
{S}(\tau_R,\tau_L,b_{1},b_{2})=&g^2 w^2 C_F\mu^{2\epsilon}\nu^{\eta}\int de_s^R\, e^{-e_s^R \tau_R}\int de_s^L\, e^{-e_s^L \tau_L}  \int  \frac{d^dk}{(2\pi)^d}\delta^{(+)}(k^2)\frac{|\overline{n}.k-n.k|^{-\eta}}{\overline{n}.k \; n.k}\nn \\
 & \times  \left[ \theta(k^3)\delta\left(e_s^R-\frac{|k_t|}{Q}\right)\delta(e_s^L)e^{i\vec{b}_{1}\cdot\vec{k}_{t} }+\theta(-k^3) \delta\left(e_s^R\right)\delta\left(e_s^L-\frac{|k_t|}{Q}\right)e^{i\vec{b}_{2}\cdot \vec{k}_{t} }\right]  \nn \\
S(\tau_R,\tau_L,b_{1},b_{2})=&\frac{g^2 w^2C_F\Omega_{2-2\epsilon}}{4(2\pi)^{3-2\epsilon}}\Gamma(-\eta-2\epsilon)\frac{2^{1-\eta}\Gamma(\frac12-\frac{\eta}{2}) \Gamma\left(\frac{\eta}{2}\right)}{\sqrt{\pi}}\\
&\times \left[\left(\frac{\nu\tau_R}{Q}\right)^{\eta}\left(\frac{\mu^2\tau_R^2}{Q^2}\right)^{\epsilon}{}_2F_1\left(\frac{-\eta-2\epsilon}{2},\frac{1-\eta-2\epsilon}{2};1-\epsilon;-\frac{b_{1}^2Q^2}{\tau_R^2}\right) \right. \nn \\
& \left.\,\,\;\;+\left(\frac{\nu\tau_L}{Q}\right)^{\eta}\left(\frac{\mu^2\tau_L^2}{Q^2}\right)^{\epsilon}{}_2F_1\left(\frac{-\eta-2\epsilon}{2},\frac{1-\eta-2\epsilon }{2};1-\epsilon;-\frac{b_{2}^2Q^2}{\tau_L^2}\right)\right]\nn.
\end{align}
The renormalized soft function to NLO can be obtained by taking the finite part of Eq.\ref{Sbt}
\begin{align}
\nonumber S^{\rm{R}}(\tau_R,\tau_L,b_{1},b_{2})=\frac{g^2C_F}{4\pi^2}&\left[-\ln^2\left(\frac{e^{\gamma_E}\mu (\tau_R+\sqrt{b_{1}^2Q^2+\tau_R^2})}{2Q}\right)+\ln\frac{\mu^2}{\nu^2}\ln\left(\frac{e^{\gamma_E}\mu (\tau_R+\sqrt{b_{1}^2Q^2+\tau^2})}{2Q}\right) \right.\\
&\left. +\text{Li}_2\Big(\frac{\tau_R-\sqrt{b_{1}^2Q^2+\tau_R^2}}{\tau_R+\sqrt{b_{1}^2Q^2+\tau_R^2}}\Big)-\frac{5\pi^2}{12}\right]+(b_{1}\leftrightarrow b_{2},\tau_R \leftrightarrow \tau_L)
\end{align}

\subsection{Renormalization Group Equations}
From the divergent part of theEq.\ref{Sbt}, we can get the renormalization factor
\begin{align}\label{Zs}
Z_S(\tau_R,\tau_L,b_{1},b_{2})&=1+\frac{g^2w^2C_F}{8\pi^2\epsilon^2}+\frac{g^2w^2C_F}{8\pi^2\epsilon}\ln\Big(\frac{\mu^2}{\nu^2}\Big)\\
&-\frac{g^2 w^2C_F}{4\pi^2}e^{-\epsilon\gamma_E}\frac{\Gamma(-2\epsilon)}{\Gamma(1-\epsilon)}\frac{1}{\eta}\left[\Big(\frac{\mu\tau_R e^{\gamma_E}}{Q}\Big)^{2\epsilon}{}_2F_1\left(-2\epsilon,\frac{1-2\epsilon}{2};1-\epsilon;-\frac{b_{1}^2Q^2}{\tau_R^2}\right)\right]\nn\\
&-\frac{g^2w^2C_F}{4\pi^2}e^{-\epsilon\gamma_E}\frac{\Gamma(-2\epsilon)}{\Gamma(1-\epsilon)}\frac{1}{\eta}\left[\Big(\frac{\mu\tau_L e^{\gamma_E}}{Q}\Big)^{2\epsilon}{}_2F_1\left(-2\epsilon,\frac{1-2\epsilon}{2};1-\epsilon;-\frac{b_{2}^2Q^2}{\tau_L^2}\right)\right],\nn
\end{align}
The $\mu$-anomalous dimensions follow in standard fashion, and  the $\nu$-anomalous dimensions can be obtained by using  Eqs.\ref{con} and \ref{nu-beta}
\begin{align}
\gamma_{S}^{\mu}(\mu,\nu)&=\frac{\alpha_sC_F}{\pi}\ln\frac{\mu^2}{\nu^2}\\
\gamma_{S}^{\nu}(\tau_R,\tau_L,,b_{1},b_{2},\mu)&=-\frac{2\alpha_sC_F}{\pi}\left[\ln\left(\frac{\mu\, e^{\gamma_E} (\tau_R+\sqrt{b_{1}^2Q^2+\tau_R^2})}{2Q}\right)+\ln\left(\frac{\mu\, e^{\gamma_E}  (\tau_L+\sqrt{b_{2}^2Q^2+\tau_L^2})}{2Q}\right)\right]
\end{align}
To eliminate the large logarithms we may run in both $\mu$ and $\nu$ to some fixed scale. That is we may write
\bea
S(\mu,\nu)&=&  U_S(\mu,\mu_S;\nu_S) V_S(\nu,\nu_S;\mu) S(\mu_S,\nu_S)
\eea
where
\begin{align}
V_S(\nu,\nu_s;\mu)=&\rm{Exp}\Big[-\frac{2\alpha_sC_F}{\pi}\ln\Big(\frac{\nu}{\nu_S}\Big)\ln\Big(\frac{\mu\, e^{\gamma_E}}{2Q}(\tau_R+\sqrt{b_{1}^2Q^2+\tau_R^2})\Big)\Big]\nn \\
&\times\rm{Exp}\Big[-\frac{2\alpha_sC_F}{\pi}\ln\Big(\frac{\nu}{\nu_S}\Big)\ln\Big(\frac{\mu \, e^{\gamma_E}}{2Q}(\tau_L+\sqrt{b_{2}^2Q^2+\tau_L^2})\Big)\Big]\nn \\
=&\left(\frac{\mu \, e^{\gamma_E}(\tau_R+\sqrt{b_{1}^2Q^2+\tau_R^2})}{2Q}\right)^{-\omega_s}
\left(\frac{\mu\, e^{\gamma_E }(\tau_L+\sqrt{b_{2}^2Q^2+\tau_L^2})}{2Q}\right)^{-\omega_s}  \\
U_S(\mu,\mu_S,\nu)=&\rm{Exp}\left\{ -2 \left[ \frac{\Gamma_{0}}{2\beta_{0}}\Bigg(\frac{4\pi}{\alpha_{s}(\nu)}\Big(\text{ln}\Big[\frac{\alpha_s(\mu)}{\alpha_s(\mu_S)}\Big]-\frac{\alpha_s(\mu_S)}{\alpha_s(\mu)}-1\Big)\right.\right. \\
&\left.\left.\,\,\,\,+\Big(\frac{\Gamma_1}{\Gamma_0}-\frac{\beta_1}{\beta_0}\Big)\Big(\frac{\alpha_s(\mu)}{\alpha_s(\mu_S)}-\ln\Big[\frac{\alpha_s(\mu)}{\alpha_s(\mu_S)}\Big]-1\Big)-\frac{\beta_1}{2\beta_0}\ln^2\Big[\frac{\alpha_s(\mu)}{\alpha_s(\mu_S)}\Big]\Bigg)
 \right]\right\}
\end{align}
and 
\beq
{\omega_s} (\nu, \nu_S)= \frac{2\alpha_sC_F}{\pi}\ln\Big(\frac{\nu}{\nu_S}\Big).
\eeq
The hard function, with no running in $nu$, has the standard evolution, 
\begin{align}
H(Q,\mu)=U(\mu&,\mu_H)H(Q,\mu_H),\\
U_H(\mu,\mu_H)=\rm{Exp}&\left\{ 2 \left[ \frac{\Gamma_{0}}{2\beta_{0}}\Bigg(\frac{4\pi}{\alpha_{s}(Q)}\Big(\text{ln}\Big[\frac{\alpha_s(\mu)}{\alpha_s(\mu_H)}\Big]-\frac{\alpha_s(\mu_H)}{\alpha_s(\mu)}-1\Big)\right.\right. \\
&\left.\left.\,\,\,\,+\Big(\frac{\Gamma_1}{\Gamma_0}-\frac{\beta_1}{\beta_0}\Big)\Big(\frac{\alpha_s(\mu)}{\alpha_s(\mu_H)}-\ln\Big[\frac{\alpha_s(\mu)}{\alpha_s(\mu_H)}\Big]-1\Big)-\frac{\beta_1}{2\beta_0}\ln^2\Big[\frac{\alpha_s(\mu)}{\alpha_s(\mu_H)}\Big]\Bigg)
 \right]\right\} \nn.
\end{align}

 \subsection{Resummation at NLL}
Resummed broadening distribution to NLL in the Laplace space including both tradition RG and RRG depends on tree-level hard, jet, and soft function, plus two-loop cusp and one-loop non-cusp hard anomalous dimension, together with one-loop soft $\nu$-anomalous dimension. The hard function, with no rapidity divergence and therefore no $\nu$ running, are as standard hard function.

The resummed left and right broadening spectrum to NLL can be written accordingly as
\begin{align}\label{JB-btau-NLL}
\frac{1}{\sigma_0}\frac{{\dd}^2 \sigma^{\rm NLL}}{\dd \tau_R\dd \tau_L}=&U_H(\mu,\mu_H)H^{(0)}(Q,\mu_H)\\
&\times \int_0^{\infty}db_1\int_0^{\infty}db_2
J^{(0)}_n(\tau_R,b_1)J^{(0)}_{\bar n}(\tau_L,b_2) V_S(\nu,\nu_S,\mu)S^{(0)}(\tau_R,\tau_L,b_1,b_2) \nn \\
\end{align}
in which the jet function in $b-\tau$ space as defined in Eq.\ref{Jbt} calculated at  tree level as 
\begin{align}
{J}^{(0)}_n(\tau_R,b_1)
&=\frac{Q^{2} \tau_R \,b_1}{(\tau_R^2+b_1^2Q^2)^{3/2}}.
\end{align}
We can then write Eq.\ref{JB-btau-NLL} as 
\begin{align}
\frac{1}{\sigma_0}\frac{{\dd}^2 \sigma}{\dd \tau_R\dd \tau_L}
=&U_H(\mu,\mu_H)H^{(0)}(Q,\mu_H)\int_0^{\infty}db_1\int_0^{\infty}db_2\frac{Q^2\tau_R b_1}{(\tau_R^2+b_1^2Q^2)^{3/2}} \frac{Q^2\tau_L b_2}{(\tau_L^2+b_2^2Q^2)^{3/2}}\nn \\ 
&\times \left(\frac{\mu \, e^{\gamma_E}(\tau_R+\sqrt{b_{1}^2Q^2+\tau_R^2})}{2Q}\right)^{-\omega_s}
\left(\frac{\mu\, e^{\gamma_E }(\tau_L+\sqrt{b_{2}^2Q^2+\tau_L^2})}{2Q}\right)^{-\omega_s} \\
=& H(Q,\mu)  \left(2 \frac{{}_2F_1(1,2,2+\omega_s,-1)}{1+\omega_s}  \right)^2 \left( \frac{\mu \tau_R}{Q}\right)^{-\omega_S}\left( \frac{\mu \tau_L}{Q}\right)^{-\omega_S}
\end{align}
The left and right broadening distribution can then be calculated by performing inverse Laplace transformation
\begin{align}
\frac1{\sigma_0}\frac{\dd^2\sigma}{\dd e_{R} \dd e_L} =&\int_{-i\infty+\gamma}^{i\infty+\gamma}\frac{d\tau_R}{2\pi i} e^{e_R\tau_R}\int_{-i\infty+\gamma}^{i\infty+\gamma}\frac{d\tau_L}{2\pi i} e^{e_L\tau_L}\frac{1}{\sigma_0}\frac{{\dd}^2 \sigma}{\dd \tau_R\dd \tau_L}
 \\
=&H(Q,\mu)\frac{4 e^{-2\gamma_E{\omega_s}}}{\Gamma^2({\omega_s})}\frac{Q^2}{\mu^2}\Big[\Big(\frac{\mu}{e_R Q}\Big)^{1-{\omega_s}}\Big]_+  \Big[\Big(\frac{\mu}{e_L Q }\Big)^{1-{\omega_s}}\Big]_+\left(\frac{{}_2F_1(1,2,2+\omega_s,-1)}{1+\omega_s} \right)^2. \nn
\end{align}
For total jet broadening, we have
\begin{align}
\frac1{\sigma_0}\frac{\dd\sigma}{\dd e} =&\int_{-i\infty+\gamma}^{i\infty+\gamma}\frac{d\tau_R}{2\pi i} e^{e_R\tau_R}\int_{-i\infty+\gamma}^{i\infty+\gamma}\frac{d\tau_L}{2\pi i} e^{e_L\tau_L} \delta(e-e_L-e_R) \frac{1}{\sigma_0}\frac{{\dd}^2 \sigma}{\dd \tau_R\dd \tau_L}\nn \\
=& \int_{-i\infty+\gamma}^{i\infty+\gamma}\frac{d\tau}{2\pi i} e^{e\tau}H(Q,\mu)  \left(2 \frac{{}_2F_1(1,2,2+\omega_s,-1)}{1+\omega_s}  \right)^2\left( \frac{\mu \tau}{Q}\right)^{-2\omega_S} \nn \\
=&H(Q,\mu)\frac{4 e^{-2\gamma_E{\omega_s}}}{\Gamma(2{\omega_s})}\left(\frac{{}_2F_1(1,2,2+\omega_s,-1)}{1+\omega_s} \right)^2\frac{1}{ e}\Big(\frac{\mu}{ e Q}\Big)^{-2{\omega_s}}.
\end{align}
To compare with standard the Total Jet Broadening ($B_T$) definition\cite{Catani:1992jc}, where $B_T=\frac12 e$, we write the broadening distribution as
\begin{align}\label{eq:JB}
\frac{1}{\sigma_0}\frac{\dd \sigma}{{\dd }B_T}
&=H(Q,\mu)\frac{e^{-2\gamma_E{\omega_s}}}{\Gamma(2{\omega_s})}\frac{1}{ B_T}\Big(\frac{\mu}{ B_T Q}\Big)^{-2{\omega_s}}\frac4{2^{-2{\omega_s}}}\left(\frac{{}_2F_1(1,2,2+\omega,-1)}{1+\omega_s} \right)^2
\end{align}
equivalent to that is derived in the physical ${\rm p_T}$-$e$ space using the relationship between Hypergeometric and incomplete Beta function.

\section{Transforms}
We collect here some transforms needed for resummations or deriving plus distribution properties for both jet broadening and higgs.
\subsubsection{Fourier Transforms For Higgs spectrum}
\begin{align}
\label{Higgs_FT_relation}
\int d\Omega_{d}e^{ib.k}&=\Gamma\left(\frac{d}{2}\right)2^{\frac{d}{2}}\Omega_{d}(bk)^{-\frac{d-2}{2}}J_{\frac{d-2}{2}}(bk)\\
\int\frac{d^2\vec{p}_{\perp}}{(2\pi)^2}e^{i\vec{b}_t.\vec{p}_{\perp}}\frac{1}{\mu^2}\Big(\frac{\mu^2}{p_{\perp}^2}\Big)^{1+\alpha}&=-\frac{e^{-2\alpha\gamma_E}}{4\pi\alpha}\frac{\Gamma(1-\alpha)}{\Gamma(1+\alpha)}\Big(\frac{b^2\mu^2e^{2\gamma_E}}{4}\Big)^{\alpha}
\end{align}
Expanding both sides in $\alpha$ and identifying powers of $\alpha$ gives the transforms of the plus distributions $L_{n}(\mu,p_{\perp})$.
\subsubsection{Transforms For Jet Broadening}
Transforms:
\begin{align}
\int_{0}^{\infty}dx\,x^\alpha J_{\beta}(x b)e^{-x\tau}&=2^{-\beta}b^{\beta}\tau^{-\alpha-\beta-1}\frac{\Gamma(1+\alpha+\beta)}{\Gamma(1+\beta)}{}_{2}F_{1}\left(\frac{1}{2}(1+\alpha+\beta),\frac{1}{2}(2+\alpha+\beta);1+\beta;-\frac{b^2}{\tau^2}\right)
\end{align}
Inverse Laplace Transforms:
\begin{align} \label{eq:inv-L-transforms}
\int_{\gamma-i\infty}^{\gamma+i\infty}\frac{d\tau}{2\pi i} e^{e\tau}(\tau+\sqrt{b^2+\tau^2})^\omega&=\theta(e)\omega b^{\omega}\frac{J_{-\omega}(b e)}{e}
\end{align}
Inverse Bessel Transform:
\begin{align}\label{eq:inv-F-transform}
\int_{0}^{\infty}db\,\, b^{1-\omega}J_{0}(b p)J_{\omega}(b e)&=\frac{\theta(e-p)}{\Gamma(\omega)}\frac{2^{1-\omega}e^{-\omega}}{(e^2-p^2)^{1-\omega}}
\end{align}

\section{Plus-Distributions over Vector Domains}\label{plus_distr_appendix}
Both the jet broadening and Higgs transverse momentum cross-sections involve convolutions in vector quantities. Expressing the renormalized functions directly in momentum space requires plus distributions that automatically perform the subtractions necessary to render convolutions finite. We give definitions for these plus distributions, to make transparent the renormalization scale dependence of the functions, and give useful identities for their manipulation.  For the purposes of this paper, we consider mostly one class of these distributions:
\begin{align}
\label{log_distr_def}\mathcal{L}_{n}^{\alpha}\Big(\mu,\vec{p};\lambda\Big)&=\frac{1}{2\pi\mu^2}\Bigg[\Big(\frac{\mu^2}{\vec{p}^{\,2}}\Big)^{1+\alpha}\ln^n\Big(\frac{\mu^2}{\vec{p}^{\,2}}\Big)\Bigg]_{+}^{\lambda}
\end{align}
Intuitively, these are distributions that render integrals over a vector domain convergent when weighted with a well-behaved function. This is accomplished by subtracting away an integral over a disc of radius $\lambda\mu$ (formally, $\mathcal{D}_{\lambda\mu}=\{\vec{p}:\,|\vec{p}|<\lambda\mu\}$) about the origin in the $\vec{p}$-space. Thus:
\begin{align}
\label{integrate_distr_to_boundary}\int_{\mathcal{D}_{\lambda\mu}}\frac{d^2\vec{q}}{(2\pi)^2}\mathcal{L}_{n}^{\alpha}\Big(\mu,\vec{q};\lambda\Big)=0.
\end{align}
From this class, there are important limits that appear frequently, and are related by derivatives. First we define the notation
\begin{eqnarray}
\label{ratio_distr_def}\mathcal{L}^{\alpha}\Big(\mu,\vec{p};\lambda\Big)&=&\mathcal{L}_{0}^{\alpha}\Big(\mu,\vec{p};\lambda\Big)\,,\\\nn
\label{simp_frac_distr_def}\mathcal{L}_{n}\Big(\mu,\vec{p};\lambda\Big)&=&\mathcal{L}_{n}^{0}\Big(\mu,\vec{p};\lambda\Big)\,. \nn
\end{eqnarray}
Then we have the relation
\beq
\mathcal{L}_{n}\Big(\mu,\vec{p};\lambda\Big)=\lim_{\beta\rightarrow 0}\frac{d^n}{d\beta^n}\mathcal{L}^{\beta}\Big(\mu,\vec{p};\lambda\Big)\,.
\eeq
Henceforth, we will also assume the notation that when the argument $\lambda$ is absent it  implies it is set equal to 1.

In what follows, we will give a definition based on dimensional regularization since it is extremely useful to perform multidimensional integrals. Any other intermediate regularization is also fine, since all divergent behavior cancels in integrals with well-behaved test functions, and so any such regulator will cancel after performing the integral. We have checked, for instance, that the use of a limit based definition with a mass regulator leads to the same expressions. 

\subsection{Definition in Dimensional Regularization}
One simple way to define a plus-distribution with vector arguments is dimensional regularization. This is especially appealing given the well-developed nature of the technology and its familiarity. Assume $g(\vec{k})$ is a reasonable test function, that is, has no singularities at the origin, and falls off at infinity fast enough. For physical applications, these criteria are easily met. The plus-distribution, $\Big[f(\mu,\vec{p})\Big]_{+}^{\lambda}$, for a function $f$ that has at most a power-like singularity at the origin and the boundary condition on the disc ${\mathcal D}_{\lambda\mu}$, is defined as:
\begin{multline}
\label{dim_reg_plus_function_def}\int \frac{d^2\vec{p}}{(2\pi)^2}g(\vec{p})\Big[f\Big(\mu,\vec{p}\Big)\Big]_{+}^{\lambda}=\lim_{\epsilon\rightarrow0^{+}}\,\mu^{-2\epsilon}\int \frac{d^{2+2\epsilon}\vec{p}}{(2\pi)^{2+2\epsilon}}g(\vec{p})\Bigg\{f\Big(\mu,\vec{p}\Big)-\mu^{2\epsilon}\mathbb{I}_{\vec{p}}\,B_{\epsilon}[f;\mu,\lambda]\Bigg\}\,.
\end{multline}
We use the notation:
\begin{align}\label{eq:boundary term}
B_{\epsilon}[f;\mu,\lambda]&=\mu^{-2\epsilon}\int_{\mathcal{D}_{\lambda\mu}}\frac{d^{2+2\epsilon}\vec{q}}{(2\pi)^{2+2\epsilon}}f\Big(\mu,\vec{q}\Big)\,,\\
B_{\epsilon}[f;\mu]&=B_{\epsilon}[f;\mu,1]\,.
\end{align}
Formally, all integrations are in two dimensions as long as the $[\cdot]_+$ symbols are used, though implicitly the integrations are in $2+2\epsilon$ dimensions where analytical continuation on $\epsilon$ is assumed as always. $\mu^{-2 \epsilon}$ that appears on right hand side of eqn. (\ref{dim_reg_plus_function_def}) is present only to formally control the dimensions\footnote{Any other momentum scale will equivalently serve the purpose.}. The use of $\mathbb{I}_{\vec{p}}$ makes the transition between these measures less cumbersome. Put simply, $\mathbb{I}_{\vec{p}}=(2\pi)^2\delta^{(2)}(\vec{p})$ or $(2\pi)^{2+2\epsilon}\delta^{(2+2\epsilon)}(\vec{p})$ depending on the context. Further, restriction to discs about the origin is not necessary, but merely convenient.
Any simply connected region containing the origin can serve to define the plus-distribution. Finally, the boundary term, $B_\epsilon$, by definition gives the following identity:
\begin{align}
\label{integrate_to_boundary}\int_{\mathcal{D}_{\lambda\mu}}\frac{d^2\vec{q}}{(2\pi)^2}\Big[f\Big(\mu,\vec{q}\Big)\Big]_{+}^{\lambda}=0\,.
\end{align}
when we replace the test function $g(\vec q) = {\mathcal D}_{\lambda \mu}(\vec q)$.
Two different boundary conditions for the plus-distributions are related by:
\begin{align}
\label{relating_boundary_conditions}\Big[f\Big(\mu,\vec{p}\Big)\Big]_{+}^{\lambda_1}-\Big[f\Big(\mu,\vec{p}\Big)\Big]_{+}^{\lambda_2}&=-\mathbb{I}_{\vec{p}}\,\Big(B_{\epsilon}[f;\mu,\lambda_1]-B_{\epsilon}[f;\mu,\lambda_2])\Big)\, ,
\end{align}

For the distributions in  \eqref{ratio_distr_def}, we can give explicit expressions for the boundary terms in dimensional regularization\footnote{Here we have pulled an explicit $1/2\pi$ compared to definition (\ref{eq:boundary term}) for convenience.}:
\begin{align}
\label{ratio_distr_with_BoundaryTerm_explicit}\mathcal{L}^{\alpha}\Big(\mu,\vec{p};\lambda\Big)&=\frac{1}{2\pi\mu^2}\Big(\frac{\mu^2}{\vec{p}^{\,2}}\Big)^{1+\alpha}- \frac{\mu^{2\epsilon} \mathbb{I}_{\vec{p}}}{2\pi}\,B_{\epsilon}[\mathcal{L}^{\alpha};\mu,\lambda]\,,\\
\label{ratio_distr_BoundaryTerm}B_{\epsilon}[\mathcal{L}^{\alpha};\mu,\lambda]&=\frac{\lambda^{-2\alpha+2\epsilon}}{(4\pi)^{1+\epsilon}\Gamma(1+\epsilon)(\epsilon-\alpha)}\,,\\
\label{log_distr_with_BoundaryTerm_explicit}\mathcal{L}_{n}\Big(\mu,\vec{p};\lambda\Big)&=\frac{1}{2\pi\vec{p}^{\,2}}\ln^n\Big(\frac{\mu^2}{\vec{p}^{\,2}}\Big)-\frac{\mu^{2\epsilon}\mathbb{I}_{\vec{p}}}{2\pi}\,B_{\epsilon}[\mathcal{L}_{n};\mu,\lambda]\,,\\
\label{log_distr_BoundaryTerm}B_{\epsilon}[\mathcal{L}_{n};\mu,\lambda]&=\frac{\Gamma(1+n)}{(4\pi)^{1+\epsilon}\Gamma(1+\epsilon)\epsilon}\lambda^{2\epsilon}\Bigg(\sum_{m=0}^{n}\frac{(-1)^{n-m}}{\Gamma(1+n-m)\epsilon^m}\ln^{n-m}(\lambda^2)\Bigg)\, ,
\end{align}
where on the right-hand side, $\epsilon$ appears explicitly and $\vec{p}$ is understood to be a vector in $2+2\epsilon$ dimensions. It is important to emphasize that the limit $\epsilon\rightarrow 0$ can only be taken in expressions where the limit is manifestly finite. In most applications this occurs only after evaluating integrals like \eqref{dim_reg_plus_function_def}, or for instance in the difference between two different boundary terms as in \eqref{relating_boundary_conditions}.

The power-law distribution $\mathcal{L}^{\alpha}(\mu,\vec{p};\lambda)$  becomes especially simple with the $\lambda=\infty$ boundary condition. In this case, the boundary term vanishes, and we have the identification
\begin{align}
\frac{1}{\mu^2}\Big[\Big(\frac{\mu^2}{\vec{p}^{\,2}}\Big)^{1+\alpha}\Big]_{+}^{\infty}=\frac{1}{\mu^2}\Big(\frac{\mu^2}{\vec{p}^{\,2}}\Big)^{1+\alpha}.\,
\end{align}
It is useful to give the expansion of this distribution as a power series in $\alpha$
\begin{align}
\frac{1}{2\pi\mu^2}\Big[\Big(\frac{\mu^2}{\vec{p}^{\,2}}\Big)^{1+\alpha}\Big]_{+}^{\infty}=-\frac{\mathbb{I}_{\vec{p}}}{8\pi^2\alpha}  +  \sum_{n=0}^{\infty}\frac{\alpha^{n}}{n!}\mathcal{L}_{n}\Big(\mu,\vec{p}\Big) \, .
\end{align}

\subsection{Rescaling}
We exhibit the scaling identities analogous to those defined for scalar domain distributions, for example see \cite{Ligeti:2008ac}.  For the distributions $\mathcal{L}^{\alpha}\Big(\mu,\vec{p}\Big)$ and $\mathcal{L}_{n}\Big(\mu,\vec{p}\Big)$, these are essentially the same as in the scalar case and take the following form in our notation:
\begin{align}
\mathcal{L}^{\alpha}\Big(\rho\mu,\vec{p}\Big)&=\rho^{2\alpha}\mathcal{L}^{\alpha}\Big(\mu,\vec{p}\Big)-\frac{\mathbb{I}_{\vec{p}}}{2\pi}\,\frac{\rho^{2\alpha}-1}{4\pi\alpha} \, ,\\
\mathcal{L}_{n}\Big(\rho\mu,\vec{p}\Big)&=\sum_{m=0}^{n} {}_{n}C_{m}\ln^m(\rho^2)\mathcal{L}_{n-m}\Big(\mu,\vec{p}\Big)-\frac{\mathbb{I}_{\vec{p}}}{2\pi}\,\frac{\ln^{n+1}( \rho^2)}{4\pi(n+1)} \, ,
\end{align}
where ${}_{n}C_{m} =\frac{\Gamma(1+n)}{\Gamma(1+m)\Gamma(1+n-m)}$.

\subsection{Derivatives and Integrals}
 A few important derivative identities are:
\begin{align}
\mu^2\frac{d}{d\mu^2}\mathcal{L}_{0}\Big(\mu,\vec{p}\Big)&=-\frac{\mathbb{I}_{\vec{p}}}{(2\pi)^2}\\
\mu^2\frac{d}{d\mu^2}\mathcal{L}_{n}\Big(\mu,\vec{p}\Big)&=n\mathcal{L}_{n-1}\Big(\mu,\vec{p}\Big) \, .
\end{align}

We can easily integrate the distribution $\mathcal{L}^{\alpha}(\mu,\vec{p})$ over a disc of radius $\lambda\mu$ by using the relations \eqref{integrate_to_boundary} and \eqref{relating_boundary_conditions}:
\begin{align}
\int_{\mathcal{D}_{\lambda\mu}}\frac{d^2\vec{p}}{(2\pi)^2}\mathcal{L}^{\alpha}\Big(\mu,\vec{p}\Big)&=\int_{\mathcal{D}_{\lambda\mu}}\frac{d^2\vec{p}}{(2\pi)^2}\Bigg\{\mathcal{L}^{\alpha}\Big(\mu,\vec{p};\lambda\Big)-\frac{\mathbb{I}_{\vec{p}}}{2\pi}\,\Big(B_{\epsilon}(\mathcal{L}^{\alpha};\mu)-B_{\epsilon}(\mathcal{L}^{\alpha};\mu;\lambda)\Big)\Bigg\}\\
&=-\frac{1}{8\pi^2\alpha}\Bigg(\lambda^{-2\alpha}-1\Bigg)
\end{align}
Similarly:
\begin{align}
\int_{\mathcal{D}_{\lambda\mu}}\frac{d^2\vec{p}}{(2\pi)^2}\mathcal{L}_{n}\Big(\mu,\vec{p}\Big)&=\frac{(-1)^{n}}{8\pi^2(n+1)}\ln^{n+1}(\lambda^2)  \, .
\end{align}

\subsection{Convolutions}
We have for the convolution of $\mathcal{L}^{\alpha}\Big(\mu,\vec{p}\Big)$ and $\mathcal{L}^{\beta}\Big(\mu,\vec{p}\Big)$:
\begin{align}
\label{master_convo}
 \int\frac{d^{2}\vec{p}}{(2\pi)^2}\mathcal{L}^{\alpha}\Big(\mu,\vec{k}-\vec{p}\Big)\mathcal{L}^{\beta}\Big(\mu, \vec{p}\Big)
 &= \frac{U(\alpha,\beta)}{2\pi}\mathcal{L}^{\alpha+\beta}\Big(\mu,\vec{k}\Big)-\frac{B[\mathcal{L}^{\alpha};\mu]}{2\pi}\mathcal{L}^{\beta}\Big(\mu,\vec{k}\Big)
-  \frac{B[\mathcal{L}^{\beta};\mu]}{2\pi} \mathcal{L}^{\alpha}\Big(\mu,\vec{k}\Big) \nn \\
& + \frac{\mathbb{I}_{\vec{k}}}{(2 \pi)^2}\,\Big\{U(\alpha,\beta)B[\mathcal{L}^{\alpha+\beta};\mu]-B[\mathcal{L}^{\alpha};\mu]B[\mathcal{L}^{\beta};\mu]\Big\}\\
U(\alpha,\beta)=&\frac{\Gamma(1+\alpha+\beta)}{4\pi\Gamma(1+\alpha)\Gamma(1+\beta)}\frac{\Gamma(-\alpha)\Gamma(-\beta)}{\Gamma(-\alpha-\beta)} \, .
\end{align}
Using the fact that logarithmic distributions are related to power-law ones via derivatives, one can use this convolution formula \eqref{master_convo} to derive identities for the convolution of logarithmic distributions. However, care must be taken in the limit of vanishing power parameters.


\section{Structure of divergence for an integral in sec. \ref{sec:lab-frame}} \label{app:B-decay}
We wish to find the divergence of the following integral from sec. \ref{sec:lab-frame},
\begin{align}\label{eq:left-lab-app}
I_{\text{left}} & =\frac{e^{\epsilon  \gamma _E}\mu ^{2\epsilon }}{(4\pi )^{\epsilon }}\nu ^{\eta }\int \frac{d^dk}{(2\pi )^d}\frac{\left|2k_3\right|{}^{-\eta }}{\left[k^2-\ell ^+ k^- + i0\right]\left[k^2-m^2+ i0\right]\left[-p^-k^+ + i0\right]} \, .
\end{align}
Consider the auxiliary integral
\begin{align}\label{eq:overlap}
I_{\text{left}}^{\rm (aux)}=\frac{e^{\epsilon  \gamma _E}\mu ^{2\epsilon }}{(4\pi )^{\epsilon }}\nu ^{\eta }\int \frac{d^dk}{(2\pi )^d}\frac{\left|2k_3\right|{}^{-\eta }}{\left[-\ell ^+ k^- + i0\right]\left[k^2-m^2 + i0 \right]\left[-p^-k^+ + i0\right]} \, .
\end{align}
For the difference, $I_{\rm left} - I_{\rm left}^{\rm (aux)}$, with some effort it can be shown that
\begin{align}
I_{\rm left} - I_{\rm left}^{\rm (aux)} & = \frac{- i m^2}{16\pi ^2p^-\left(\ell ^+\right)^2}\frac{e^{\epsilon  \gamma _E}\mu ^{2\epsilon }\nu ^{\eta }}{2^{\eta }}\Bigg [ \int _0^{\infty }dk_3 \, d k_t^2  \\
& \quad\quad \quad \frac{\left(k_t^2\right)^{-\epsilon }k_3^{-\eta }}{\left(k_t^2+m^2\right)\sqrt{k_3^2+k_t^2+m^2} \left[\sqrt{k_3^2+k_t^2+m^2} -k_3-\frac{m^2}{\ell ^+}-i 0\right]}+\eta\text{-finite} \Bigg ] \nn \, ,
\end{align}
where the integral shown is the only integral that has a rapidity divergence. The divergence appears  when $k_3 \to \infty$. So, as far as the rapidity divergences are concerned, it has the same asymptotic properties as the integral
\begin{align}
I_{\rm asymp} & = \frac{ i}{16\pi ^2p^-\ell ^+}\frac{e^{\epsilon  \gamma _E}\mu ^{2\epsilon }\nu ^{\eta }}{2^{\eta }}  \int _0^{\infty }dk_3 \, d k_t^2  
 \frac{\left(k_t^2\right)^{-\epsilon }k_3^{-\eta }}{\left(k_t^2+m^2\right)\sqrt{k_3^2+k_t^2+m^2} } \nn \\
 & = \frac{i}{16\pi ^2p^-\ell ^+}\frac{e^{\epsilon  \gamma _E}}{2^{\eta }}\frac{ \Gamma \left(\frac{1}{2}-\frac{\eta }{2}\right) }{\sqrt{\pi }}\frac{ \Gamma \left(\epsilon +\frac{\eta }{2}\right)}{\eta }\left(\frac{\mu ^2}{m^2}\right)^{\epsilon }\left(\frac{\nu }{m}\right)^{\eta } \, .
\end{align}
In other words, $I_{\rm left} - I_{\rm left}^{\rm (aux)} - I_{\rm asymp}$ does not have any rapidity divergence. Since, $I_{\rm left}^{\rm (aux)}$ is the same as $I_{\rm center}$ of eqn.~(\ref{eq:sym-integrals}) we conclude
\begin{align}
I_{\rm left} = \frac{-i}{16\pi ^2p^-\ell ^+}\left[\frac{ e^{\epsilon  \gamma _E} \Gamma (\epsilon )}{\eta }\left(\frac{\mu ^2}{m^2}\right)^{\epsilon }+ \eta\text{-finite} \right]\, .
\end{align}


\bibliographystyle{jhep}
\bibliography{../Master}

\providecommand{\href}[2]{#2}\begingroup\raggedright\begin{thebibliography}{10}

\bibitem{Sterman:1995fz}
G.~F. Sterman, {\it {Partons, Factorization And Resummation, TASI 95}},
  \href{http://arXiv.org/abs/hep-ph/9606312}{{\tt hep-ph/9606312}}.

\bibitem{Dokshitzer:1998kz}
Y.~L. Dokshitzer, A.~Lucenti, G.~Marchesini, and G.~P. Salam, {\it {On the
  {QCD} analysis of jet broadening}},  {\em JHEP} {\bf 01} (1998) 011,
  [\href{http://arXiv.org/abs/hep-ph/9801324}{{\tt hep-ph/9801324}}].

\bibitem{Catani:1992jc}
S.~Catani, G.~Turnock, and B.~R. Webber, {\it {Jet broadening measures in
  $e^{+} e^{-}$ annihilation}},  {\em Phys. Lett.} {\bf B295} (1992) 269--276.

\bibitem{Bauer:2002aj}
C.~W. Bauer, D.~Pirjol, and I.~W. Stewart, {\it {Factorization and endpoint
  singularities in heavy to light decays}},  {\em Phys.Rev.} {\bf D67} (2003)
  071502, [\href{http://arXiv.org/abs/hep-ph/0211069}{{\tt hep-ph/0211069}}].

\bibitem{Manohar:2006nz}
A.~V. Manohar and I.~W. Stewart, {\it {The Zero-Bin and Mode Factorization in
  Quantum Field Theory}},  {\em Phys.Rev.} {\bf D76} (2007) 074002,
  [\href{http://arXiv.org/abs/hep-ph/0605001}{{\tt hep-ph/0605001}}].

\bibitem{Collins:1992tv}
J.~C. Collins and F.~Tkachov, {\it {Breakdown Of Dimensional Regularization In
  The Sudakov Problem}},  {\em Phys.Lett.} {\bf B294} (1992) 403--411,
  [\href{http://arXiv.org/abs/hep-ph/9208209}{{\tt hep-ph/9208209}}].

\bibitem{Collins:1981uk}
J.~C. Collins and D.~E. Soper, {\it {Back-To-Back Jets in QCD}},  {\em Nucl.
  Phys.} {\bf B193} (1981) 381.

\bibitem{Collins:2008ht}
J.~Collins, {\it {Rapidity divergences and valid definitions of parton
  densities}},  {\em PoS} {\bf LC2008} (2008) 028,
  [\href{http://arXiv.org/abs/0808.2665}{{\tt arXiv:0808.2665}}].

\bibitem{Grinstein:1997gv}
B.~Grinstein and I.~Z. Rothstein, {\it {Effective field theory and matching in
  nonrelativistic gauge theories}},  {\em Phys.Rev.} {\bf D57} (1998) 78--82,
  [\href{http://arXiv.org/abs/hep-ph/9703298}{{\tt hep-ph/9703298}}].

\bibitem{Bauer:2000ew}
C.~W. Bauer, S.~Fleming, and M.~E. Luke, {\it {Summing Sudakov logarithms in $B
  \to X_s\gamma$ in effective field theory}},  {\em Phys. Rev. D} {\bf 63}
  (2000) 014006, [\href{http://arXiv.org/abs/hep-ph/0005275}{{\tt
  hep-ph/0005275}}].

\bibitem{Bauer:2000yr}
C.~W. Bauer, S.~Fleming, D.~Pirjol, and I.~W. Stewart, {\it An effective field
  theory for collinear and soft gluons: Heavy to light decays},  {\em Phys.
  Rev. D} {\bf 63} (2001) 114020,
  [\href{http://arXiv.org/abs/hep-ph/0011336}{{\tt hep-ph/0011336}}].

\bibitem{Bauer:2001yt}
C.~W. Bauer, D.~Pirjol, and I.~W. Stewart, {\it Soft-collinear factorization in
  effective field theory},  {\em Phys. Rev. D} {\bf 65} (2002) 054022,
  [\href{http://arXiv.org/abs/hep-ph/0109045}{{\tt hep-ph/0109045}}].

\bibitem{Bauer:2003mga}
C.~W. Bauer, D.~Pirjol, and I.~W. Stewart, {\it {On power suppressed operators
  and gauge invariance in SCET}},  {\em Phys. Rev.} {\bf D68} (2003) 034021,
  [\href{http://arXiv.org/abs/hep-ph/0303156}{{\tt hep-ph/0303156}}].

\bibitem{Luke:1999kz}
M.~E. Luke, A.~V. Manohar, and I.~Z. Rothstein, {\it {Renormalization group
  scaling in nonrelativistic QCD}},  {\em Phys.Rev.} {\bf D61} (2000) 074025,
  [\href{http://arXiv.org/abs/hep-ph/9910209}{{\tt hep-ph/9910209}}].

\bibitem{Rothstein:2003mp}
I.~Z. Rothstein, {\it {TASI lectures on effective field theories}},
  \href{http://arXiv.org/abs/hep-ph/0308266}{{\tt hep-ph/0308266}}.

\bibitem{Lee:2006nr}
C.~Lee and G.~F. Sterman, {\it {Momentum Flow Correlations from Event Shapes:
  Factorized Soft Gluons and Soft-Collinear Effective Theory}},  {\em
  Phys.Rev.} {\bf D75} (2007) 014022,
  [\href{http://arXiv.org/abs/hep-ph/0611061}{{\tt hep-ph/0611061}}].

\bibitem{Idilbi:2007yi}
A.~Idilbi and T.~Mehen, {\it {Demonstration of the equivalence of soft and
  zero-bin subtractions}},  {\em Phys.Rev.} {\bf D76} (2007) 094015,
  [\href{http://arXiv.org/abs/0707.1101}{{\tt arXiv:0707.1101}}].

\bibitem{Idilbi:2007ff}
A.~Idilbi and T.~Mehen, {\it {On the equivalence of soft and zero-bin
  subtractions}},  {\em Phys.Rev.} {\bf D75} (2007) 114017,
  [\href{http://arXiv.org/abs/hep-ph/0702022}{{\tt hep-ph/0702022}}].

\bibitem{Collins:1989bt}
J.~C. Collins, {\it {Sudakov form-factors}},  {\em Adv.Ser.Direct.High Energy
  Phys.} {\bf 5} (1989) 573--614,
  [\href{http://arXiv.org/abs/hep-ph/0312336}{{\tt hep-ph/0312336}}].

\bibitem{Jain:2011xz}
A.~Jain, M.~Procura, and W.~J. Waalewijn, {\it {Parton Fragmentation within an
  Identified Jet at NNLL}},  {\em JHEP} {\bf 1105} (2011) 035,
  [\href{http://arXiv.org/abs/1101.4953}{{\tt arXiv:1101.4953}}].

\bibitem{Beneke:2003pa}
M.~Beneke and T.~Feldmann, {\it {Factorization of heavy to light form-factors
  in soft collinear effective theory}},  {\em Nucl.Phys.} {\bf B685} (2004)
  249--296, [\href{http://arXiv.org/abs/hep-ph/0311335}{{\tt hep-ph/0311335}}].

\bibitem{Manohar:2002fd}
A.~V. Manohar, T.~Mehen, D.~Pirjol, and I.~W. Stewart, {\it {Reparameterization
  invariance for collinear operators}},  {\em Phys.Lett.} {\bf B539} (2002)
  59--66, [\href{http://arXiv.org/abs/hep-ph/0204229}{{\tt hep-ph/0204229}}].

\bibitem{Chiu:2011qc}
J.-y. Chiu, A.~Jain, D.~Neill, and I.~Z. Rothstein, {\it {The Rapidity
  Renormalization Group}},  \href{http://arXiv.org/abs/1104.0881}{{\tt
  arXiv:1104.0881}}.

\bibitem{Becher:2003qh}
T.~Becher, R.~J. Hill, and M.~Neubert, {\it {Soft collinear messengers: A New
  mode in soft collinear effective theory}},  {\em Phys.Rev.} {\bf D69} (2004)
  054017, [\href{http://arXiv.org/abs/hep-ph/0308122}{{\tt hep-ph/0308122}}].

\bibitem{Korchemsky}
G.~P. Korchemsky, {\it {Sudakov Form Factor in QCD}},  {\em Phys. Lett.} {\bf
  B220} (1989) 629.

\bibitem{Bauer:2010cc}
C.~W. Bauer, B.~O. Lange, and G.~Ovanesyan, {\it {On Glauber modes in
  Soft-Collinear Effective Theory}},  {\em JHEP} {\bf 1107} (2011) 077,
  [\href{http://arXiv.org/abs/1010.1027}{{\tt arXiv:1010.1027}}].

\bibitem{Collins:1988ig}
J.~C. Collins, D.~E. Soper, and G.~F. Sterman, {\it {Soft Gluons and
  Factorization}},  {\em Nucl.Phys.} {\bf B308} (1988) 833.

\bibitem{Bodwin:1984hc}
G.~T. Bodwin, {\it {Factorization of the Drell-Yan Cross-Section in
  Perturbation Theory}},  {\em Phys.Rev.} {\bf D31} (1985) 2616. Revised
  version.

\bibitem{Idilbi:2008vm}
A.~Idilbi and A.~Majumder, {\it {Extending Soft-Collinear-Effective-Theory to
  describe hard jets in dense QCD media}},  {\em Phys.Rev.} {\bf D80} (2009)
  054022, [\href{http://arXiv.org/abs/0808.1087}{{\tt arXiv:0808.1087}}].

\bibitem{D'Eramo:2010xk}
F.~D'Eramo, H.~Liu, and K.~Rajagopal, {\it {Jet Quenching Parameter via Soft
  Collinear Effective Theory (SCET)}},  {\em Int.J.Mod.Phys.} {\bf E20} (2011)
  1610--1615, [\href{http://arXiv.org/abs/1010.0890}{{\tt arXiv:1010.0890}}].

\bibitem{Smirnov:1990rz}
V.~A. Smirnov, {\it {Asymptotic expansions in limits of large momenta and
  masses}},  {\em Commun.Math.Phys.} {\bf 134} (1990) 109--137.

\bibitem{Chiu:2009yx}
J.-y. Chiu, A.~Fuhrer, A.~H. Hoang, R.~Kelley, and A.~V. Manohar, {\it
  {Soft-Collinear Factorization and Zero-Bin Subtractions}},  {\em Phys.Rev.}
  {\bf D79} (2009) 053007, [\href{http://arXiv.org/abs/0901.1332}{{\tt
  arXiv:0901.1332}}].

\bibitem{Chiu:2009mg}
J.-y. Chiu, A.~Fuhrer, R.~Kelley, and A.~V. Manohar, {\it {Factorization
  Structure of Gauge Theory Amplitudes and Application to Hard Scattering
  Processes at the LHC}},  {\em Phys.Rev.} {\bf D80} (2009) 094013,
  [\href{http://arXiv.org/abs/0909.0012}{{\tt arXiv:0909.0012}}].

\bibitem{Gatheral:1983cz}
J.~G.~M. Gatheral, {\it {Exponentiation Of Eikonal Cross-Sections In Nonabelian
  Gauge Theories}},  {\em Phys. Lett.} {\bf B133} (1983) 90.

\bibitem{Frenkel:1984pz}
J.~Frenkel and J.~C. Taylor, {\it {Nonabelian Eikonal Exponentiation}},  {\em
  Nucl. Phys.} {\bf B246} (1984) 231.

\bibitem{Laenen:2008gt}
E.~Laenen, G.~Stavenga, and C.~D. White, {\it {Path integral approach to
  eikonal and next-to-eikonal exponentiation}},  {\em JHEP} {\bf 03} (2009)
  054, [\href{http://arXiv.org/abs/0811.2067}{{\tt arXiv:0811.2067}}].

\bibitem{Korchemsky:1987wg}
G.~P. Korchemsky and A.~V. Radyushkin, {\it {Renormalization of the Wilson
  Loops Beyond the Leading Order}},  {\em Nucl. Phys.} {\bf B283} (1987)
  342--364.

\bibitem{Manohar:2003vb}
A.~V. Manohar, {\it {Deep inelastic scattering as x --$>$ 1 using
  soft-collinear effective theory}},  {\em Phys. Rev.} {\bf D68} (2003) 114019,
  [\href{http://arXiv.org/abs/hep-ph/0309176}{{\tt hep-ph/0309176}}].

\bibitem{Arnesen:2008fb}
C.~Arnesen, I.~Z. Rothstein, and J.~Zupan, {\it {Smoking Guns for On-Shell New
  Physics at the LHC}},  {\em Phys.Rev.Lett.} {\bf 103} (2009) 151801,
  [\href{http://arXiv.org/abs/0809.1429}{{\tt arXiv:0809.1429}}].

\bibitem{Becher:2010tm}
T.~Becher and M.~Neubert, {\it {Drell-Yan production at small $q_T$, transverse
  parton distributions and the collinear anomaly}},  {\em Eur. Phys. J.} {\bf
  C71} (2011) 1665, [\href{http://arXiv.org/abs/1007.4005}{{\tt
  arXiv:1007.4005}}].

\bibitem{Mantry:2009qz}
S.~Mantry and F.~Petriello, {\it {Factorization and Resummation of Higgs Boson
  Differential Distributions in Soft-Collinear Effective Theory}},  {\em
  Phys.Rev.} {\bf D81} (2010) 093007,
  [\href{http://arXiv.org/abs/0911.4135}{{\tt arXiv:0911.4135}}].

\bibitem{Gao:2005iu}
Y.~Gao, C.~S. Li, and J.~J. Liu, {\it {Transverse momentum resummation for
  Higgs production in soft-collinear effective theory}},  {\em Phys.Rev.} {\bf
  D72} (2005) 114020, [\href{http://arXiv.org/abs/hep-ph/0501229}{{\tt
  hep-ph/0501229}}].

\bibitem{Idilbi:2005er}
A.~Idilbi, X.-d. Ji, and F.~Yuan, {\it {Transverse momentum distribution
  through soft-gluon resummation in effective field theory}},  {\em Phys.Lett.}
  {\bf B625} (2005) 253--263, [\href{http://arXiv.org/abs/hep-ph/0507196}{{\tt
  hep-ph/0507196}}].

\bibitem{Collins:1984kg}
J.~C. Collins, D.~E. Soper, and G.~F. Sterman, {\it {Transverse Momentum
  Distribution in Drell-Yan Pair and W and Z Boson Production}},  {\em Nucl.
  Phys.} {\bf B250} (1985) 199.

\bibitem{PhysRevD.38.3475}
I.~Hinchliffe and S.~F. Novaes, {\it Transverse-momentum distribution of higgs
  bosons at the superconducting super collider},  {\em Phys. Rev. D} {\bf 38}
  (Dec, 1988) 3475--3480.

\bibitem{PhysRevD.44.1415}
R.~P. Kauffman, {\it Higgs-boson $pt$ in gluon fusion},  {\em Phys. Rev. D}
  {\bf 44} (Sep, 1991) 1415--1425.

\bibitem{deFlorian:2000pr}
D.~de~Florian and M.~Grazzini, {\it {Next-to-next-to-leading logarithmic
  corrections at small transverse momentum in hadronic collisions}},  {\em
  Phys. Rev. Lett.} {\bf 85} (2000) 4678--4681,
  [\href{http://arXiv.org/abs/hep-ph/0008152}{{\tt hep-ph/0008152}}].

\bibitem{deFlorian:2001zd}
D.~de~Florian and M.~Grazzini, {\it {The structure of large logarithmic
  corrections at small transverse momentum in hadronic collisions}},  {\em
  Nucl. Phys.} {\bf B616} (2001) 247--285,
  [\href{http://arXiv.org/abs/hep-ph/0108273}{{\tt hep-ph/0108273}}].

\bibitem{Catani:2010pd}
S.~Catani and M.~Grazzini, {\it {QCD transverse-momentum resummation in gluon
  fusion processes}},  {\em Nucl. Phys.} {\bf B845} (2011) 297--323,
  [\href{http://arXiv.org/abs/1011.3918}{{\tt arXiv:1011.3918}}].

\bibitem{ATLAS}
{\bf ATLAS} Collaboration, {\it {Combination of Higgs Boson Searches with up to
  4.9~fb$^{-1}$ of $pp$ Collision Data Taken at $\sqrt{s} = 7$~TeV with the
  ATLAS Experiment at the LHC}},  {\em ATLAS-CONF-2011-163} (2011).

\bibitem{CMS}
{\bf CMS} Collaboration, {\it {Combination of CMS searches for a Standard Model
  Higgs boson}},  {\em CMS PAS HIG-11-032} (2011).

\bibitem{Inami:1982xt}
T.~Inami, T.~Kubota, and Y.~Okada, {\it {Effective Gauge Theory And The Effect
  Of Heavy Quarks In Higgs Boson Decays}},  {\em Z.Phys.} {\bf C18} (1983) 69.

\bibitem{Djouadi:1991tka}
A.~Djouadi, M.~Spira, and P.~Zerwas, {\it {Production of Higgs bosons in proton
  colliders: QCD corrections}},  {\em Phys.Lett.} {\bf B264} (1991) 440--446.

\bibitem{Bauer:2002nz}
C.~W. Bauer, S.~Fleming, D.~Pirjol, I.~Z. Rothstein, and I.~W. Stewart, {\it
  {Hard scattering factorization from effective field theory}},  {\em Phys.
  Rev.} {\bf D66} (2002) 014017,
  [\href{http://arXiv.org/abs/hep-ph/0202088}{{\tt hep-ph/0202088}}].

\bibitem{Cherednikov:2007tw}
I.~O. Cherednikov and N.~G. Stefanis, {\it {Renormalization, Wilson lines, and
  transverse-momentum dependent parton distribution functions}},  {\em Phys.
  Rev.} {\bf D77} (2008) 094001, [\href{http://arXiv.org/abs/0710.1955}{{\tt
  arXiv:0710.1955}}].

\bibitem{Cherednikov:2008ua}
I.~Cherednikov and N.~Stefanis, {\it {Wilson lines and transverse-momentum
  dependent parton distribution functions: A Renormalization-group analysis}},
  {\em Nucl.Phys.} {\bf B802} (2008) 146--179,
  [\href{http://arXiv.org/abs/0802.2821}{{\tt arXiv:0802.2821}}].

\bibitem{Cherednikov:2009wk}
I.~Cherednikov and N.~Stefanis, {\it {Renormalization-group properties of
  transverse-momentum dependent parton distribution functions in the light-cone
  gauge with the Mandelstam-Leibbrandt prescription}},  {\em Phys.Rev.} {\bf
  D80} (2009) 054008, [\href{http://arXiv.org/abs/0904.2727}{{\tt
  arXiv:0904.2727}}].

\bibitem{Collins:2003fm}
J.~C. Collins, {\it {What exactly is a parton density?}},  {\em Acta Phys.
  Polon.} {\bf B34} (2003) 3103,
  [\href{http://arXiv.org/abs/hep-ph/0304122}{{\tt hep-ph/0304122}}].

\bibitem{Ji:2004wu}
X.-d. Ji, J.-p. Ma, and F.~Yuan, {\it {QCD factorization for semi-inclusive
  deep-inelastic scattering at low transverse momentum}},  {\em Phys.Rev.} {\bf
  D71} (2005) 034005, [\href{http://arXiv.org/abs/hep-ph/0404183}{{\tt
  hep-ph/0404183}}].

\bibitem{Hautmann:2007uw}
F.~Hautmann, {\it {Endpoint singularities in unintegrated parton
  distributions}},  {\em Phys.Lett.} {\bf B655} (2007) 26--31,
  [\href{http://arXiv.org/abs/hep-ph/0702196}{{\tt hep-ph/0702196}}].

\bibitem{Meissner:2008xs}
S.~Meissner, A.~Metz, and M.~Schlegel, {\it {Generalized transverse momentum
  dependent parton distributions of the nucleon}},
  \href{http://arXiv.org/abs/0807.1154}{{\tt arXiv:0807.1154}}.

\bibitem{Pasquini:2008ax}
B.~Pasquini, S.~Cazzaniga, and S.~Boffi, {\it {Transverse momentum dependent
  parton distributions in a light-cone quark model}},  {\em Phys.Rev.} {\bf
  D78} (2008) 034025, [\href{http://arXiv.org/abs/0806.2298}{{\tt
  arXiv:0806.2298}}].

\bibitem{Jain:2011iu}
A.~Jain, M.~Procura, and W.~J. Waalewijn, {\it {Fully-Unintegrated Parton
  Distribution and Fragmentation Functions at Perturbative $k_T$}},
  \href{http://arXiv.org/abs/1110.0839}{{\tt arXiv:1110.0839}}.

\bibitem{Stewart:2009yx}
I.~W. Stewart, F.~J. Tackmann, and W.~J. Waalewijn, {\it {Factorization at the
  LHC: From PDFs to Initial State Jets}},  {\em Phys.Rev.} {\bf D81} (2010)
  094035, [\href{http://arXiv.org/abs/0910.0467}{{\tt arXiv:0910.0467}}].

\bibitem{Frixione:1998dw}
S.~Frixione, P.~Nason, and G.~Ridolfi, {\it {Problems in the resummation of
  soft gluon effects in the transverse momentum distributions of massive vector
  bosons in hadronic collisions}},  {\em Nucl.Phys.} {\bf B542} (1999)
  311--328, [\href{http://arXiv.org/abs/hep-ph/9809367}{{\tt hep-ph/9809367}}].

\bibitem{Harlander:2001is}
R.~V. Harlander and W.~B. Kilgore, {\it {Soft and virtual corrections to proton
  proton ---> H + x at NNLO}},  {\em Phys.Rev.} {\bf D64} (2001) 013015,
  [\href{http://arXiv.org/abs/hep-ph/0102241}{{\tt hep-ph/0102241}}].

\bibitem{Kauffman:1991jt}
R.~P. Kauffman, {\it {Higgs boson p(T) in gluon fusion}},  {\em Phys.Rev.} {\bf
  D44} (1991) 1415--1425.

\bibitem{Aybat:2011zv}
S.~Aybat and T.~C. Rogers, {\it {TMD Parton Distribution and Fragmentation
  Functions with QCD Evolution}},  {\em Phys.Rev.} {\bf D83} (2011) 114042,
  [\href{http://arXiv.org/abs/1101.5057}{{\tt arXiv:1101.5057}}].

\bibitem{Becher:2011dz}
T.~Becher and G.~Bell, {\it {Analytic Regularization in Soft-Collinear
  Effective Theory}},  \href{http://arXiv.org/abs/1112.3907}{{\tt
  arXiv:1112.3907}}.

\bibitem{Collins:2011ca}
J.~Collins, {\it {New definition of TMD parton densities}},
  \href{http://arXiv.org/abs/1107.4123}{{\tt arXiv:1107.4123}}.

\bibitem{Dixon:2008gr}
L.~J. Dixon, L.~Magnea, and G.~F. Sterman, {\it {Universal structure of
  subleading infrared poles in gauge theory amplitudes}},  {\em JHEP} {\bf
  0808} (2008) 022, [\href{http://arXiv.org/abs/0805.3515}{{\tt
  arXiv:0805.3515}}].

\bibitem{Bozzi:2007pn}
G.~Bozzi, S.~Catani, D.~de~Florian, and M.~Grazzini, {\it {Higgs boson
  production at the LHC: Transverse-momentum resummation and rapidity
  dependence}},  {\em Nucl.Phys.} {\bf B791} (2008) 1--19,
  [\href{http://arXiv.org/abs/0705.3887}{{\tt arXiv:0705.3887}}]. This paper is
  dedicated to the memory of Jiro Kodaira, great friend and distinguished
  colleague.

\bibitem{Collins:2011zzd}
J.~Collins, {\it {Foundations of perturbative QCD}}, .

\bibitem{Stewart:SCET2009}
I.~W. Stewart and I.~Z. Rothstein, {\it {Glauber Gluons in SCET}},  {\em SCET
  workshop 2010} (2010).

\bibitem{GarciaEchevarria:2011rb}
M.~Garcia-Echevarria, A.~Idilbi, and I.~Scimemi, {\it {Factorization Theorem
  For Drell-Yan At Low q$_T$ And Transverse Momentum Distributions
  On-The-Light-Cone}},  \href{http://arXiv.org/abs/1111.4996}{{\tt
  arXiv:1111.4996}}.

\bibitem{Dasgupta:2001sh}
M.~Dasgupta and G.~Salam, {\it {Resummation of nonglobal QCD observables}},
  {\em Phys.Lett.} {\bf B512} (2001) 323--330,
  [\href{http://arXiv.org/abs/hep-ph/0104277}{{\tt hep-ph/0104277}}].

\bibitem{Kelley:2011ng}
R.~Kelley, M.~D. Schwartz, R.~M. Schabinger, and H.~X. Zhu, {\it {The two-loop
  hemisphere soft function}},  {\em Phys.Rev.} {\bf D84} (2011) 045022,
  [\href{http://arXiv.org/abs/1105.3676}{{\tt arXiv:1105.3676}}].

\bibitem{Hornig:2011iu}
A.~Hornig, C.~Lee, I.~W. Stewart, J.~R. Walsh, and S.~Zuberi, {\it {Non-global
  Structure of the $O({\alpha}_s^2)$ Dijet Soft Function}},  {\em JHEP} {\bf
  1108} (2011) 054, [\href{http://arXiv.org/abs/1105.4628}{{\tt
  arXiv:1105.4628}}].

\bibitem{Kluth:2006bw}
S.~Kluth, {\it {Tests of Quantum Chromo Dynamics at e+ e- Colliders}},  {\em
  Rept.Prog.Phys.} {\bf 69} (2006) 1771--1846,
  [\href{http://arXiv.org/abs/hep-ex/0603011}{{\tt hep-ex/0603011}}].

\bibitem{Berger:2003pk}
C.~F. Berger and G.~F. Sterman, {\it {Scaling rule for nonperturbative
  radiation in a class of event shapes}},  {\em JHEP} {\bf 0309} (2003) 058,
  [\href{http://arXiv.org/abs/hep-ph/0307394}{{\tt hep-ph/0307394}}].

\bibitem{Farhi:1977sg}
E.~Farhi, {\it {A QCD Test for Jets}},  {\em Phys. Rev. Lett.} {\bf 39} (1977)
  1587--1588.

\bibitem{Catani:1989ne}
S.~Catani and L.~Trentadue, {\it {Resummation of the QCD Perturbative Series
  for Hard Processes}},  {\em Nucl.Phys.} {\bf B327} (1989) 323.

\bibitem{Hornig:2009vb}
A.~Hornig, C.~Lee, and G.~Ovanesyan, {\it {Effective Predictions of Event
  Shapes: Factorized, Resummed, and Gapped Angularity Distributions}},  {\em
  JHEP} {\bf 0905} (2009) 122, [\href{http://arXiv.org/abs/0901.3780}{{\tt
  arXiv:0901.3780}}].

\bibitem{Abbate:2010xh}
R.~Abbate, M.~Fickinger, A.~H. Hoang, V.~Mateu, and I.~W. Stewart, {\it {Thrust
  at $N^3LL$ with Power Corrections and a Precision Global Fit for
  alphas(mZ)}},  \href{http://arXiv.org/abs/1006.3080}{{\tt arXiv:1006.3080}}.

\bibitem{Fleming:2007qr}
S.~Fleming, A.~H. Hoang, S.~Mantry, and I.~W. Stewart, {\it {Jets from massive
  unstable particles: Top-mass determination}},  {\em Phys.Rev.} {\bf D77}
  (2008) 074010, [\href{http://arXiv.org/abs/hep-ph/0703207}{{\tt
  hep-ph/0703207}}].

\bibitem{Rakow:1981qn}
P.~E. Rakow and B.~Webber, {\it {Transverse Momentum Moments Of Hadron
  Distributions In Qcd Jets}},  {\em Nucl.Phys.} {\bf B191} (1981) 63.

\bibitem{Bauer:2003di}
C.~W. Bauer, C.~Lee, A.~V. Manohar, and M.~B. Wise, {\it {Enhanced
  nonperturbative effects in Z decays to hadrons}},  {\em Phys. Rev. D} {\bf
  70} (2004) 034014, [\href{http://arXiv.org/abs/hep-ph/0309278}{{\tt
  hep-ph/0309278}}].

\bibitem{Achard:2004sv}
{\bf L3 Collaboration} Collaboration, P.~Achard {\em et.~al.}, {\it {Studies of
  hadronic event structure in $e^{+} e^{-}$ annihilation from 30-GeV to 209-GeV
  with the L3 detector}},  {\em Phys.Rept.} {\bf 399} (2004) 71--174,
  [\href{http://arXiv.org/abs/hep-ex/0406049}{{\tt hep-ex/0406049}}].

\bibitem{Becher:2011pf}
T.~Becher, G.~Bell, and M.~Neubert, {\it {Factorization and Resummation for Jet
  Broadening}},  \href{http://arXiv.org/abs/1104.4108}{{\tt arXiv:1104.4108}}.

\bibitem{Chiu:2007dg}
J.-y. Chiu, F.~Golf, R.~Kelley, and A.~V. Manohar, {\it {Electroweak
  Corrections in High Energy Processes using Effective Field Theory}},  {\em
  Phys.Rev.} {\bf D77} (2008) 053004,
  [\href{http://arXiv.org/abs/0712.0396}{{\tt arXiv:0712.0396}}].

\bibitem{Geshkenbein:1982zs}
B.~V. Geshkenbein and M.~V. Terentev, {\it {The Enhanced Power Correction To
  The Asymptotics Of The Pion Form-Factor}},  {\em Phys. Lett.} {\bf B117}
  (1982) 243--246.

\bibitem{Chernyak:1983ej}
V.~L. Chernyak and A.~R. Zhitnitsky, {\it {Asymptotic Behavior of Exclusive
  Processes in QCD}},  {\em Phys. Rept.} {\bf 112} (1984) 173.

\bibitem{Akhoury:1993uw}
R.~Akhoury, G.~F. Sterman, and Y.~P. Yao, {\it {Exclusive semileptonic decays
  of B mesons into light mesons}},  {\em Phys. Rev.} {\bf D50} (1994) 358--372.

\bibitem{Bpipi}
C.~W. Bauer, D.~Pirjol, I.~Z. Rothstein, and I.~W. Stewart, {\it {B ---$>$:
  M(1) M(2): Factorization, charming penguins, strong phases, and
  polarization}},  {\em Phys.Rev.} {\bf D70} (2004) 054015,
  [\href{http://arXiv.org/abs/hep-ph/0401188}{{\tt hep-ph/0401188}}].

\bibitem{BKpi}
C.~W. Bauer, I.~Z. Rothstein, and I.~W. Stewart, {\it {SCET analysis of B --$>$
  K pi, B --$>$ K anti-K, and B --$>$ pi pi decays}},  {\em Phys. Rev.} {\bf
  D74} (2006) 034010, [\href{http://arXiv.org/abs/hep-ph/0510241}{{\tt
  hep-ph/0510241}}].

\bibitem{Lepage:1980fj}
G.~Lepage and S.~J. Brodsky, {\it {Exclusive Processes in Perturbative Quantum
  Chromodynamics}},  {\em Phys.Rev.} {\bf D22} (1980) 2157.

\bibitem{Korchemsky:1992xv}
G.~Korchemsky and G.~Marchesini, {\it {Structure function for large x and
  renormalization of Wilson loop}},  {\em Nucl.Phys.} {\bf B406} (1993)
  225--258, [\href{http://arXiv.org/abs/hep-ph/9210281}{{\tt hep-ph/9210281}}].

\bibitem{Hoang:2001rr}
A.~H. Hoang, A.~V. Manohar, and I.~W. Stewart, {\it {The Running Coulomb
  Potential and Lamb Shift in QCD}},  {\em Phys. Rev.} {\bf D64} (2001) 014033,
  [\href{http://arXiv.org/abs/hep-ph/0102257}{{\tt hep-ph/0102257}}].

\bibitem{Ligeti:2008ac}
Z.~Ligeti, I.~W. Stewart, and F.~J. Tackmann, {\it Treating the $b$ quark
  distribution function with reliable uncertainties},  {\em Phys. Rev. D} {\bf
  78} (Dec, 2008) 114014.

\end{thebibliography}\endgroup
\end{document}